\documentclass[]{aa}

\usepackage[latin1]{inputenc}

\usepackage{lscape}
\usepackage{graphicx}
\usepackage{epstopdf}
\usepackage{epsfig}
\usepackage{natbib}
\usepackage{longtable}
\usepackage{supertabular}
\usepackage{booktabs}
\bibpunct{(}{)}{;}{a}{}{,}
\usepackage{txfonts}
\begin{document}

   \title{The X-ray lightcurve of Sgr A* over the past 150 years inferred from Fe-K$\alpha$ line reverberation 
in Galactic Centre molecular clouds}
   \titlerunning{The X-ray lightcurve of Sgr A* over the past 150 years}
   \authorrunning{R. Capelli et al.}

   \author{R. Capelli\inst{1}, R.S. Warwick\inst{2},  D. Porquet\inst{3}, S. Gillessen\inst{1} and P. 
Predehl\inst{1}}

   \offprints{R. Capelli,\\ e-mail: capelli@mpe.mpg.de}

   \institute{Max-Planck-Institut f\"ur extraterrestrische Physik, Postfach 1312, Giessenbachstr., 
   85741 Garching, Germany
\and
	      Department of Physics and Astronomy, University of 
	      Leicester, Leicester LE1 7RH, UK
\and
	      Observatoire Astronomique de Strasbourg, Universit\'e de Strasbourg, CNRS, UMR 7550, 11 rue de
 l'Universit\'e, 67000 Strasbourg, France}

   \date{Received ...; ...}

\abstract
{The spatial distribution and variability of Fe-K$\alpha$ emission from 
molecular clouds in the Galactic Centre region may provide an  important key to the
understanding of the recent history of Sgr A*.  A very plausible interpretation is that
this variability represents an echo in the reflected radiation from the clouds of a
past episode of high activity in Sgr A*.}
{We examine the temporal and spectral properties of nine Fe-K$\alpha$ 
bright molecular clouds within about 30 pc of Sgr A*, in order to understand and constrain
the primary energising  source of the Fe fluorescence.}
{The variability of the Fe-K$\alpha$ line at 6.4-keV was
investigated by spectrally fitting the data derived from the EPIC MOS cameras, after 
subtracting a modelled background.  We have also studied the reflection imprints in 
time-averaged pn-spectra of each cloud, by measuring the equivalent width (EW) of the 6.4 keV line
and the optical depth of the Fe-K absorption edge at 7.1 keV.}
{Significant Fe-K$\alpha$ variability was detected, with a spatial and temporal pattern 
consistent with that reported in previous studies. 
The main breakthrough that sets our paper apart from earlier contributions
on this topic is the direct measurement of the column density and the Fe abundance
of the MCs in our sample.
%All the spectra were characterised
%by a high EW of the Fe-K$\alpha$ line and the presence of absorption at the Fe-K edge. 
We used the EW measurements
to infer the average Fe abundance within the clouds to be 1.6$\pm$0.1 times solar.
The cloud column densities derived from the spectral analysis were typically of the
order of 10$^{23}$ cm$^{-2}$, which is significantly higher than previous estimates.
This in turn has a significant impact on the
inferred geometry and time delays within the cloud system.
%The measured cloud parameters were used to set constraints on
%the past activity of Sgr A* and to investigate whether a contribution to the
%Fe fluorescence by cosmic-ray bombardment is plausible.
}  
{
Past X-ray activity of Sgr A* is the most likely source of ionisation within the
molecular clouds in the innermost 30 pc of the Galaxy. In this scenario, the X-ray
luminosity required to excite these reflection nebulae is of the order of 
10$^{37}$-10$^{38}$ erg s$^{-1}$, significantly lower than that estimated for
the Sgr B2 molecular cloud. Moreover, the inferred Sgr A* lightcurve over the past
150 years shows a long-term downwards trend punctuated by occasional counter-trend brightening episodes
of at least 5 years duration.
Finally, we found that contributions to the Fe fluorescence by X-ray transient binaries and cosmic-ray
bombardment are very likely, and suggest possible ways to study this latter
phenomenon in the near future.}

\keywords{Galaxy: center -- X-rays: ISM -- ISM: clouds, cosmic rays}

\maketitle

% =============================================================================
\section{Introduction}\label{intro}

The  Galactic Centre (hereafter  GC) region is  a unique  environment within  the local
Universe, which provides  many tests of our understanding  of fundamental issues
in astrophysics.  The region hosts the  nearest Super Massive  Black Hole (SMBH), Sgr A*,
with a mass of 4$\times$10$^{6}$M$_{\odot}$ \citep[][]{2002Natur.419..694S, 2008ApJ...689.1044G}.
It is  also a region in  which high energy phenomena abound.  For example, in X-rays the Sgr B2
and Sgr C molecular complexes, located at projected distances of 90 pc and 70 pc
from Sgr A*,  shine brightly through 6.4-keV Fe-K$\alpha$ line emission
\citep[][]{1996PASJ...48..249K, 2000ApJ...534..283M, 2009PASJ...61S.233N}, consistent with 
the prediction of \citet{1993ApJ...407..606S}. 

The  physical mechanism  responsible for  the Fe-K$\alpha$
emission  from these molecular  clouds near the GC  is the fluorescence of cold,
neutral or near-neutral matter irradiated by  high  energy 
particles or X-ray photons.  So far several hypotheses have been  proposed as to the
nature and origin of the  primary source of this irradiation. The possibilities
include: the  X-ray reflection nebulae  model \citep[XRN,][]{1998MNRAS.297.1279S}, heating
by low-energy cosmic-rays (CR) 
\citep[][]{2002ApJ...568L.121Y}, shock mechanisms \citep[][]{1997AIPC..410.1027Y},
and electron bombardments \citep[][]{2003AN....324...73P}. 
A recent suggestion is that subrelativistic  protons can be created via accretion
of stellar debris onto the central black hole, thus explaining both the observed
X-ray   continuum    and   the   6.4    keV   line   emission   from    the   GC
\citep[][]{2009PASJ...61..901D}. Other particle-like candidates can be CR
electrons originating in  supernova events \citep[][]{2000ApJ...543..733V}.

If the fluorescence observed in molecular complexes is the result of irradiation by 
X-ray  photons, a luminous  localised source of X-rays must be invoked to power
the observed  emission,  since  the  diffuse hot plasma which  permeates the  central
regions of  the Galactic plane  produces a factor  ten less photons than is
required to account for the observed Fe-K$\alpha$ flux. 
It has been estimated that, depending on its position relatively to  the Sgr B2 and
Sgr C  molecular clouds, this  source should have  a 2-10 keV luminosity  of the
order of 10$^{39}$erg~s$^{-1}$.  Presently no  persistent sources in  the GC
region  have such  a high  luminosity; however  a past  transient outburst  in a
source, which has now returned to a relatively quiescent state, might well match
the requirement. Sgr  A* itself is arguably the best candidate, since although its activity
is  currently rather  weak  \citep[the  brightest  flare  measured  with
\textit{XMM-Newton}   has  a   luminosity  of   the   order  of   a  few   10$^{35}$ erg~s$^{-1}$,][]{2003A&A...407L..17P}, 
the SMBH  could  have  undergone a  period  of  high-state
activity in  the past \citep[][]{1996PASJ...48..249K}.
In  fact, the projected light travel time to Sgr B2 implies an outburst 
roughly 75-150  years ago, according to a recent estimate \citep[][] {2010ApJ...719..143T}.
Since there are a number of filaments nearer to Sgr A* 
which emit at 6.4-keV, the same argument would require Sgr  A* to also have flared 
more recently, say within the last 100 years, depending on the relative position of the clouds along the line of sight.

One method of distinguishing  between the
point source and the particle hypothesis is to investigate the lightcurve of the
filaments which emit the 6.4-keV line. If we are dealing with a transient source as an
engine of the fluorescent emission,  one might expect the temporal evolution of
the observed line flux to follow the light curve of the outburst, subject to 
blurring arising from the spread in light-travel delays inherent in the 
source to cloud geometry. 
Such effects have been found in the Sgr B2 molecular cloud where
the brightest peak of the Fe-K$\alpha$ emission observed
in 1995 by ASCA \citep[][]{1996PASJ...48..249K,2000ApJ...534..283M}
maintained the same  brightness level when remeasured in 2000 with
Chandra \citep[][]{2001ApJ...558..687M} but then, some five years later
when observed by Suzaku, showed a marked decline to approximately
half the peak value \citep[][]{2008PASJ...60S.201K,2009PASJ...61S.241I}.
Similarly, 
\citet[][]{2009PASJ...61S.233N} discovered variability of the Fe-K$\alpha$ flux 
from Sgr C on the basis of recent Sukaku observations. 
Moreover,  \citet[][]{2007ApJ...656L..69M}  have reported the  evolution  in intensity  and
morphology of the 4-8 keV continuum emission in two filamentary regions located
close to Sgr A*.  The latter changes occurred on parsec scales
(in projection), which requires a brightening/fading of the illuminating source
over a 2-3  year period, with an inferred 2--8 keV luminosity of at least 10$^{37}$ erg~s$^{-1}$. 
Recently, \citet[][]{2010ApJ...714..732P} showed  that the 6.4
keV  line   flux  from  the molecular   filaments with 15 arcmin of Sgr A* 
exhibit a  complex  pattern  of
variability.  If these molecular clouds/knots have a particular distribution
along the  line of sight, then it is possible to argue that they were all energised
by the same outburst on Sgr A*, consistent with
the XRN scenario (\citealt{2010ApJ...714..732P}).
However, given the observational complexities, it is very possible that this is not the
complete story; indeed, very recently \citet[][]{2011A&A...530A..38C} studied the 
Fe-K$\alpha$ line emission from the MCs in the Arches cluster region (about 20 pc 
in projection from Sgr A*), showing that the XRN/Sgr A* scenario can hardly describe 
the spectral and temporal properties of those clouds.

The above results open once more the question whether Sgr A* has exhibited AGN activity
in the past and, if so, what is the exact nature of that activity.
It is in this context that we reassess in this paper, the morphology, variability and
spectral properties of the 6.4-keV emitting clouds within 15\arcmin of Sgr A*, 
using the extensive set of \textit{XMM-Newton} observations targeted at this region.
Our goal is to investigate both the past role of Sgr A* (or other transient sources)
in illuminating the GC molecular clouds and also to seek evidence for a
contribution from alternative mechanisms, such as CR bombardment, in the
excitation of the Fe fluorescence which characterises the  GC environment.
Throughout this work, the distance to the GC has been taken to be 8 
kpc \citep[][]{2009ApJ...692.1075G}.

\onltab{1}
{
\begin{table*}
\caption{Specifications for the selected OBSIDs: MODE/FILTER combination used for each of the pointings, 
and GTI compared to the total exposure for each instrument. F=Full Frame MODE; E=Extended Full Frame MODE; 
T=Thick filter; M=Medium Filter. OBSID 0506291201 has PN in Timing MODE.} 
\label{obs_table}
\centering
\begin{tabular}{|c|c|ccc|ccc|}
\hline
 & & & Instrument specifics & & & GTI \& Exposure (ks) & \\
\hline
OBSID & Obs Date & PN & MOS1 & MOS2 & PN & MOS1 & MOS2 \\
 & yyyy-mm-dd & mode/filter & mode/filter & mode/filter & GTI/exp & GTI/exp & GTI/exp \\
\hline
\hline
0111350101 & 2002-02-26 & F/T & F/M & F/M & 38.590/40.030 & 42.262/52.105 & 41.700/52.120 \\
%0111350301 & 2002-10-03 & F/T & F/M & F/M & 7.037/15.377 & 7.479/16.960 & 7.857/16.996 \\
%0112970501 & 2000-09-21 & E/M & F/M & F/M & 4.980/21.119 & 13.977/24.914 & 14.351/24.911 \\
%0112972101 & 2001-09-04 & E/M & F/M & F/M & 18.840/21.687 & 21.697/26.039 & 21.472/26.055 \\
0202670501 & 2004-03-28 & E/M & F/M & F/M & 13.320/101.170 & 33.070/107.784 & 30.049/108.572 \\
0202670601 & 2004-03-30 & E/M & F/M & F/M & 25.680/112.204 & 32.841/120.863 & 35.390/122.521 \\
0202670701 & 2004-08-31 & F/M & F/M & F/M & 59.400/127.470 & 80.640/132.469 & 84.180/132.502 \\
0202670801 & 2004-09-02 & F/M & F/M & F/M & 69.360/130.951 & 94.774/132.997 & 98.757/133.036 \\
0402430301 & 2007-04-01 & F/M & F/M & F/M & 61.465/101.319 & 61.002/93.947 & 62.987/94.022 \\
0402430401 & 2007-04-03 & F/M & F/M & F/M & 48.862/93.594 & 40.372/97.566 & 41.317/96.461 \\
0402430701 & 2007-03-30 & F/M & F/M & F/M & 32.337/32.338 & 26.720/33.912 & 27.685/33.917 \\
%0504940201 & 2007-09-06 & F/M & F/M & F/M & 0.180/11.092 & 8.880/12.649 & 9.263/12.652 \\
0505670101 & 2008-03-23 & F/M & F/M & F/M & 74.216/96.601 & 73.662/97.787 & 74.027/97.787 \\
%0506291201 & 2007-02-27 & - & F/M & F/M & - & 22.857/38.616 & 24.597/38.621 \\
0554750401 & 2009-04-01 & F/M & F/M & F/M & 30.114/38.034 & 32.567/39.614 & 33.802/39.619  \\
0554750501 & 2009-04-03 & F/M & F/M & F/M & 36.374/42.434 & 41.376/44.016 & 41.318/44.018 \\
0554750601 & 2009-04-05 & F/M & F/M & F/M & 28.697/32.837 & 37.076/38.816 & 36.840/38.818\\
\hline
\end{tabular}
\end{table*}
}
%-----------------------------Figure Start------------------------------

\begin{figure}[!Ht]
\begin{center}
% un-comment the following line to include your fig1a.eps postscript file
\includegraphics[width=0.4\textwidth,angle=-90]{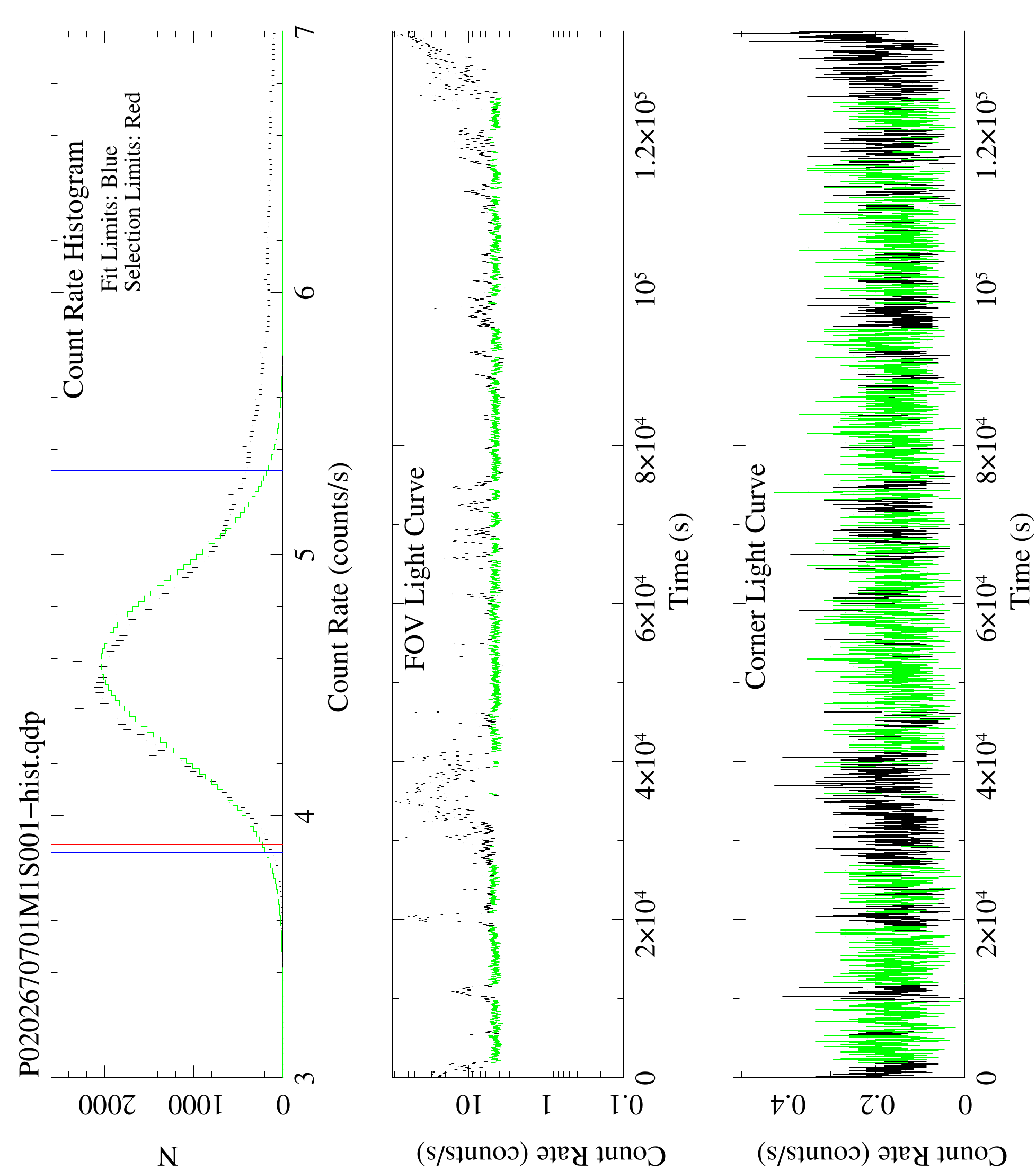}
\end{center}
\caption{\footnotesize
{Soft proton flare filtering of the MOS1 dataset from OBSID 0202670701. \textit{Upper panel}: 
2.5-12 keV count rate histogram. The blue lines mark the region used in the Gaussian fit,
the green line represents the best fit Gaussian and the red lines show the bounds used
to filter the data. \textit{Mid panel}: 2.5-8.5 keV lightcurve of the IN FOV region. 
\textit{Lower panel}: 2.5-8.5 keV lightcurve of the corner data. In both the Mid and the 
Lower panels, the points coloured green in the light-curves correspond to the selected 
GTI intervals (see text).}}
\label{espfilt}
\end{figure}
%-----------------------------Figure End--------------------------------

\section{Observations and Data Reduction}\label{obs}

We checked  in the HEASARC archive  for \textit{XMM-Newton} observations  of the GC 
region and selected observations  with a nominal pointing position within
15 \arcmin from Sgr A* (R.A.=17$^{h}$45$^{m}$40.045$^{s}$,
DEC.=-29$^{\circ}$0'27.9''). Data  from the EPIC cameras, consisting of  one PN back
illuminated  CCD  detector  \citep[][]{2001A&A...365L..18S}  and  two  MOS  front
illuminated CCD detectors  \citep[][]{2001A&A...365L..27T}, have been reprocessed
using the tasks  EMPROC and EPPROC in the Science  Analysis Software SAS version
9.0. For  our  purposes it  is  necessary to  rigorously  select  the Good  Time
Intervals (GTI), representing the periods in which the internal background
of the cameras was relatively quiescent. For this
reason we  used the SAS task ESPFILT  to screen data with a  high particle-induced
background.  This task  was  originally developed  inside  the Extended  Sources
Analysis Software (ESAS, \citealt[][]{2004ApJ...610.1182S}) and then converted into
a specific procedure in the SAS. This  task fits a Gaussian peak to the
distribution of count rates,
and  creates  a  GTI for  those  time  intervals  with  count rates  within  the
thresholds, defined  to be  $\pm$1.5$\sigma$ from the  mean count rate.  The GTI
filtering  has been  done  separately for  all  the three  EPIC cameras.  Figure
\ref{espfilt} shows the results for the screening done on the MOS1 data of OBSID
0202670701, and how  the task operates. The upper panel  shows the histogram for
the full  Field Of View (FOV)  lightcurve in the  2.5-12 keV band, while  in the
middle and the lower panels are, respectively, the 2.5-8.5 keV lightcurve of the
IN FOV region and the corner data. The green time intervals are those selected by the
filtering  process  and  used for  the  further  analysis.  The results  of  the
screening done  on all the  selected data sets  together with the  specifics of
each observation are reported in Table \ref{obs_table}.

%-----------------------------Figure Start------------------------------
\begin{figure*}[ht]
\begin{center}
% un-comment the following line to include your fig1a.eps postscript file
\includegraphics[width=0.9\textwidth]{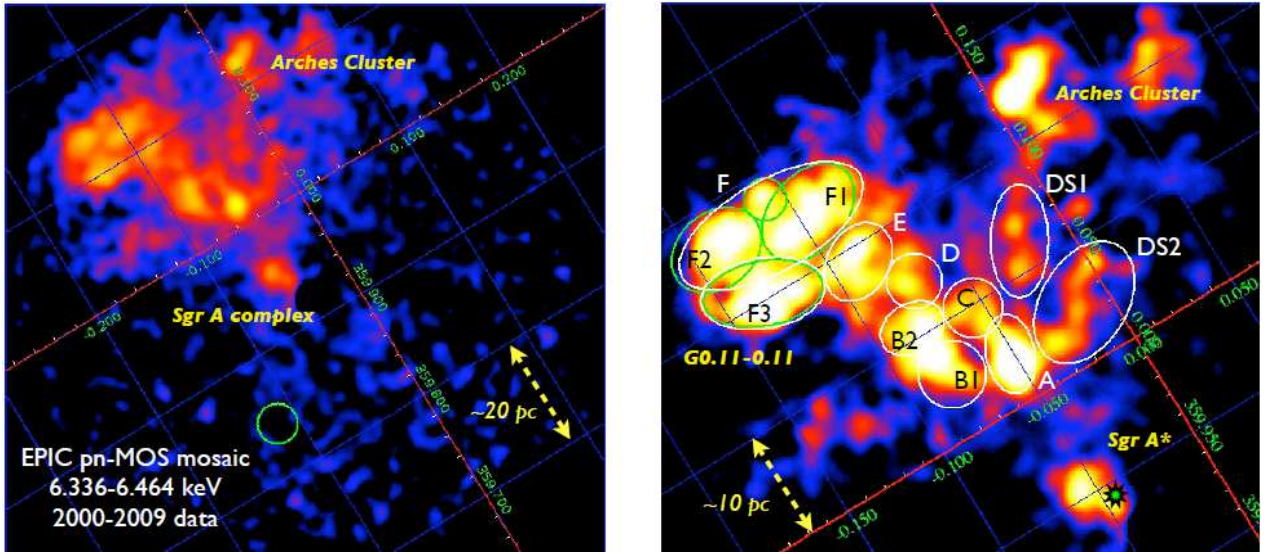}
\end{center}
\caption{\footnotesize
{\textit{Left panel}: Fluence map (6.336-6.464 keV) of the Fe-K$\alpha$ emission 
from the filaments in the GC. The image has been smoothed with a 8-pixel diameter
top-hat filter.
The grid lines define the Galactic Coordinate frame, where the green circle shows the background region selected for the 
spectral analysis of the clouds marked in the right panel (Section \ref{imprint}). 
\textit{Right panel}: A zoom-in on the region of the bright filaments. The white (green) 
ellipses show the regions (sub-regions) selected for the spectral analysis 
(see Table \ref{capelli_reg}); region B has been divided into two 
subregions B1 and B2; region F (G0.11-0.11) into three subregions, F1 (ellipse plus small circle), F2 and F3.
The position of Sgr A* is marked by a green star at l=-0.056 deg and b=-0.046 deg.}}
\label{fluence}
\end{figure*}
%-----------------------------Figure End--------------------------------

We  performed a different  GTI selection for OBSIDs 0402430301, 0402430401 and 0505670101, 
which were contaminated by very  high radiation levels towards the end 
of the respective observations.  Indeed, the filtering approach of selecting  the count  rate inside
$\pm$1.5$\sigma$ from the  mean count rate can lead  to a bias in the case of  a very strong
flare  during  an observation  (especially  for the  PN  camera;  this does  not
influence the  MOS cameras).  Moreover when  removing the count  rates below the
mean  count  rate-1.5$\sigma$  (see   Fig.\ref{espfilt})  all  the  genuine  low
background exposure times  in the observation are removed.  The ESPFILT filtered
PN  event files  for OBSIDs  0402430301, 0402430401  and 0505670101  have  a GTI
exposure of  0.3, 4.3 and  9.2 ks, respectively.  We built the  10-12 keV
lightcurve of the full FOV for these OBSIDs, and selected those intervals with a
count rate  lower than 1.5 counts~s$^{-1}$  (for OBSIDs 0402430301  and 0402430401) and
1.25 counts~s$^{-1}$ (OBSID 0505670101). The total exposures in the PN cameras of these
OBSIDs  with  the   new  filtering  process  are  61.5,   48.9  and  74.2  ks,
respectively.

Throughout our analysis we have only selected single and double events
(PATTERN$\le$4)  for  PN   and  up  to  quadruple  events (PATTERN$\le$12)  for  MOS1  and  MOS2 
cameras. For  all the instruments a further  screening for good
events was done selecting only events marked as real X-rays (FLAG==0).

\section{Spatial and spectral characterisation of the Fe-K$\alpha$ emission}\label{data_analysis}

%*****************************************************************************
\begin{table*}[!Ht]
\begin{center}
\caption{The central position, sizes (radius or semi-axis) and projected
distance from Sgr A* of the regions and sub-regions selected for the spectral
analysis - see Fig.\ref{fluence}. Note that region F (G0.11-0.11)  and also
sub-region F1 are comprised of two separate components. A cross-reference to the
cloud   designation   employed by \citet[][]{2010ApJ...714..732P} is also provided.} \label{capelli_reg}
\begin{tabular}{|cccccc|}
\hline
Region & Ponti & R.A.(J2000) & DEC.(J2000) & size (arcmin) & $\rm{d_{proj}}$ (pc)\\
\hline
The 9 clouds & & & & & \\
A  & MC1 & 17:45:52.398 & -28:56:38.15 & 0.65x1.08 & 10.9 \\
%B & 17:46:01.784 & -28:56:41.27 & 1.69x1.14 \\
B1 & MC2 & 17:45:59.592 & -28:57:10.09 & 0.91x0.91 & 12.6 \\
B2 & 1\&2* & 17:46:04.394 & -28:55:56.74 & 0.90x0.68 & 16.3 \\
C & new & 17:45:57.026 & -28:55:22.91 & 0.8x0.8 & 14.7 \\
D & 3\&4* & 17:46:04.252 & -28:54:39.64 & 0.75x0.75 & 18.3 \\
E & 5* & 17:46:10.997 & -28:54:09.50 & 1.12x0.83 & 21.6 \\
F & G0.11-0.11 & 17:46:21.627 & -28:53:11.36 & 2.75x1.22 & 27.2 \\
 & & 17:46:22.615 & -28:55:00.14 & 1.68x0.95 & \\
DS1 & new & 17:45:51.294 & -28:53:35.76 & 0.77x1.52 & 17.0 \\
DS2 & new & 17:45:43.377 & -28:55:14.99 & 1.78x1.16 & 12.3\\
\hline
The 3 F subregions & & & & &\\
F1 & & 17:46:17.084 & -28:52:49.45 & 1.49x1.01 & - \\
 & & 17:46:22.443 & -28:52:26.99 & 0.59x0.59 & - \\
F2 & & 17:46:28.297 & -28:53:50.10 & 1.32x1.01 & - \\
F3 & & 17:46:22.615 & -28:55:00.14 & 1.68x0.95 & - \\
\hline
Large regions & & & & & \\
EDE & & 17:45:53.128 & -28:52:26.19 & 10.95x7.09 & - \\
EDW & & 17:45:13.233 & -29:06:46.85 & 10.95x7.09 & - \\
\hline
\hline
\end{tabular}
\tablefoot{The clouds marked with a star are subregions of the \textit{Bridge}
feature identified by \citet[][]{2010ApJ...714..732P}}
\end{center}
\end{table*}
%**************************************************************************

\subsection{Imaging}

\subsubsection{Method}
First we built a fluence (integrated flux over time) map for the Fe-K$\alpha$
line  in a  very  narrow  energy window  in  order to  identify  and locate  the
brightest  Fe  fluorescence regions.  We  proceeded  by  merging all  the  GTI
filtered  event files for  the three  EPIC cameras  (where available)  for every
OBSID selected  (see Table  \ref{obs_table}), using  the SAS
task EMOSAIC. In building the image  of the Fe-K$\alpha$ emission, we assumed
a value  of E/$\Delta$E=50  for the  spectral resolution (FWHM)  of both  PN and
MOS1\&2 at 6.4  keV, corresponding to a bandpass for  the Fe fluorescence signal
of  6336-6464  eV.

An  important  issue  for  our  analysis  is  the  subtraction of instrument
background and, where appropriate, the continuum underlying the line emission. 
To quantify  the percentage of counts in  the narrow band 6336-6464
eV due to both components,  we selected the  energy range 3904-6208 eV  in the
total  event file  and  produced 18  adjacent images  of  the whole  FOV, each 128 eV
broad.  The spectral  window thereby  encompassed  is free  of contamination  by
strong emissions lines, including both the Fe-K$\alpha$ emission from neutral
or near-neutral  ions and  the Fe-K emission from  high ionisation  lines which
characterise the GC high-energy thermal components. We then fit the distribution
of counts as a function of the energy with a powerlaw of the form E$^{-\Gamma}$, and further used  this best fit to infer
the number of counts in the continuum  for the images in the ranges 6208-6336 eV
and 6336-6464 eV. Finally, we used the 6272 eV centered image properly scaled
as background  for the Fe-K$\alpha$ map.

\subsubsection{Results}

The background- and continuum-subtracted narrow-band (6336-6464 eV) image produced 
by the method
outlined in the  previous section, in effect, represents  a Fe-K$\alpha$ line
fluence map (with the provision that the effective exposure varies somewhat over
the field). This  image is shown in Fig.\ref{fluence}. The  morphology of the Fe
K$\alpha$ emission  evident in  this image is  consistent with  that reported
earlier   by  other   X-ray  satellites  including  ASCA
\cite[][]{1996PASJ...48..249K},   Chandra   \cite[][]{2007ApJ...656..847Y},  and
Suzaku  \cite[][]{2006JPhCS..54...95K}. 
The  main differences  arise  because of
instrumental effects  (different angular resolution and sensitivity  in the hard
X-ray domain)  or different  exposure times. Our results are also in good 
agreement with those reported recently by \cite{2010ApJ...714..732P}, based on
an independent analysis of a broadly similar set of \textit{XMM-Newton} observations.

The  maps clearly
show  an asymmetry  in the  distribution of  the Fe-K$\alpha$  emission with
respect to the position  of Sgr A*, with all the bright filaments located to the
east of   Sgr A*, at negative  Galactic latitudes.  On the western side  of the  GC  region, the
closest molecular complex bright in Fe  fluorescent lines is Sgr C, located at a
projected distance from  Sgr A* of approximately 70 pc. Looking at
the radio  maps of  the GC region  in CO \cite[][]{1998ApJS..118..455O}  and CS
\cite[][]{1999ApJS..120....1T}  rotational lines, an  asymmetry in
the distribution of molecular matter  is very noticeable, with more concentrated
to the east of Sgr A*. Nevertheless, a certain amount of cold matter not shining
in the  6.4-keV line is  present along the whole  plane of the  Galaxy, and this
should be  taken into account  in explaining the  origin of the  Fe-K$\alpha$
emission in the filaments between Sgr A* and the GC feature known as the Radio Arc. 
Looking at the left panel in Fig.\ref{fluence}, the presence of low
surface brightness 6.4-keV line  emission is evident across the  whole region 
between  the Radio Arc (l$\approx$0.2$^{\circ}$) and Sgr A*  
(coded as diffuse blue in Fig.\ref{fluence}). In contrast, this low-level emission
is largely absent at negative  longitudes. We emphasise that this is not  an image artifact.
In fact, the 6.4-keV spatial distribution at faint levels  coincides remarkably well 
with the  contours  of TeV $\gamma$-ray emission measured by HESS
\cite[][]{2006Natur.439..695A}, a connection first noted by \cite{2007ApJ...656..847Y}
using the 6.4-keV line equivalent-width image measured by Chandra.  Thanks  
to  the  much  higher  effective  area  of \textit{XMM-Newton},  we  are now  able  
to establish that this strong correlation  apparently extends to regions outside of those
occupied by very dense molecular clouds (see Section \ref{low_sb}).

% THis is for two columns
%*****************************************************************************
\begin{table*}[!Ht]
\begin{center}
\begin{tabular}{|c|cccccc|}
\hline
Region & Feb02 & Mar04 & Sep04 & Apr07 & Mar08 & Apr09 \\ 
\hline
A & 2.5$\pm$0.3 & 2.2$\pm$0.2 & 2.1$\pm$0.1 & 2.2$\pm$0.2 & 2.2$\pm$0.2 & 2.2$\pm$0.2 \\
B1 & 2.3$\pm$0.3 & 1.7$\pm$0.2 & 1.5$\pm$0.1 & 1.6$\pm$0.1 & 1.3$\pm$0.2 & 1.2$\pm$0.1 \\
B2 & 0.9$\pm$0.2 & 0.7$\pm$0.2 & 1.1$\pm$0.1 & 2.1$\pm$0.2 & 2.1$\pm$0.2 & 2.1$\pm$0.2 \\
C & 0.8$\pm$0.3 & 0.7$\pm$0.2 & 0.9$\pm$0.2 & 1.2$\pm$0.2 & 1.1$\pm$0.3 & 1.2$\pm$0.2 \\
D & 0.5$\pm$0.2 & 0.7$\pm$0.2 & 0.8$\pm$0.1 & 1.0$\pm$0.1 & 1.1$\pm$0.2 & 2.1$\pm$0.2 \\
E & 2.4$\pm$0.4 & 2.2$\pm$0.3 & 1.8$\pm$0.2 & 2.1$\pm$0.2 & 2.2$\pm$0.3 & 2.1$\pm$0.2 \\
F & 13.1$\pm$1.0 & 11.2$\pm$0.8 & 11.4$\pm$0.5 & 10.0$\pm$0.5 & 9.4$\pm$0.7 & 9.1$\pm$0.5 \\
F1 & 4.4$\pm$0.6 & 4.0$\pm$0.5 & 4.4$\pm$0.3 & 3.4$\pm$0.3 & 3.0$\pm$0.4 & 3.4$\pm$0.3 \\
F2 & 4.1$\pm$0.6 & 3.1$\pm$0.4 & 3.6$\pm$0.3 & 3.7$\pm$0.3 & 3.3$\pm$0.4 & 2.6$\pm$0.3 \\
F3 & 5.2$\pm$0.6 & 4.2$\pm$0.5 & 3.8$\pm$0.3 & 3.5$\pm$0.3 & 3.4$\pm$0.4 & 3.4$\pm$0.3 \\
DS1 & 1.5$\pm$0.3 & 1.5$\pm$0.3 & 1.4$\pm$0.1 & 1.4$\pm$0.1 & 1.4$\pm$0.2 & 1.4$\pm$0.2 \\
DS2 & 2.8$\pm$0.4 & 2.3$\pm$0.3 & 2.4$\pm$0.2 & 2.1$\pm$0.2 & 2.4$\pm$0.2 & 2.5$\pm$0.2 \\
\hline
\end{tabular}
\end{center}
\caption{\footnotesize
{Fluxes of the Fe-K$\alpha$ line in units of 10$^{-5}$photons~cm$^{-2}$~s$^{-1}$. 
The values are the weighted mean of the MOS1\&2 measured fluxes. The 
different columns refer to different datasets (see text). The Apr07, 
Mar08 and Apr 09 fluxes for the regions D, DS1 and DS2 are based on  MOS 2 data only, as a
consequence of the damaged sustained by CCD6 in MOS 1 on 9 March 2005
\cite[][]{2006ESASP.604..943A}.
}}
\label{fluxes_final}
\end{table*}
%**************************************************************************

The bright region in Fig.\ref{fluence} located  at  l$\sim$0.12$^{\circ}$,
b$\sim$0.05$^{\circ}$  is the  Arches  Cluster; for the study of the 6.4-keV line emission
from the MCs in this region we cross refer to the recent work by \citet[][]{2011A&A...530A..38C}. 
We have excluded from our study the bright feature  at
l$\sim$-0.05$^{\circ}$. This is  the Sgr A complex, containing both Sgr A* itself
and the Sgr A East Supernova Remnant (SNR). In Table \ref{capelli_reg} we report
the  coordinates  and the  sizes  of the  selected regions of elliptical shape 
upon which our spectral analysis is based.  We selected these sky  regions solely
on the basis of Fe-K$\alpha$ morphology without reference to potential
radio molecular  counterparts (that could help to identify spatially 
and physically connected  regions). In doing  so, we neglected the  
possibility that some of the filaments which are bright in the Fe-K$\alpha$ line
could be part  of a single, more extended, molecular  complex.

\subsection{Spectroscopy: studying the Fe-K$\alpha$ variability}\label{timing}

\subsubsection{Method}

We carried out a spectroscopic  analysis of the Fe-K$\alpha$ bright filaments
shown in Fig.\ref{fluence},  with the objective of measuring  the photon flux in
the 6.4  keV line and searching  for any variations  in this quantity over  an 
eight-year timeframe. The  molecular complexes bright in the  Fe fluorescent 
line are embedded in strong X-ray emission due to the presence along the line of 
sight of unresolved thermal X-ray emission. This thermal emission, which permeates 
the whole GC  region can  be  characterised  as a  two-temperature  plasma with  
typically kT$_{1}$$\approx$1 keV and kT$_{2}$$\approx$6.5 keV
\cite[][]{2007PASJ...59S.245K}.  For our  purposes, everything  but the  
6.4-keV line  emission is considered as background radiation. Unfortunately due
to the relatively low surface brightness of the fluorescent emission and the spatial 
variations inherent in the foreground emission, the subtraction of a ``local 
background'' introduces many systematics in the resultant spectrum. 
Therefore, we adopted the approach of modelling the Non X-ray Background (NXB) 
rather than  subtracting it. In Appendix \ref{background_appendix} we describe the
different ways of dealing with  NXB in an \textit{XMM-Newton} dataset, and specify the
methodology   we actually employed. One consequence of the  use of the background modelling
technique was that for this part of analysis we were restricted to the use of the
MOS data only.

We built response and ancillary files with  the SAS tasks 
RMFGEN v1.55.1 and ARFGEN v1.76.4. No point source excision has been applied. 
The  need for  a systematic analysis with spectra encompassing
very different exposure times and net counts, led us to  use the Cash  
statistic \cite[][]{1979ApJ...228..939C}  rather  than $\chi^{2}$.
We grouped the  channels in each spectrum with the GRPPHA tool to ensure
at least 1  count/bin.

We fitted all the spectra with a model  which accounts  for all of the  
source  and instrumental  components.  The former includes the two thermal
components noted above, a power-law continuum plus 
an associated 6.4-keV Fe line (representing reflection and fluorescence in
the clouds). We use a second power-law component to model a defined
contribution from the instrument background, which is not subject to the
instrument response (see Appendix A). 
%[RENZO - I THINK THIS IS CORRECT??
%- I HAVE BEEN MEANING TO ASK YOU HOW TO DO THIS IN XSPEC?? - HAVE WE FORGOTTEN 
%TO MENTION THE CXB?? -- YES IT IS OK!]
For a more detailed description of these spectral
components see \citet[][]{2011A&A...530A..38C}. Note that in this analysis we fixed the
temperature of the hot plasma in every spectrum to 6.5 keV \citep[][]{2007PASJ...59S.245K}
since this is critical in obtaining a consistent measurement of the Fe-K$\alpha$ 
line flux across all the spectra (different instruments/epochs).

\begin{figure}[!Ht]
\begin{center}
% un-comment the following line to include your fig1a.eps postscript file
\includegraphics[width=0.3\textwidth,angle=-90]{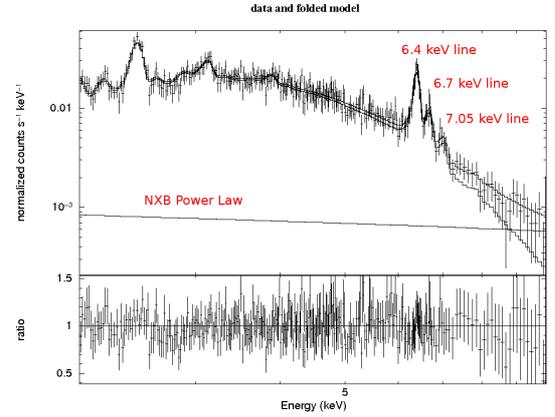}
\end{center}
\caption{\footnotesize {$\textit{Upper panel}$: The spectrum derived for Region B1
from the September 2004 MOS1 dataset together with the best-fitting spectral model.
The lower solid line shows the NXB power-law component. The two curves show the net 
X-ray source spectrum and the source plus NXB best fit model (top curve).
Three Fe lines are
labelled; neutral Fe-K$\alpha$ and K$\beta$ lines at 6.4  and 7.05 keV, 
and the Fe XXV K$\alpha$ line at 6.7  keV. 
$\textit{Lower panel}$: Ratio between  the spectral points and 
the best fit model.}}
\label{spec}
\end{figure}

Throughout our spectral  analysis we have considered only observations
with an exposure  longer than 30 ks.  If different datasets were  part of a set
of observations carried out over a couple of days, the  associated 
spectra were added together with  the ftool MATHPHA, with the same done 
for  the response files using the  ftools ADDARF  and ADDRMF.  
Errors have been  considered at  the 90\% confidence  level.  
The resulting temporal sampling was as follows:  the  OBSIDs  0111350101  (Feb02), 0202670501  +
0202670601 (Mar04),  0202670701 + 0202670801  (Sep04), 0402430301 +  402430401 +
402430701  (Apr07),  0505670101 (Mar08),  054750401  +  0554750501 +  0554750601
(Apr09).
A typical result from this spectral modelling process is illustrated in
Fig.\ref{spec}.

%-----------------------------Figure Start------------------------------
%-----------------------------Figure End--------------------------------

%-----------------------------Figure Start------------------------------
\begin{figure*}[ht]
\begin{center}
% un-comment the following line to include your fig1a.eps postscript file
\includegraphics[width=1.0\textwidth,height=19cm]{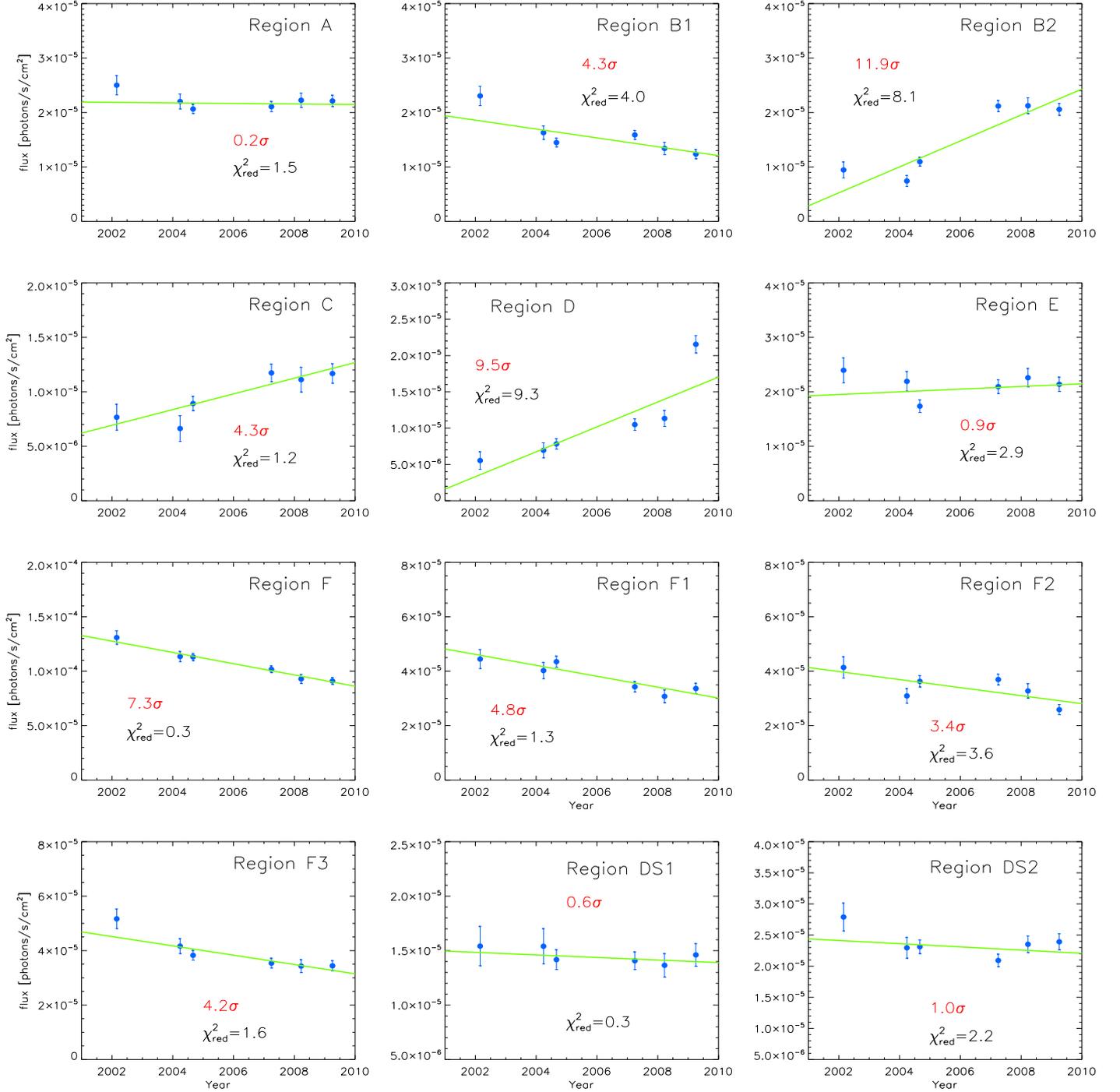}
\end{center}
\caption{\footnotesize {Lightcurves  of  the  Fe-K$\alpha$  flux  from  the
filaments    in    the   GC.    The    flux    is    plotted   in    units    of
10$^{-5}$photons~cm$^{-2}$~s$^{-1}$.  The  green line  represents  the  best fit  linear
function   to    the   data,   whose   parameters   are    reported   in   Table
\ref{number_sigma}.  In each  panel we  present the  significance 
of the measured gradient (B/$\sigma_{B}$) together with the $\chi^{2}_{red}$ 
for the best fit linear function.}}
\label{pmulti_capelli}
\end{figure*}
%-----------------------------Figure End--------------------------------

\subsubsection{Results}

Our goal is  to study the temporal behaviour  of the Fe fluorescence in the 
filaments with a view to constraining the nature and
location of the energising source.  
For example,  a clear spatial pattern to the variations might, in
principle,  help  us  identify a  preferred  direction  for  the  incoming
photons/particles. The 6.4-keV line fluxes derived from our spectral analysis
for the selected regions in the different observing periods are presented
in Table \ref{fluxes_final}. Here
the measurements are calculated as  the weighted  mean of  the
MOS1\&2 values with the errors representing the 90\%  confidence range.
Fig.\ref{pmulti_capelli} shows the resulting Fe-K$\alpha$ flux lightcurves for the 
different regions covering the period 2002-2009. 
In each of  the selected regions the surface brightness  of the 6.4-keV
line  lies in  the range  5-10$\times$10$^{-6}$ photons~cm$^{-2}$~s$^{-1}$~arcmin$^{-2}$; from  the lightcurves
it   is  clear  that  some  regions show evidence of
substantial variability,  while some others appear to remain 
roughly constant.  
To quantify  this, we  fitted the 6.4-keV 
line fluxes as a linear function of time $t$:

\[
\mathrm{F_{64}=A+B\cdot t},
\]

\noindent   where   B$>$0  indicates   an   increasing   Fe-K$\alpha$   flux (F$\rm{_{64}}$).  
Table \ref{number_sigma} reports the  results for the
selected regions and in  Fig.\ref{pmulti_capelli} we show the  best fit linear
functions  overplotted on  the lightcurves, where the error bars represent (for this case only) 
the 1$\sigma$ uncertainty as used for the calculation of the $\chi^{2}_{red}$. From  this it is clear that 
the temporal behaviour of the Fe-K$\alpha$ flux differs markedly
from filament to filament. In brief, the different regions exhibit 
the following behaviour:

\vspace{0.1cm}

\noindent  $\textit{Region  A}$:  the  Fe-K$\alpha$ flux  measured  in  this
molecular complex does  not show any   significant variability on a timescale of 8 years.  
The weighted mean flux across the full set 
of observations is 2.16$\pm$0.07$\times$10$^{-5}$  photons~cm$^{-2}$~s$^{-1}$,  
which translates to  a  surface brightness of 
9.9$\pm$0.03$\times$10$^{-6}$ photons~cm$^{-2}$~s$^{-1}$~arcmin$^{-2}$.

%*****************************************************************************
\begin{table}[!Ht]
\begin{center}
\begin{tabular}{|c|cccc|}
\hline
Region & B (10$^{-6}$) & $\sigma_{B}$ (10$^{-6}$) & $\#_{\sigma}$ (B/$\sigma_{B}$) & $\chi^{2}_{red}$ \\
\hline
A & -0.05 & 0.21 & 0.2 & 1.5   \\
B1 & -0.81 & 0.19 & 4.3 & 4.0   \\
B2 & 2.38 & 0.20 & 11.9 & 8.1  \\
C & 0.72 & 0.17 & 4.3 & 1.2   \\
D & 1.71 & 0.18 & 9.5 & 9.3   \\
E & 0.25 & 0.27 & 0.9 & 2.9   \\ 
F & -5.18 & 0.71 & 7.3 & 0.3   \\
F1 & -2.00 & 0.42 & 4.8 & 1.3  \\
F2 & -1.48 & 0.43 & 3.4 & 3.6   \\
F3 & -1.71 & 0.41 & 4.2 & 1.6   \\
DS1 & -0.12 & 0.21 & 0.6 & 0.3 \\
DS2 & -0.26 & 0.26 & 1.0 & 2.2 \\

\hline
\end{tabular}
\end{center}
\caption{\footnotesize {Results for the study of the Fe-K$\alpha$ variability
in the  selected regions  and sub-regions (first column).  The second  and third  columns give
the value of the gradient obtained for the best-fit linear function 
together with its uncertainty (in  units of 10$^{-6}$  photons~cm$^{-2}$~s$^{-1}$~year$^{-1}$). 
The fourth  and the fifth columns report,  respectively, the number of 
sigma  for the deviation from the constant behaviour of the best-fit function 
and the reduced $\chi^{2}$ for the linear fit.}}
\label{number_sigma}
\end{table}
%**************************************************************************

\vspace{0.1cm}

\noindent $\textit{Region B1}$: we have  measured a decrease  of the
6.4-keV flux  from  the sub-region  B1 at  a  confidence of  4.3$\sigma$, although a linear decrease does not 
provide a satisfactorily description of the light curve ($\chi^{2}_{red}$ = 1.46, see Table \ref{number_sigma}).  
The rate of decrease 
of  the  Fe-K$\alpha$  line  flux  in this  cloud  is
8.1$\pm$3.1$\times$10$^{-7}$ photons~cm$^{-2}$~s$^{-1}$~year$^{-1}$.

\vspace{0.1cm}

\noindent $\textit{Region B2}$: we have measured a strong increase of the Fe
K$\alpha$ flux  from the region B2 with  a 11.9$\sigma$  confidence,
although the high  value  of the  $\chi^{2}_{red}$  (8.1)  for the fit suggests  
that this rise may not be truly linear. The measured rate of increase of 
the Fe-K$\alpha$ line flux is
2.4$\pm$0.3$\times$10$^{-6}$ photons~cm$^{-2}$~s$^{-1}$~year$^{-1}$.

\vspace{0.1cm}

\noindent $\textit{Region C}$:  the Fe-K$\alpha$ flux from  region C has been
measured   to  have   an  increase  at  4.3$\sigma$  confidence with the overall
$\chi^{2}_{red}$ for the best fit linear function being 1.2. The rate of increase 
of the  Fe-K$\alpha$ line flux is 7$\pm$4$\times$10$^{-7}$ 
photons~cm$^{-2}$~s$^{-1}$~year$^{-1}$. Looking at the lightcurve in greater detail, 
we notice that there is a step-wise change at   roughly the mid-way
point, with the weighted means for two segments being, respectively,   0.8$\pm$0.1 and 1.2$\pm$0.1$\times$10$^{-5}$ 
photons~cm$^{-2}$~s$^{-1}$. Region C encompasses the position of a known X-ray
transient, XMMU J174554.4-285456, \citep[][]{2005A&A...430L...9P},
located at the north-west edge of the circle region. This spatial coincidence and the
possibility that the transient is the illumination source for the Region C cloud
will be discussed later (\S 7.1.1). 

\vspace{0.1cm}

\noindent $\textit{Region D}$: we have measured  a rapid increase of the 6.4-keV
flux from region  D with a confidence of  9.5 $\sigma$. The rate of growth of 
the Fe-K$\alpha$   line   flux  in   this   region is 
1.6$\pm$0.3$\times$10$^{-6}$  photons~cm$^{-2}$~s$^{-1}$~year$^{-1}$.   In  the  lightcurve  of
region D three  of the six points are  not fitted by the best  fit function (see
Fig.\ref{pmulti_capelli},   fourth   panel);  this   translates   into  a   high
$\chi^{2}_{red}$ value  (9.3, Table  \ref{number_sigma}) suggesting the behaviour
is not well represented by a simple linear increase.

\vspace{0.1cm}

\noindent  $\textit{Region  E}$:  the  Fe-K$\alpha$ flux  measured  in  this
molecular  complex is rather constant  over a period  of  8 years;  given the high $\chi^{2}_{red}$ of
the best  fit for the  lightcurve of this region  ($\chi^{2}_{red}$=2.9), we cannot exclude minor excursions 
of the flux. The weighted mean flux over the set of observations is
2.1$\pm$0.1$\times$10$^{-5}$   photons~cm$^{-2}$~s$^{-1}$, which translates 
to   a  surface brightness of 7.24$\pm$0.03$\times$10$^{-6}$ photons~cm$^{-2}$~s$^{-1}$~arcmin$^{-2}$.

\vspace{0.1cm}

\noindent $\textit{Region F}$: this is the region for which the measurements are
most precise. The Fe-K$\alpha$ flux shows a clear decrease with a 
7.3 $\sigma$  confidence; the $\chi^{2}_{red}$ value of the best
fit function is 0.3, strongly confirming the measured decrease and constraining
the functional shape of this decrease to  be linear. The rate of decrease 
of the Fe-K$\alpha$   line   flux  is 5.1$\pm$1.2$\times$10$^{-6}$  
photons~cm$^{-2}$~s$^{-1}$~year$^{-1}$.  To  improve the  spatial
resolution, we have divided this region into three
sub-regions and studied these separately.

\vspace{0.1cm}

\noindent $\textit{Sub-regions F1-F2-F3}$: we have  measured a decrease  of the
6.4-keV line flux from all three sub-regions with a confidence of 4.8, 3.4 and 4.2
$\sigma$  respectively.  The   corresponding  rates of decrease  are  2.1$\pm$0.7,
1.5$\pm$0.7   and   1.7$\pm$0.7$\times$10$^{-6}$  photons~cm$^{-2}$~s$^{-1}$~year$^{-1}$.   The
$\chi^{2}_{red}$  values  of the  best  fit  linear  functions are, respectively, 
1.3, 3.6 and 1.6; confirming the decrease is a linear function of time. 
Moreover we point  out that the rates of decrease from the three
sub-regions are  compatible with  each other.  The agreement  between the
temporal behaviour of both the whole  complex and the individual subregions 
suggests a common origin  for the Fe-K$\alpha$ emission within 
a single large  molecular structure, \textit{i.e.} the G0.11-0.11 molecular
cloud (\citealt{2002ApJ...568L.121Y}).  \cite{1997ApJ...481..263T} have show
that this molecular complex  has a shell-like structure that may be interacting
with the western  vertical filaments of the Radio Arc. Thus,  G0.11-0.11 
could  host an extended high energy particle accelerator/decelerator.

%-----------------------------Figure Start------------------------------
\begin{figure}[!Ht]
\begin{center}
\includegraphics[width=0.5\textwidth]{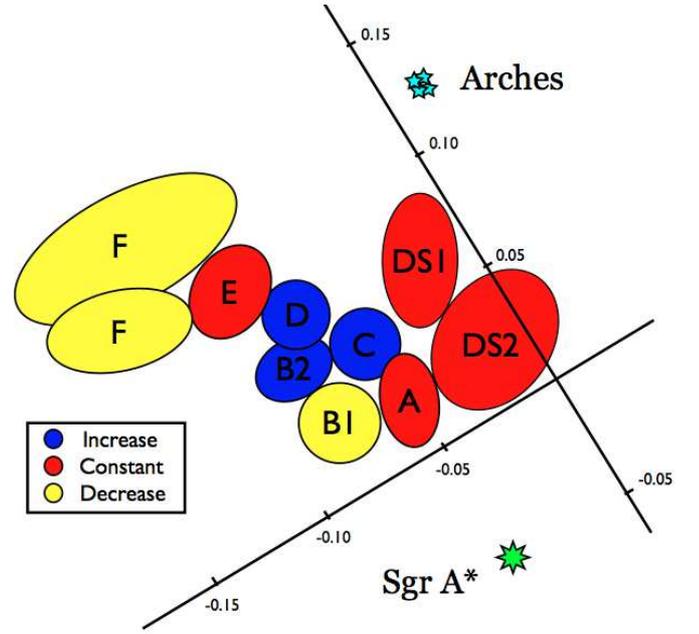}
\end{center}
\caption{Colour-coded sketch of the molecular clouds and Fe-K$\alpha$ line flux
variability studied in this work. The circles and ellipses represent the regions
selected for the timing and spectral studies (see also the right panel of
Fig.\ref{fluence}). The colours are blue (increase of the 6.4-keV line flux), 
red (constant) and yellow (decrease). A Galactic coordinate grid, and the positions of Sgr A* 
(green star) and the Arches cluster (cyan stars) are also indicated.}
\label{cartoon}
\end{figure}
%-----------------------------Figure End--------------------------------

% Two columns format
%-----------------------------Figure Start------------------------------
\begin{table*}[!Ht]
\caption{Results from the analysis of the PN time-averaged spectra  of  the  Fe-K$\alpha$
bright  filaments. The spectral parameters reported
are the column density along the line of sight (N$_{H}$, in units of 10$^{22}$ cm$^{-2}$), 
the temperatures (in keV) of the warm and the hot plasma components and their normalisations
(in units of 10$^{-17}$ and 10$^{-18}$$\int$n$_{e}$n$_{H}$dV/4$\pi$D$^{2}$ 
respectively), the column densities intrinsic to the MCs (N$\rm{_{H_{C}}}$ in units of 10$^{22}$ cm$^{-2}$), 
the central energy and the intrinsic width $\sigma$ of the
Fe-K$\alpha$ line (in keV), the 6.4-keV line flux (in units of 10$^{-5}$ 
photons~cm$^{-2}$~s$^{-1}$), the EW of the 6.4-keV line (in keV), the excess optical depth $\tau$ at the 
Fe-K edge, the power law normalisation (in units of 10$^{-4}$ photons cm$^{2}$ s$^{-1}$ keV$^{-1}$ at 1 keV)
and the C-stat value and the degree of freedom for the best fit model.}
\label{edge_parameters}
\begin{center}
\begin{tabular}{|c|ccccccccc|}
\hline
&\hspace{-0.2cm} A &\hspace{-0.2cm} B1 &\hspace{-0.2cm}   B2 &\hspace{-0.2cm}   C &\hspace{-0.2cm}   D &\hspace{-0.2cm}   E &\hspace{-0.2cm}   F &\hspace{-0.2cm}   DS1 &\hspace{-0.2cm}   DS2 \\
\hline
N$_{H}$ &\hspace{-0.2cm}   7.0$^{+0.6}_{-0.4}$ &\hspace{-0.2cm}   8.0$\pm$0.4 &\hspace{-0.2cm}    8.1$^{+0.4}_{-0.5}$ &\hspace{-0.2cm}   8.0$\pm$0.5 &\hspace{-0.2cm}   6.8$\pm$0.5 &\hspace{-0.2cm}    6.7$^{+0.2}_{-0.3}$ &\hspace{-0.2cm}   7.8$\pm$0.3 &\hspace{-0.2cm}   8.0$\pm$0.6 &\hspace{-0.2cm}   7.1$^{+0.7}_{-0.6}$  \\

kT$_{\rm{warm}}$ &\hspace{-0.2cm}   0.9$\pm$0.1 &\hspace{-0.2cm} 0.9$\pm$0.1 &\hspace{-0.2cm} 0.89$^{+0.04}_{-0.07}$ &\hspace{-0.2cm} 0.79$^{+0.04}_{-0.06}$ &\hspace{-0.2cm} 1.0$\pm$0.1 &\hspace{-0.2cm}  1.01$^{+0.05}_{-0.03}$ &\hspace{-0.2cm} 1.08$^{+0.01}_{-0.03}$ &\hspace{-0.2cm}   0.7$\pm$0.1 &\hspace{-0.2cm}   0.7$\pm$0.1 \\

norm &\hspace{-0.2cm} 3.5$^{+1.6}_{-0.6}$ &\hspace{-0.2cm} 5.5$^{+0.9}_{-0.8}$ &\hspace{-0.2cm} 4.4$^{+1.1}_{-0.8}$ &\hspace{-0.2cm} 5.4$^{+1.6}_{-1.3}$ &\hspace{-0.2cm} 2.7$^{0.6}_{-0.5}$ &\hspace{-0.2cm} 4.7$^{+0.7}_{-0.5}$ &\hspace{-0.2cm} 24.5$^{+2.5}_{-1.9}$ &\hspace{-0.2cm}   6.4$^{+3.0}_{-1.8}$ &\hspace{-0.2cm}   6.5$^{+4.7}_{-1.8}$ \\

kT$_{\rm{hot}}$ &\hspace{-0.2cm} 8.9$\pm$1.2 &\hspace{-0.2cm} 6.5 fix &\hspace{-0.2cm} 5.3$^{+1.8}_{-1.9}$ &\hspace{-0.2cm} 6.5 fix &\hspace{-0.2cm} 6.5 fix &\hspace{-0.2cm} 6.5 fix &\hspace{-0.2cm} -- &\hspace{-0.2cm}   7.4$^{+1.2}_{-1.4}$ &\hspace{-0.2cm}   7.5$^{+0.8}_{-1.0}$  \\

norm &\hspace{-0.2cm} 3.1$^{+0.5}_{-0.4}$ &\hspace{-0.2cm} 2.0$\pm$0.3 &\hspace{-0.2cm} 1.1$\pm$0.3 &\hspace{-0.2cm} 1.4$\pm$0.3 &\hspace{-0.2cm} 0.6$\pm$0.3 &\hspace{-0.2cm} 0.8$^{+0.3}_{-0.2}$ &\hspace{-0.2cm} -- &\hspace{-0.2cm}   3.3$\pm$0.6 &\hspace{-0.2cm}   6.1$\pm$0.7  \\

N$_{\rm{H_{C}}}$ &\hspace{-0.2cm} 18.4$^{+1.5}_{-2.7}$ &\hspace{-0.2cm} 10.2$^{+2.3}_{-2.0}$ &\hspace{-0.2cm} 12.3$^{+3.0}_{-2.7}$ &\hspace{-0.2cm} 5.8$^{+2.3}_{-1.9}$ &\hspace{-0.2cm} 13.2$^{+5.0}_{-4.7}$ &\hspace{-0.2cm} 9.6$^{+1.7}_{-1.4}$ &\hspace{-0.2cm} 9.2$^{+2.7}_{-2.3}$ &\hspace{-0.2cm}   15.5$^{+3.9}_{-3.3}$ &\hspace{-0.2cm}   14.5$\pm$2.3 \\

E$_{\rm{64}}$ &\hspace{-0.2cm} 6.420$^{+0.007}_{-0.004}$ &\hspace{-0.2cm}  6.416$^{+0.007}_{-0.006}$ &\hspace{-0.2cm} 6.415$^{+0.003}_{-0.004}$ &\hspace{-0.2cm} 6.421$^{+0.006}_{-0.013}$ &\hspace{-0.2cm} 6.417$^{+0.011}_{-0.007}$ &\hspace{-0.2cm} 6.417$\pm$0.008 &\hspace{-0.2cm} 6.411$\pm$0.005 &\hspace{-0.2cm}    6.414$^{+0.010}_{-0.009}$ &\hspace{-0.2cm}   6.412$\pm$0.007  \\

$\sigma$ &\hspace{-0.2cm} 0.03$\pm$0.01 &\hspace{-0.2cm} 0.05$\pm$0.01 &\hspace{-0.2cm} 0.039$^{+0.005}_{-0.008}$ &\hspace{-0.2cm} 0.042$^{+0.015}_{-0.011}$ &\hspace{-0.2cm} 0.043$^{+0.012}_{-0.020}$ &\hspace{-0.2cm} 0.030$^{+0.018}_{-0.014}$ &\hspace{-0.2cm} 0.029$^{+0.009}_{-0.012}$ &\hspace{-0.2cm}   0.041$^{+0.014}_{-0.019}$ &\hspace{-0.2cm}   0.053$^{+0.009}_{-0.012}$  \\

F$\rm{_{64}}$ &\hspace{-0.2cm} 2.5$\pm$0.1 &\hspace{-0.2cm} 1.7$\pm$0.1 &\hspace{-0.2cm} 2.0$\pm$0.1 &\hspace{-0.2cm} 1.1$\pm$0.1 &\hspace{-0.2cm} 1.1$\pm$0.1 &\hspace{-0.2cm} 2.0$\pm$0.1 &\hspace{-0.2cm} 9.1$^{+0.3}_{-0.5}$ &\hspace{-0.2cm}   1.5$\pm$0.1 &\hspace{-0.2cm}   2.6$\pm$0.1 \\

EW &\hspace{-0.2cm} 0.9$\pm$0.1 &\hspace{-0.2cm} 0.9$\pm$0.1 &\hspace{-0.2cm} 1.5$^{+0.3}_{-0.2}$ &\hspace{-0.2cm} 1.0$^{+0.2}_{-0.1}$ &\hspace{-0.2cm}  1.3$^{+0.4}_{-0.3}$ &\hspace{-0.2cm} 1.4$\pm$0.2 &\hspace{-0.2cm} 1.7$\pm$0.2 &\hspace{-0.2cm}   0.9$\pm$0.2 &\hspace{-0.2cm}    0.9$\pm$0.1   \\

$\tau$ &\hspace{-0.2cm} 0.13 &\hspace{-0.2cm} 0.07 &\hspace{-0.2cm} 0.08 &\hspace{-0.2cm} 0.04 &\hspace{-0.2cm} 0.09 &\hspace{-0.2cm} 0.07 &\hspace{-0.2cm} 0.06 &\hspace{-0.2cm}   0.11 &\hspace{-0.2cm}   0.10 \\

norm$_{\rm{pow}}$ &\hspace{-0.2cm} 9.3$^{+0.7}_{-0.6}$ &\hspace{-0.2cm} 6.2$^{+0.5}_{-0.4}$ &\hspace{-0.2cm} 4.4$^{+0.5}_{-0.4}$ &\hspace{-0.2cm} 3.7$\pm$0.3 &\hspace{-0.2cm} 2.8$^{+0.5}_{-0.4}$ &\hspace{-0.2cm} 5.0$^{+0.6}_{-0.2}$ &\hspace{-0.2cm} 18.7$^{+1.6}_{-1.3}$ &\hspace{-0.2cm}   6.0$\pm$0.8 &\hspace{-0.2cm}   10.5$^{+0.9}_{-0.8}$  \\

chi$^{2}$/dof &\hspace{-0.2cm} 521.62/1283 &\hspace{-0.2cm} 546.36/1316 &\hspace{-0.2cm} 403.12/1246 &\hspace{-0.2cm} 432.87/1215 &\hspace{-0.2cm} 364.12/1130 &\hspace{-0.2cm} 473.89/1254 &\hspace{-0.2cm} 588.24/1551 &\hspace{-0.2cm}   477.09/1336 &\hspace{-0.2cm}    596.67/1503 \\   

C-stat/dof &\hspace{-0.2cm} 1612/1529 &\hspace{-0.2cm} 1654/1526 &\hspace{-0.2cm} 1521/1529 &\hspace{-0.2cm}  1649/1529 &\hspace{-0.2cm}  1644/1530 &\hspace{-0.2cm} 1654/1528 &\hspace{-0.2cm} 1718/1530 &\hspace{-0.2cm}   1571/1527 &\hspace{-0.2cm}   1548/1528 \\ 

\hline
\end{tabular}
\end{center}
\end{table*}
%-----------------------------Figure Start------------------------------

% Two columns format
%-----------------------------Figure Start------------------------------
\begin{figure*}[ht]
\begin{center}
% un-comment the following line to include your fig1a.eps postscript file
\includegraphics[width=0.25\textwidth,angle=-90]{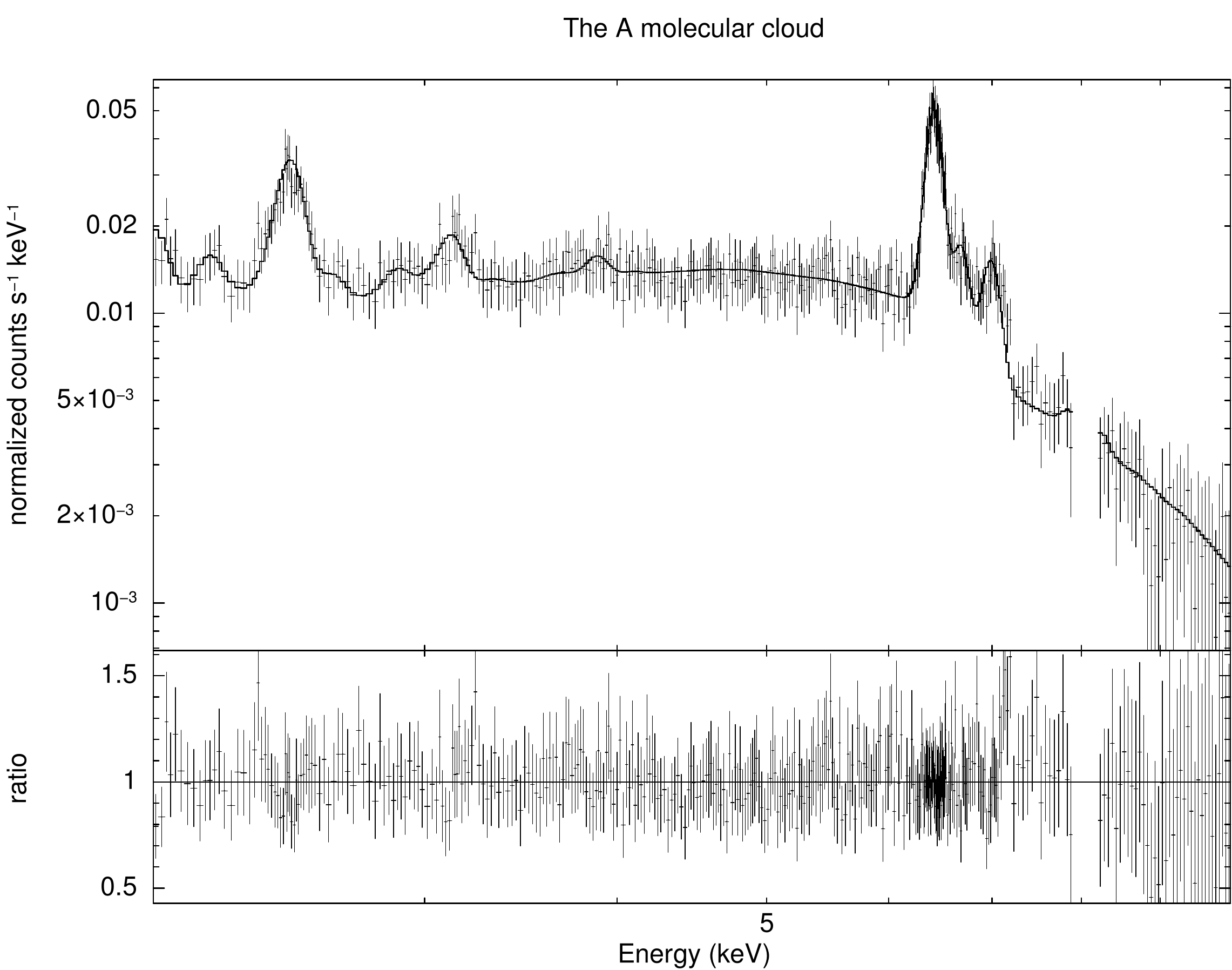}
\includegraphics[width=0.25\textwidth,angle=-90]{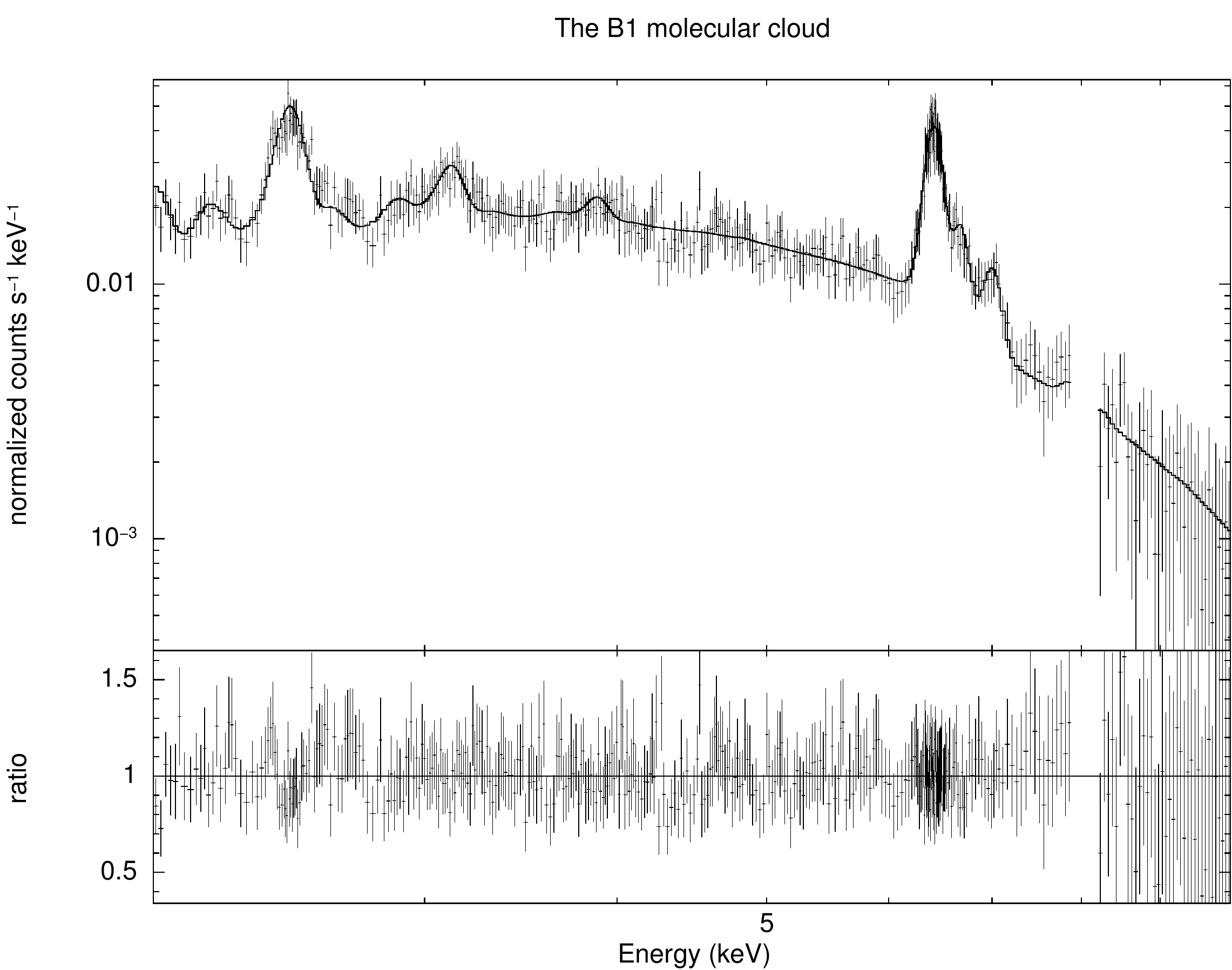}
\includegraphics[width=0.25\textwidth,angle=-90]{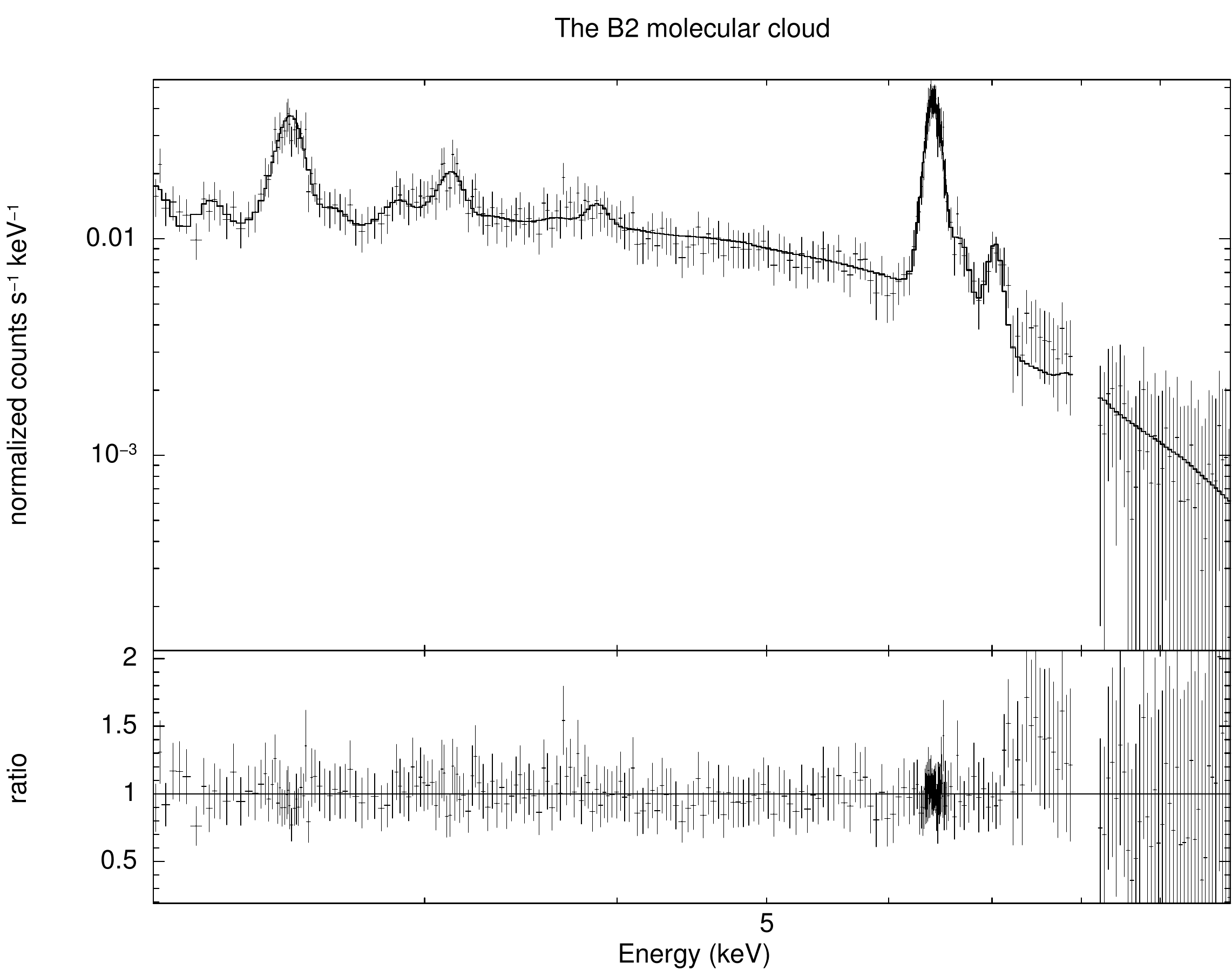}\\
\vspace{0.3cm}
\includegraphics[width=0.25\textwidth,angle=-90]{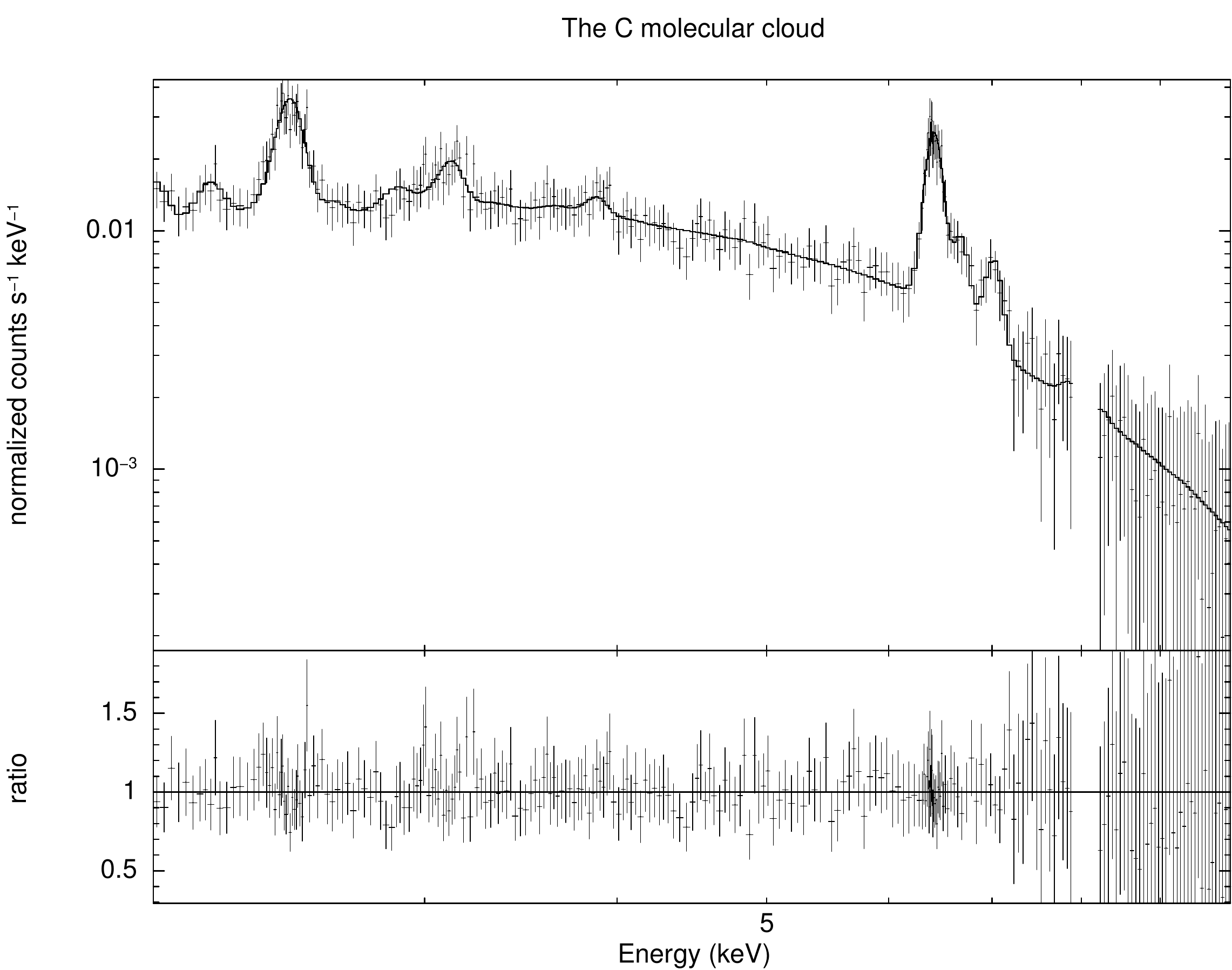}
\includegraphics[width=0.25\textwidth,angle=-90]{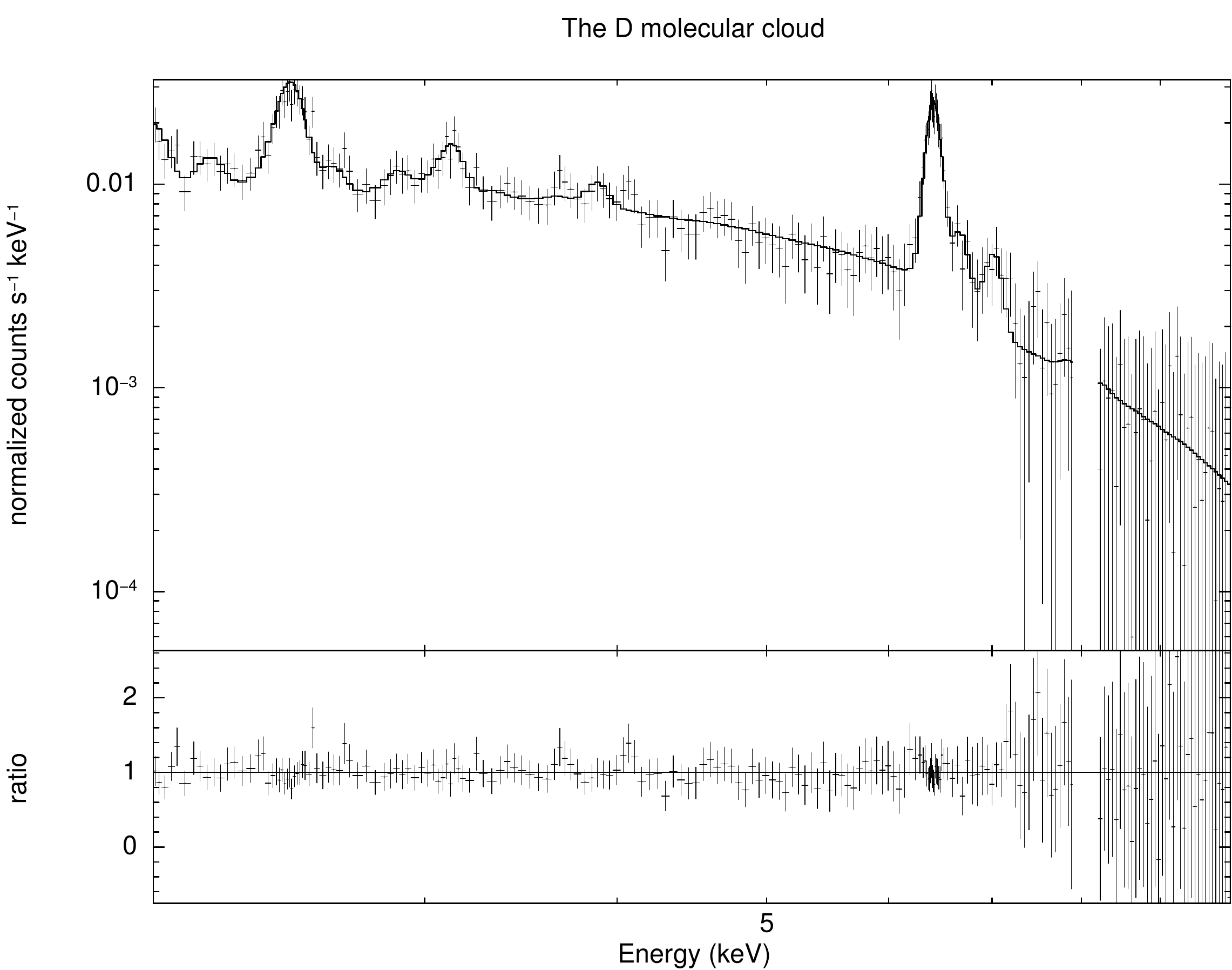}
\includegraphics[width=0.25\textwidth,angle=-90]{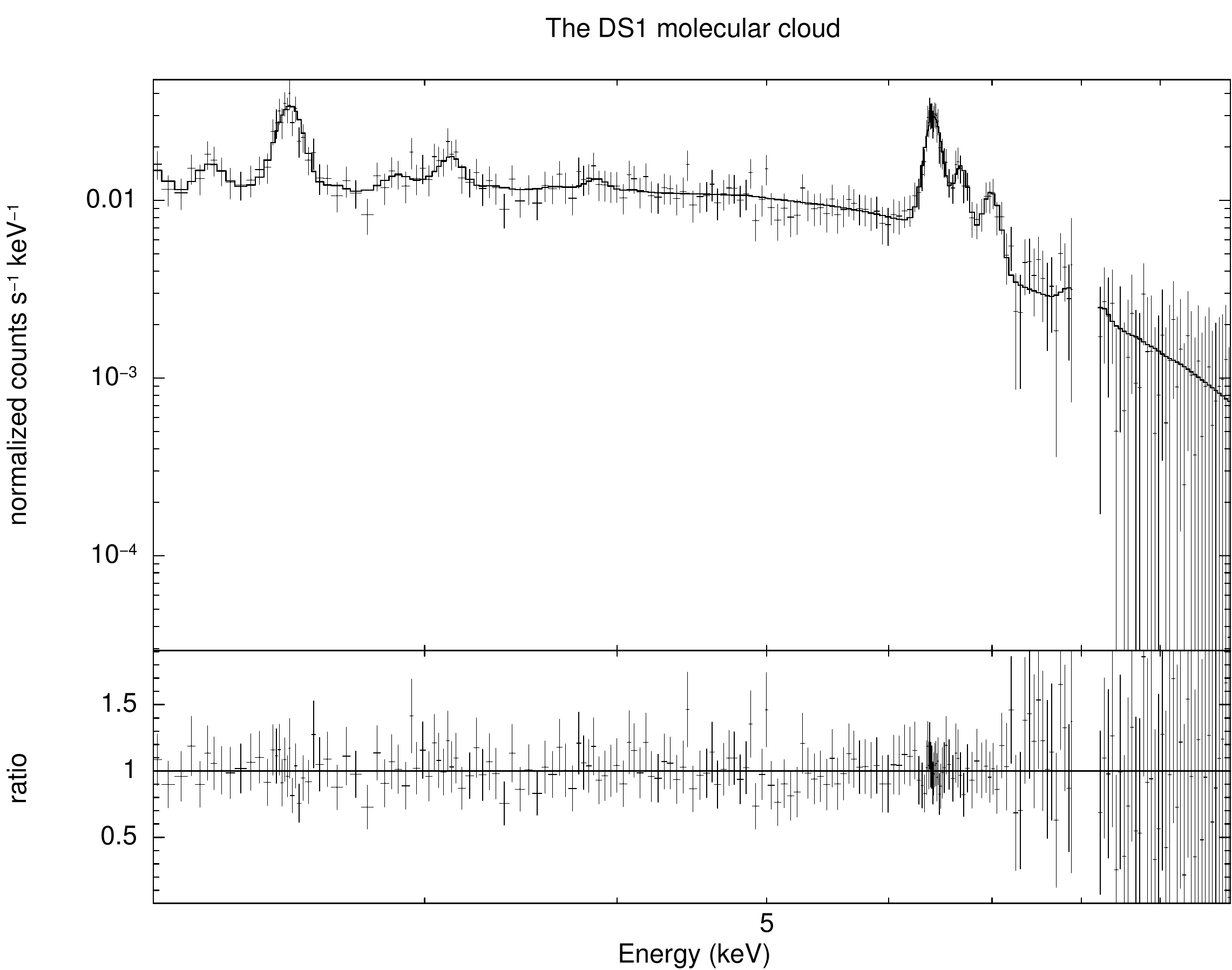}\\
\vspace{0.3cm}
\includegraphics[width=0.25\textwidth,angle=-90]{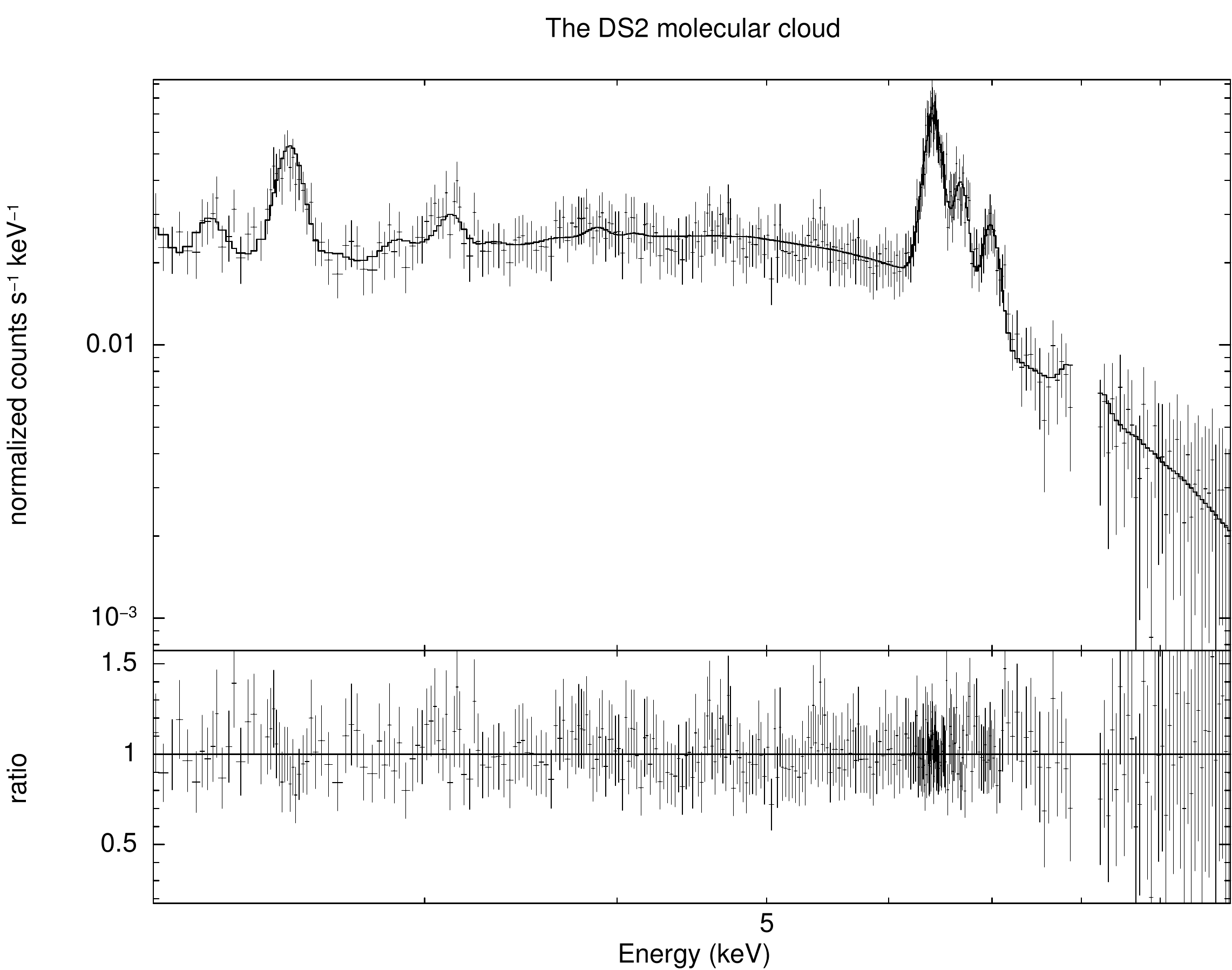}
\includegraphics[width=0.25\textwidth,angle=-90]{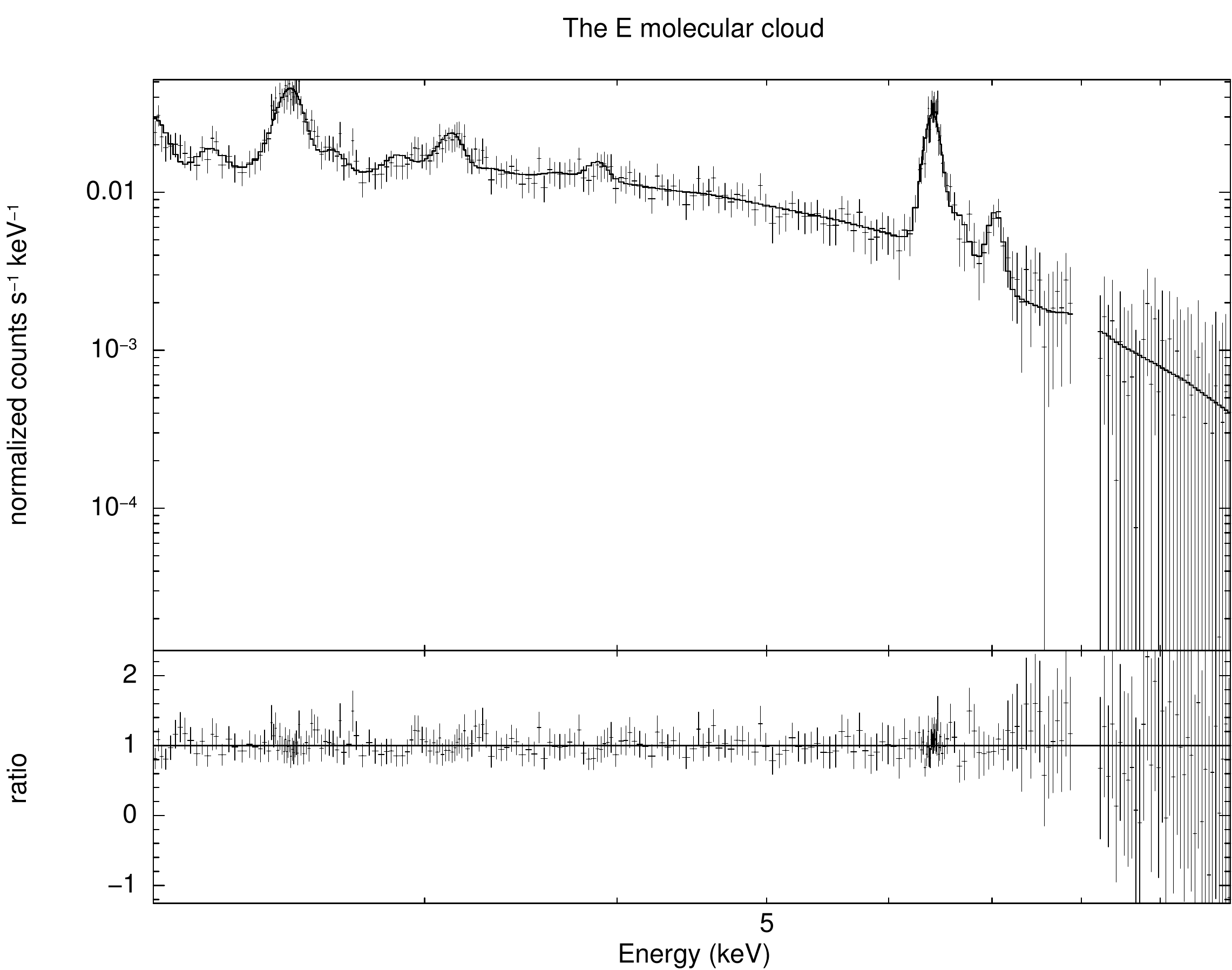}
\includegraphics[width=0.25\textwidth,angle=-90]{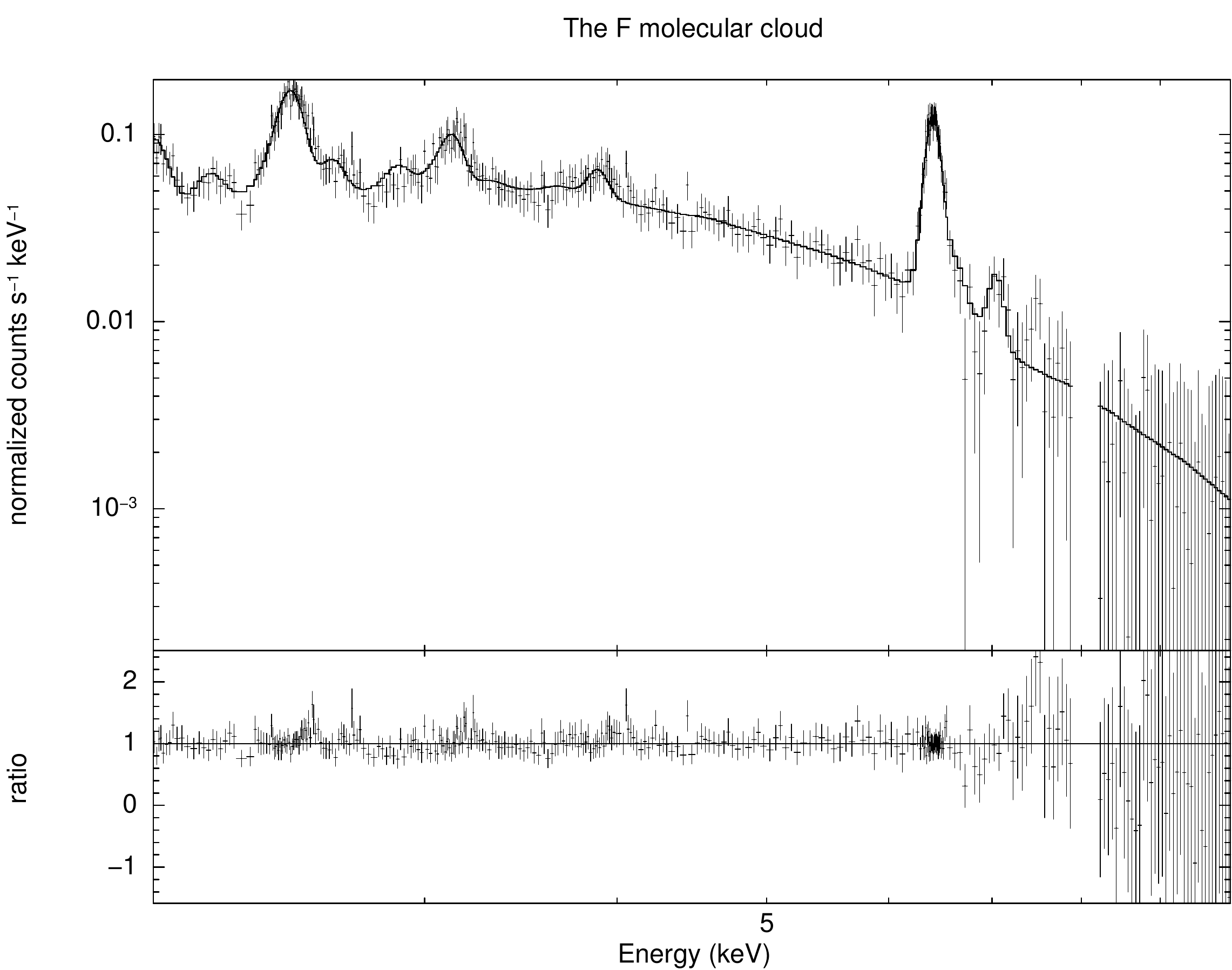}
\end{center}
\caption{Time-averaged PN spectra of the set of MCs in the present study.
Each panel shows the measured data along with the best fitting spectral model and 
(in the lower section) the ratio of these two quantities. The spectral channels encompassing
the Cu-K$\alpha$ line at 8.05 keV (7.8-8.2 keV) have been excluded from the fitting.}
\label{stack}
\end{figure*}
%-----------------------------Figure End--------------------------------

\vspace{0.1cm}

\noindent $\textit{Regions DS1 and DS2}$: both the low surface brightness
regions DS1 and DS2 show a constant 6.4-keV line signal over a period of 8 years. The 
weighted mean flux within this set of observations is 1.42$\pm$0.07 and 
2.33$\pm$0.09$\times$10$^{-5}$ photons~cm$^{-2}$~s$^{-1}$ respectively, which translates 
into a  surface brightness of 3.8$\pm$0.2 and 3.6$\pm$0.1$\times$10$^{-6}$ 
photons~cm$^{-2}$~s$^{-1}$~arcmin$^{-2}$. The $\chi^{2}_{red}$ values make us confident 
of the constant nature of the light curves although, as for region E, we found some little excursions in the
lightcurve of the region DS2.

\vspace{0.1cm}

To summarize, we have independently established a pattern of Fe-K$\alpha$ 
variability similar to what already found by \citet[][]{2010ApJ...714..732P} 
for the clouds A, B2, D, E and F.  For cloud B1, our analysis implies  
a decrease of the line flux at a confidence level of 2.6$\sigma$, 
whereas  \citet[][]{2010ApJ...714..732P} report a constant line flux for
this cloud (designated MC2 in their study); we attribute this significant difference to the use of
the background modelling. This result has implications for the location
of the cloud relative to the leading and trailing edges of the ionizing wavefront.
We have also studied three lower-surface brightness regions (C, DS1 and DS2),
the first of which shows an interesting step-wise increase in its fluorescent
line flux, with the latter two having constant light curves.

In order to  investigate the spatial dependency of  the observed Fe-K$\alpha$
variability, and show the complexity of its pattern across the inner CMZ, we sketched 
the distribution of the MCs studied in this paper, with a colour-coding 
which identifies the type of variability measured in the 6.4-keV line flux 
(see Fig.\ref{cartoon}).  The most noticeable increase in the Fe-K$\alpha$
flux comes from the zone bounded by 0.05$^{\circ}$$<$l$<$0.1$^{\circ}$ and
-0.1$^{\circ}$$<$b$<$-0.05$^{\circ}$  (l  and   b  are,  respectively,  galactic
longitude and latitude), which encompasses the regions B2 and D. Indeed, 
these molecular filaments  show  the clearest evidence for
Fe-K$\alpha$  variability across the selected regions. Besides these complexes, 
region C also shows an upwards trend in its Fe-K$\alpha$ surface brightness, 
whereas the G0.11-0.11 cloud shows fading 6.4-keV line emission. 
\citet[][]{2010ApJ...714..732P} have suggested that this complex pattern of variability
can be largely explained in terms of a single flare from Sgr A* illuminating molecular
clouds at differing positions along the line of sight. However, as we show below,
detailed consideration of the spectral properties of the various MCs lead us to
conclude that this is not the complete story.

\subsection{Spectroscopy: looking for the reflection imprints}\label{imprint}

\subsubsection{Method}

In order  to study the  spectral features typical  of an XRN,  we have
considered  the spectra from the PN camera  of the  regions  highlighted  in
Fig.\ref{cartoon} (see also Table \ref{capelli_reg}). For this analysis we  selected only 
PN spectra because of the much higher effective area in the hard X-ray band ($\gtrsim$6 keV). 
To build the PN spectra for the Fe-K$\alpha$ bright
filaments, we  used the same GTI  filtered event files  as employed to
construct the fluence map in Section 2. We stacked all the spectra (of
both the source  and the background) and the response files using the
ftools MATHPHA,  ADDARF and ADDRMF,  neglecting the 6.4-keV
line   variability. The background   region   was   centred   on
R.A.=17:45:42.289, DEC.=-29:07:53.39 (circle  1 arcmin in radius, see left panel of Fig.\ref{fluence}), at the 
same Galactic latitude  as the filaments in
order to avoid  any systematics due to the  different intensity of the
Galactic Ridge emission at  different Galactic latitudes. 

The spectral model  used  in  the  fitting  comprises two thermal  plasmas 
\citep[APEC,][]{2001ApJ...556L..91S}  with a metallicity fixed  at  2 times
solar \citep[e.g.,][]{2010PASJ...62..423N}, two gaussian  emission  lines (GAUSS)  which  account for  the  % just put in the answer to the referee report
Fe-K$\alpha$  and K$\beta$  lines \citep[the ratio of the line 
fluxes has been fixed to 0.11,][]{2009PASJ...61S.255K}, and a non-thermal power-law
continuum (POW), with a slope fixed at $\Gamma$=1.9.
The latter is the spectral index obtained from a joint fit of
the stacked PN spectra of all the MCs using the model defined below
(where we assume that primary illumination source is the same for all the XRN);
specifically the best-fit value was $\Gamma$=1.9$^{+0.1}_{-0.2}$.
All the previously listed components are then subject to a common photoelectric
absorption by the line-of-sight column density, N$\rm{_{H}}$
\citep[WABS$_{1}$,][]{1983ApJ...270..119M}.

We model the absorption within the MCs in terms of an intrinsic column
density N$\rm{_{H_{C}}}$ (WABS$_{2}$) plus an 
additional contribution to the optical depth of the Fe-K absorption edge 
at 7.1 keV. The latter allows the iron abundance relative to the solar value 
(Z$\rm{_{Fe}}$) in the MC to be greater than 1
(since the WABS prescription assumes solar metal abundances).
In our spectra model the extra optical depth $\tau$ (the free parameter in EDGE,
see below) and N$\rm{_{H_{C}}}$ are tied by the relation:

\[\rm{\tau=(Z_{Fe}-1) \times \sigma_{Fe} \times A_{Fe:H} \times N_{H_{C}} = \frac{N_{H_{C}}}{10^{22}} 
\times (Z_{Fe}-1) \times 0.0116,}\]

where A$_{\rm{Fe:H}} = 3.3 \times 10^{-5}$ is the (solar abundance) ratio of Fe to H atoms and 
$\sigma_{\rm{Fe}}$=3.5$\times$10$^{-20}$ cm$^{2}$ is the photoionisation cross-section
of the Fe-K shell.

In the event we fixed Z$\rm{_{Fe}}$ at a value of 1.6 (a choice which will be justified in
section \ref{metallicity}).
Note that N$\rm{_{H_{C}}}$ (and $\tau$) account for the absorption imprinted on the
reflected continuum by the MC both up to and after the point of scattering.  
A simple simulation shows that, to zeroth order, the value of N$\rm{_{H_{C}}}$ so
determined is comparable to the total column density through the cloud (although,
of course, the exact relation will depend on the geometry).
To summarize, the model used in the spectral analysis of the clouds is (in XSPEC format):

\medskip

WABS$_{1}$*[APEC+APEC+WABS$_{2}$*(GAUSS+GAUSS+ EDGE*POW)].

\medskip

In the spectral fitting procedure we ignored the channels encompassing
the Cu-K$\alpha$ line (7.8-8.2 keV) in order to minimize the effects of the
instrumental background on the modelling of the reflection component. Also, 
we again employed the Cash statistic because of the low signal-to-noise
ratio in the spectral channels and because its use gave 
tighter contraints on the derived parameter values.

In region F the normalization of the hot thermal component was compatible with 
being zero and therefore we excluded this component from the best-fit model. 
Also, in fitting the spectra from regions B1, C, D and E, we were forced to fix
the temperature of the hot component to kT=6.5 keV \citep[][]{2007PASJ...59S.245K}
in order to make the fit converge  (see Table \ref{edge_parameters}). 
Although  we have applied the  background subtraction
technique  in  this  study,  all  the spectra  still exhibit strong
residual contributions   from  the two APEC   components (as is evident
in Fig.\ref{stack}).

The specific goal of this analysis  was to quantify the intrinsic column density
(N$\rm{_{H_{C}}}$) of the MCs and also to establish the equivalent width (EW) of
the 6.4-keV line with respect to the underlying ionising 
continuum.  These important parameters will be of use in Sections 
\ref{cloud_properties} and \ref{200years}.

\subsubsection{Results}

\cite{1998MNRAS.297.1279S}  have  shown that  in  the  XRN model,  the
6.4-keV line emission has to be accompanied by strong absorption above
7.1 keV,  the minimum energy  that a photon  must have in order  to be
able to produce a K-shell  vacancy in neutral Fe via the photoelectric
effect. Moreover, because the primary source of ionisation is not seen
directly by  the observer, the  EW of  the line  is expected  to be  very high,
i.e. $\gtrsim$ 1 keV. On the other hand, the depth of the Fe
absorption  edge at  7.1  keV  is strongly  related  to the  intrinsic
column density of  the MC, with columns in excess of 10$^{23}$cm$^{-2}$
resulting in the imprinting of very strong Fe-K edge features on the 
emergent (\textit{i.e.,} electron-scattered) continuum.    
In    Fig.\ref{stack}    and    Table
\ref{edge_parameters}, we  present the results  of our search  for the
imprints of reflection on the time averaged spectra of the selected GC
molecular filaments.  We measure a high value for the  EW of the
6.4-keV line in  all the MCs studied. Interestingly the  largest EWs 
($\sim 1.5$ keV) were those of the B2 and  F clouds, both of which
exhibit pronounced Fe-K$\alpha$ flux
variability (a decreasing flux in the case of the F cloud and an
increasing flux for B2). Another very interesting finding is that 
the values of N$\rm{_{H_{C}}}$ typically lie in the range
6 to 20$\times$10$^{22}$ cm$^{-2}$, which is significantly higher
than previous estimates/assumptions for these clouds \citep[e.g.][]{2010ApJ...714..732P}. 

A very important result of our work is the identification of the hard non-thermal component
associated with the 6.4-keV line, which in the XRN model represents
the fraction of the incident continuum Thomson scattered by 
the molecular material. In  all the observed spectra we
were able to discern an underlying hard continuum, which via
a joint-fit to all the MC spectra, was modelled as a power-law
with a photon index $\Gamma \approx 1.9$. This value is very typical
of the continuum slopes that characterise the spectra of luminous
AGN.

We have also investigated the characteristics of the Fe-K$\alpha$ line,
in terms of its energy and width. For this purpose we built a time-averaged
spectrum for the whole region, stacking all the spectra across all the
observations. In this stacked spectrum,
the intrinsic width  of the  line is 36$\pm$3 eV, 
while its peak is at 6414$\pm$2 eV, slightly higher than the
nominal value of 6403 eV for neutral iron\footnote{The nominal values for the energy 
peak of the Fe-K$\alpha$ and Cu K$\alpha$ lines have been taken 
from \textit{http://physics.nist.gov}}. The results from the 
individual spectra (see Table \ref{edge_parameters}) show the same trend,
with the measured Fe-K$\alpha$  line  energy in all cases exceeding
the theoretical value. Although the EPIC calibration is known to be
better than 10 eV, the presence of  soft protons in  the FOV might  
slightly change the energy  and the width  of the  lines \cite[][]{2010ApJ...714..732P}.  
To check whether these are instrumental effects or real features,
we measured the same quantities for the Cu instrumental line at 8.05 keV (before cutting it out).
The centroid of the  Cu  K$\alpha$   line  is   at
8047$^{+4}_{-2}$ eV, while its  width is 34$^{+6}_{-4}$ eV. The line peak 
is in good agreement with its nominal value of 8048 eV.
Although a relative shifting between the two lines is present 
(the Cu K$\alpha$ peak perfectly matches the theoretical expectation, 
whereas the Fe-K$\alpha$ is noticeably higher), taking into account the systematics 
and the statistical uncertainties, we conclude that no statistically significant 
line blueshift and broadening has been measured and that our results 
are consistent with emission from neutral, or close-to-neutral, Fe atoms 
embedded in the material of the cold molecular clouds. 
The 11 eV upwards shift of the Fe-K$\alpha$ line central energy might be the result of very modest
ionisation of the Fe atoms (Fe II to FeX); in fact, the ionisation potentials for the low ionisation Fe atoms are 
between few 10eV and few 100eV, which suggests that these ionisation states of Fe might plausibly exist in GC MCs, 
given the peculiar environment.
We also note that 
Suzaku measured the Fe-K$\alpha$ line 
energy in the molecular filaments to be 6409$\pm$1 eV with a line width of 
33$^{+2}_{-4}$ eV, the latter being marginally higher than the expected 
systematics ($\approx$30 eV) \cite[][]{2007PASJ...59S.245K}.

The results of this spectral analysis will be used in Sections \ref{cloud_properties} and \ref{200years}
in order to derive a geometrical distribution of these
nebulae in the GC region and a X-ray lightcurve for the putative energising source.

\section{The derived properties of the MCs}\label{cloud_properties}

\subsection{Relative Fe abundance in the clouds}\label{metallicity}

In  this Section we  make use  of the  measurements of  the EW  of the
Fe-K$\alpha$ line from  the spectral analysis of the stacked PN spectra 
(see Table \ref{edge_parameters}) to infer the mean Fe abundance across
the MCs which compose our sample. The underlying assumption in this
section is that the Fe fluorescence arises as a result of the X-ray
illumination of the clouds by one or more X-ray outbursts in Sgr A*.
To proceed we first require an expression for Fe-line equivalent width, 
in terms of the intensity of the X-ray illumination and the cloud properties.
As    discussed   in \citet[][]{1998MNRAS.297.1279S},   the   Fe-K$\alpha$
line flux (F$\rm{_{64}}$) emanating from the illuminated cloud can be written as:

\begin{equation}\label{sun1}
\mathrm{F_{64}\approx\frac{\Omega}{4\pi D^{2}}~Z_{Fe}~\tau_{T}~I_{8}~~~~photons ~cm^{-2}s^{-1}},
\end{equation}

In this equation $\Omega$ is the solid angle (in units of 4$\pi$) subtended by the diffuse cloud 
from the perspective of the radiation source, D is the distance to the GC, Z$_{\rm{Fe}}$
is the Fe abundance relative to solar within the MC, $\tau_{\rm{T}}$ is the 
optical depth due to Thomson-scattering through the cloud and I$\rm{_{8}}$
is the photon flux from the source at 8 keV (in units of photon $\rm{s^{-1}~keV^{-1}}$).

In addition the scattered continuum at 6.4 keV measured at an angle $\theta$ with
respect to the direction of travel of the incident radiation is:

\begin{equation}
\mathrm{S_{64}=\frac{\Omega}{4\pi D^{2}}~\tau_{T}~(1 + cos^{2}(\theta))~I_{8}~1.172~~~~photons~cm^{-2}s^{-1}keV^{-1}},
\end{equation}

We can therefore write the EW of the Fe-K$\alpha$ line with respect to the reflected continuum as:

\begin{equation}\label{EW_theory}
\mathrm{EW \approx \frac{850 ~Z_{Fe}}{[(1 + cos^{2}(\theta)]} ~ eV}
\end{equation}

%-----------------------------Figure Start------------------------------
\begin{figure}[!Ht]
\begin{center}
% un-comment the following line to include your fig1a.eps postscript file
\includegraphics[width=0.35\textwidth]{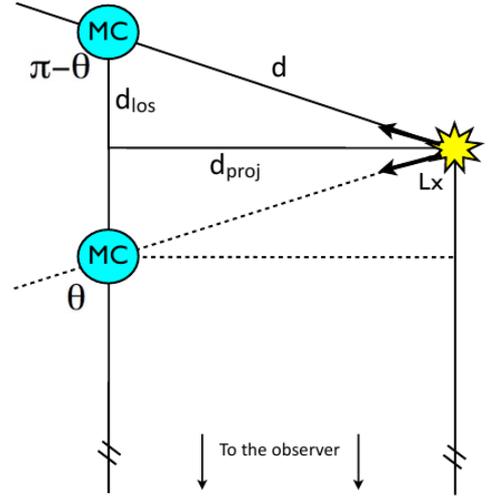}
\end{center}
\caption{The cloud-source geometry for the two cloud positions ($\theta$,$\pi-\theta$) which produce the same 
effective scattering. The total
distance from the cloud to the source (d), the projected distance on the
plane of the sky ($\rm{d_{proj}}$), and the line of sight displacement
with respect to the Sgr A* plane ($\rm{d_{los}}$) are all indicated, as are 
the scattering angles $\theta$ and $\pi-\theta$.}
\label{geometria}
\end{figure}
%-----------------------------Figure End--------------------------------

%-----------------------------Figure Start------------------------------
\begin{figure*}[!Ht]
\begin{center}
\includegraphics[width=0.45\textwidth]{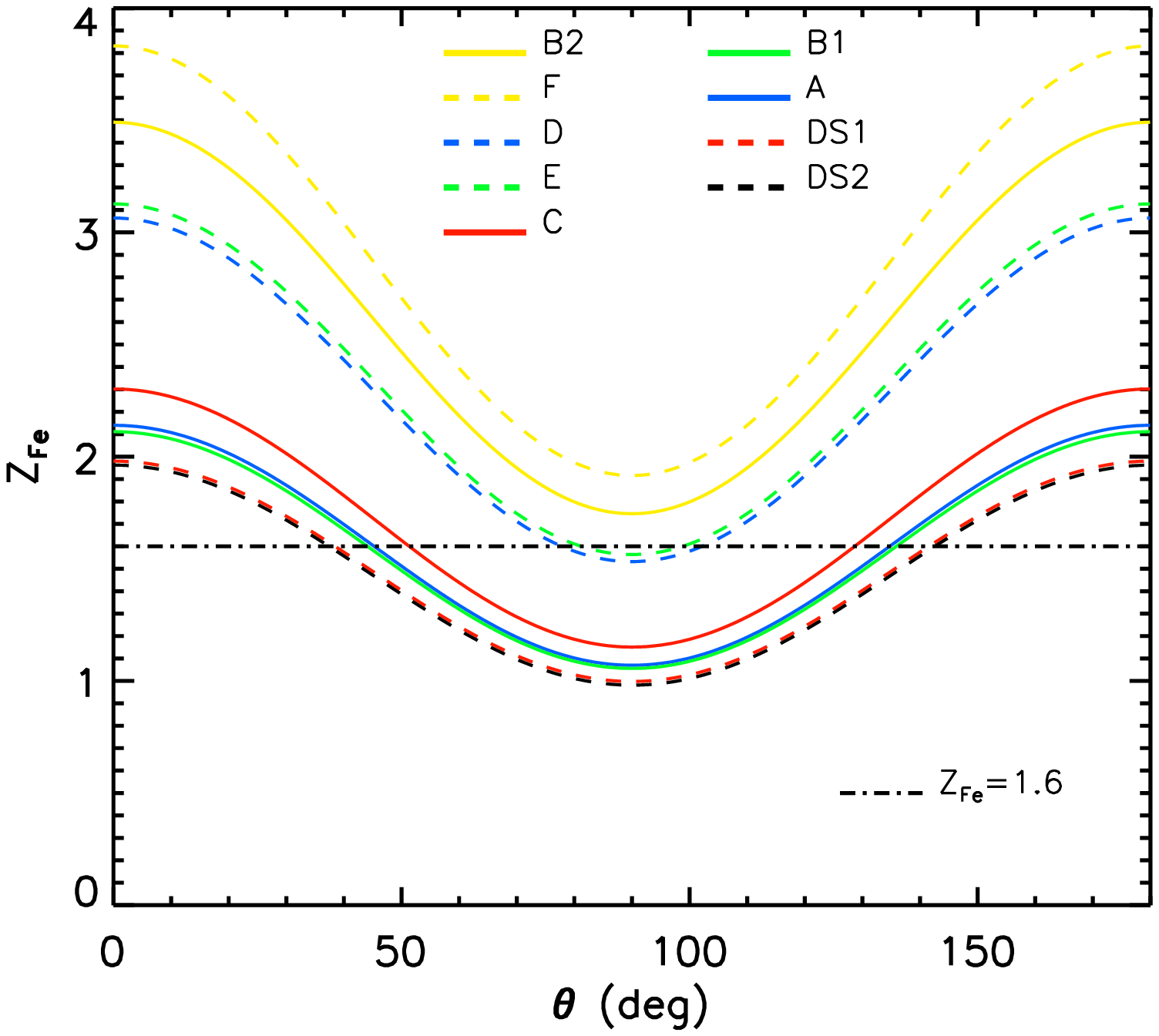} 
\includegraphics[width=0.45\textwidth]{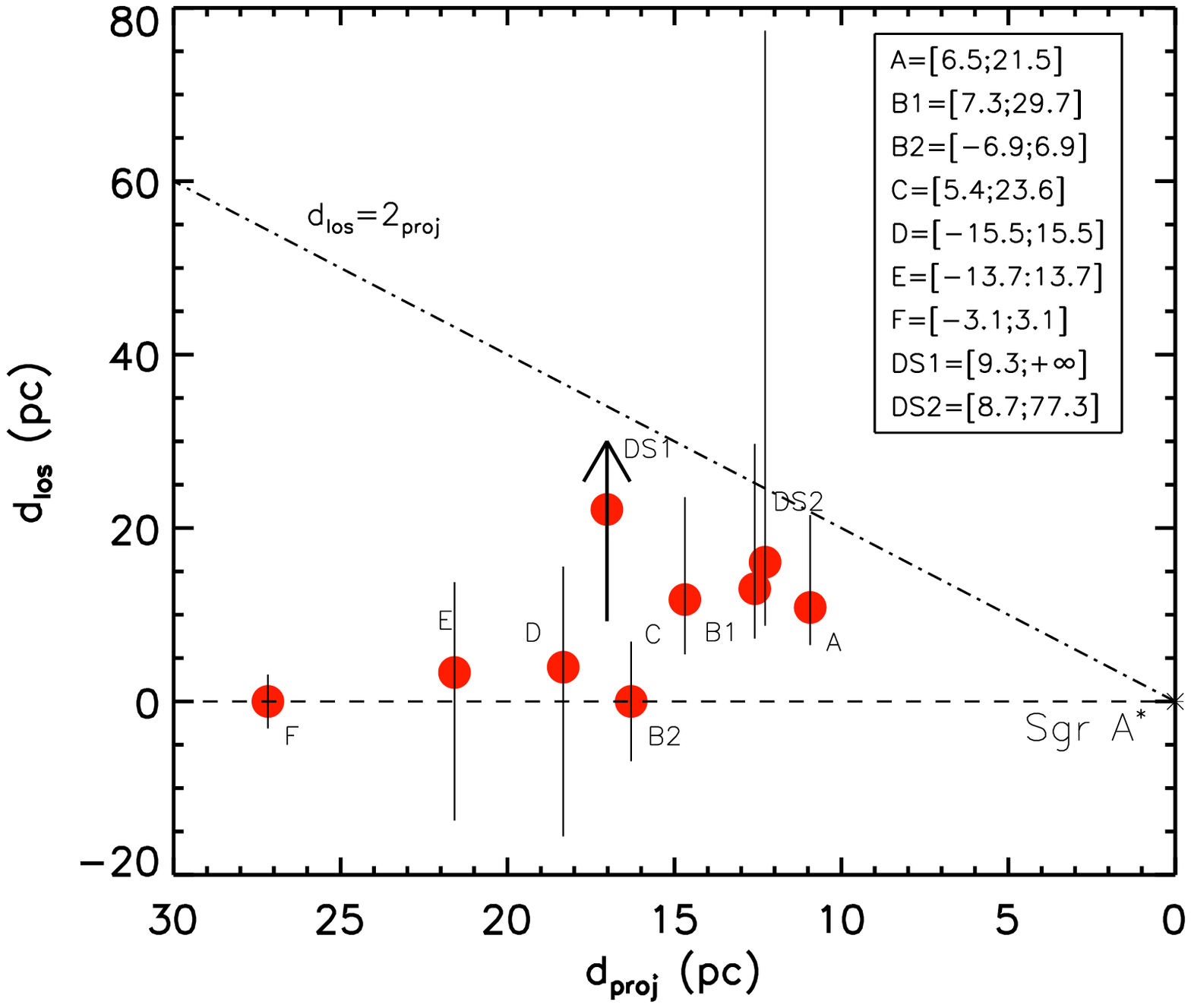} 
\end{center}
\caption{\textit{Left panel}: Relative Fe abundance  Z$\rm{_{Fe}}$ as a function of  the
scattering  angle $\theta$ for the different MCs obtained by using the measured EW and 
equation (3).  The horizontal dashed-dotted line  represents the average  metallicity 
(Z$\rm{_{Fe}}$=1.6) inferred for set of MCs. \textit{Right panel}: Line-of-sight displacement 
plotted against the projected distance d$\rm{_{proj}}$ for each MC. The d$\rm{_{los}}$ value for 
the DS1 cloud is a lower limit (d$\rm{_{los}}\geq$9.3 pc). The horizontal dashed line
represents the plane of Sgr A* at d$\rm{_{los}}$=0. The dashed-dotted  line shows the line
d$\rm{_{proj}}$=2$\times$d$\rm{_{los}}$.  We make the assumption that the clouds are
located in the region between the dashed and dashed-dotted lines. (see text).
The permitted d$_{\rm{los}}$ range (in pc) for each cloud is given in the top-right box.
Note: the method does not distinguish between whether an individual cloud lies in front or 
behind the plane of Sgr A* (\textit{i.e.,} at +ve or -ve d$\rm{_{los}}$).}
\label{geometry_ew}
\end{figure*}
%-----------------------------Figure End--------------------------------

In these calculations we assume that the cloud is not optically thick along any viewing direction. 
We also assume the photon index of the incident radiation is $\Gamma_{\rm{in}} \approx 2$; the impact of 
using $\Gamma_{\rm{in}} = 1.5$ is to increase the EW estimate by $\sim$ 20\%.

If we make the assumption  that Z$_{\rm{Fe}}$ is  constant across  the MCs
considered in this study,  then we can use equation \ref{EW_theory} to determine
the value of this parameter.
For this purpose we need to
make a geometrical assumption, namely that the clouds are distributed
such as to give a reasonably uniform distribution of scattering
angles within the range 25$^{\circ}\lesssim\theta\lesssim$155$^{\circ}$.
This is equivalent to assuming the distance of each cloud along
the line of sight, measured relative to the plane of Sgr A*,
is less than twice its projected distance from Sgr A*.
On this basis, we obtain $<$1+$\cos^{2}$($\theta$)$>$=1.33.
The mean Fe-K$\alpha$ EW based on the measurements for the 
nine  clouds reported in Table \ref{edge_parameters} is 1.0$\pm$0.1 keV,  
where we have given each measurement a weight proportional to
the inverse  of the  error squared.
Substituting this weighted  mean for the EW 
and the value for $<$1+$\cos^{2}$($\theta$)$>$ noted above
into equation \ref{EW_theory}, we obtain an estimate for
the relative abundance of Fe in the clouds of
Z$_{\rm{Fe}}$=1.6$\pm$0.1~Z$_{\odot}$.  
% [RENZO - I GET Z$_{\rm{Fe}}$=1.58$\pm$0.19~Z$_{\odot}$
% IE IS THE QUOTED ERROR CORRECT?]
Note that if we relax the geometrical constraint to allow the
clouds to be displaced along the line of sight
relative to the plane of Sgr A* by up to three times their
projected distance, then the relative abundance estimate
increases by  less than 5\%.
This  important result confirms a higher than solar metallicity
for the MCs and presumably the interstellar medium generally
in the GC region.

The geometry of the illumination process is shown in Fig.\ref{geometria}, which also defines the 
distance along the line of sight behind (or in front) the plane of Sgr A* (d$_{\rm{los}}$)
and the distance of the cloud as projected in the sky (d$_{\rm{proj}}$), as employed in our work.
In Fig.\ref{geometria}, 
two clouds are shown at the two angles $\theta$ and 180-$\theta$ for which the scattering
effects are the same. Troughout our work, we will always consider a positive d$_{\rm{los}}$ term
for the 180-$\theta$ case, and negative otherwise.

Another  interesting application  of equation  \ref{EW_theory}  deals with
the geometry  of the illumination process. In principle,
if the Fe abundance is tightly constrained, then the accurate
measurement of the EW of an individual cloud will allow 
the scattering angle $\theta$ to be determined and hence the cloud's
position in 3-dimensional space to be inferred (albeit with an ambiguity as to
whether the cloud is positioned in front of or behind the plane of Sgr A*). 
The process is illustrated in the left panel of Fig.\ref{geometry_ew},
which shows how the inferred relative abundance Z$_{\rm{Fe}}$ varies
as a function of the scattering angle $\theta$ for the set of clouds
in our sample.
The intersection of the horizontal line (corresponding to Z$_{\rm{Fe}}$=1.6)
with the plotted curve for a particular cloud determines the scattering
angle for that cloud (with the positional ambiguity arising from the
fact that the scattering angle can be $\theta$ or 180$^{\circ}$-$\theta$).
Although there is no such intersection for B2 and F clouds, 
when the errors
on the EW measurements (which are not shown in Fig.\ref{geometry_ew}) are taken into
we can also derive a range of possible angles for these two clouds.
The error range on all the scattering determinations for all the other
clouds can similarly be determined.

The results of the above analysis are plotted in the right panel in
Fig.\ref{geometry_ew}. This figure shows for each MC the 
line-of-sight displacement relative to the plane of Sgr A*
versus the projected distance from Sgr A* (on the plane of the sky)
as determined from the scattering angle constraints. 
In the figure, the MCs are positioned behind the plane of Sgr A* 
but in fact the method cannot distinguish whether
cloud is in front of or behind this plane (see above).
Fig.\ref{geometry_ew} also shows that our analysis is self-consistent in
the sense that the geometrical assumption employed when  calculating
the averaged Fe abundance across the set of clouds holds true
(\textit{i.e.,} all the MCs lie in the region bounded by the plane of Sgr A*
and the straight line d$_{\rm{los}}\leq$ 2$\times$d$_{\rm{proj}}$).
Unfortunately, given the current EW measurement errors, the scattering
angle determinations results in only weak constraints on the locations
of a few of the clouds.

In all of these calculations,  we have assumed that the MCs are illuminated
by the same powerful X-ray source,  in a pure XRN scenario. Clearly if
there is a significant contribution to the Fe-K$\alpha$ line from
an alternative process, such as CR bombardment, then this would impact on
these results. However we show in Section \ref{low_sb} 
that at least for brightest Fe-line emitting clouds, X-ray illumination
is the most likely excitation source of Fe fluorescence. Potentially, the method 
outlined above provides a powerful tool for the investigation of the illumination 
geometry of XRN, particularly when coupled with studies at radio and other
wavelengths which may serve to resolve the ambiguity in the derived geometry.

\subsection{Column densities of the clouds}\label{NH_calculation}

In the XRN scenario,  both the column density of Fe atoms within the cloud and
the distance of the cloud from the illuminating source are the crucial parameters,
when  estimating the X-ray luminosity required to produce a given
Fe-K$\alpha$ flux. Previous studies have often determined 
the former using estimates of the hydrogen column density for the cloud,
N$\rm{_{H_{C}}}$, combined with the solar
Fe:H ratio and an assumed value for Z$_{\rm{Fe}}$. 
In turn, the estimates of N$_{\rm{H_{C}}}$
have often been inferred from
intensity maps of CO, CS  or other  molecular  tracers  of  high
density  material  in  the  inner Galaxy. However, the
estimation  of   the  cloud N$_{\rm{H_{C}}}$ from molecular line
observations is notoriously difficult with different measurement techniques
often giving different results. For  example, the G0.11-0.11 cloud (our region F) has
been the subject of numerous studies over the last decade. Using, respectively, the
intensity  of   the  CS  and  the   H$^{13}$CO$^{+}$  emission  lines,
\citet[][]{2009ApJ...694..943A}   and  \citet[][]{2006ApJ...636..261H}
measured   two  extreme values for  the N$\rm{_{H_{C}}}$ of this   MC, namely
2$\times$10$^{22}$  and 10$^{24}$  cm$^{-2}$,  respectively. 

A better approach is to measure the column density of Fe atoms in
the cloud directly from the X-ray spectrum. This is essentially the approach
we have taken in the spectral fitting reported in Section \ref{imprint},
where the Fe column density can be obtained from the quoted 
N$\rm{_{H_{C}}}$ values as:

\[\rm{N_{Fe} = N_{H_{C}} \times A_{Fe:H} \times (Z_{Fe}) } =  N_{H_{C}} \times 5.3 \times 10^{-5}~~~cm^{-2} \]

In practice, the fact that we were able to set Z$\rm{_{Fe}}$ = 1.6 and
the spectral index of the underlying power-law continuum (arising from reflection)
to $\Gamma$=1.9,
resulted in a reasonable robust set of N$\rm{_{H_{C}}}$ measurements.

The N$\rm{_{H_{C}}}$ values we derived for our set of 9 clouds are summarised in Table \ref{NH}, where a comparison is made 
with the earlier estimates (based on radio molecular line measurements) by \citet[][]{2010ApJ...714..732P} 
for the same MC sample. Clearly the X-ray measurements lead to substantially higher estimates of the cloud
column densities, more similar to the values measured within the Sgr B2 complex by \citet[][]{2011ApJ...739L..52N}. 
This in turn has a significant impact on the inferred geometry and time delays 
within the cloud system (see below).
To conclude, we note that the XRN parameters we have found for the MCs within 30 pc of Sgr A* are in good agreement with the 
ones for the Sgr B2 MC found by \citet[][]{2011ApJ...739L..52N}, i.e. Z$_{\rm{Fe}}$=1.3$\pm$0.3 Z$_{\odot}$ and $\Gamma$=2.5$\pm$0.6.

%*****************************************************************************
\begin{table}[!Ht]
\caption{Comparison between the cloud density measurements. The first line 
reports the N$\rm{_{H_{C}}}$ values employed by \citet[][]{2010ApJ...714..732P},
whereas the second lists the results from this paper (see Table \ref{edge_parameters}). 
All values are in units of 10$^{22}$ cm$^{-2}$.}
\begin{center}
\begin{tabular}{|ccccccccc|}
\hline
A & B1 & B2 & C & D & E & F & DS1 & DS2 \\
\hline
4.0 & $\leq$2.0 & 9.0 & -- & 9.0 & 9.0 & 2.0 & -- & -- \\ 
18.4 & 10.2 & 12.3 & 5.8 & 13.2 & 9.6 & 9.2 & 15.5 & 14.5 \\
\hline
\end{tabular}
\end{center}
\label{NH}
\end{table}
%**************************************************************************

\section{The X-ray lightcurve of Sgr A* over the last 200 years}\label{200years}

\subsection{Placing constraints on the lightcurve of Sgr A*}

If the number of Fe atoms in a particular MC is known, it is possible to
use the observed Fe-K$\alpha$ flux to estimate the cloud's position relative to
the illumination source, assuming that the cloud remains optically
thin at the line energy. This method   has    been    recently
employed by \citet[][]{2010ApJ...714..732P}, who assumed that the MCs in their
study are all illuminated by a past outburst on Sgr A* reaching an
X-ray luminosity of $\sim$10$^{39}$ erg~s$^{-1}$. However, in the previous section (\S
\ref{NH_calculation}) we noted that the N$\rm{_{H_{C}}}$ values used by these
authors may be severe underestimates, implying
the need for a substantial revision of the cloud-source geometry.

Following the same approach as discussed by \citet[][]{1998MNRAS.297.1279S},
equation \ref{sun1} can be used to derive an expression for the
separation, d, of the MC from the ionising source as:

\begin{equation}\label{los}
\mathrm{d^{2}=\frac{10^{7}~L_{X}~\tau_{T}~R^{2}~Z_{Fe}}{16\pi ~D^{2}~F_{64}}}
\end{equation}

In this formula $\tau_{\rm{T}}$ is the optical depth of the MC to Thomson scattering
and R is the radius of the cloud. The other parameters are the X-ray luminosity L$_{\rm{X}}$
of the source (in a nominal ~ 2--10 keV bandwidth),
the relative Fe abundance Z$_{\rm{Fe}}$  within the cloud, the distance to 
the GC D and the measured Fe-K$\alpha$ line flux F$\rm{_{64}}$. In applying this
formula we make the assumption that the cloud is sufficiently symmetric such that
we can employ the values of $\tau_{\rm{T}}$ and R determined by
observation along our particular line of sight.
 
If we assume that the clouds are illuminated by a past outburst on Sgr A*
with L$_{\rm{X}}$=1.4$\times$10$^{39}$ erg~s$^{-1}$ \citep[e.g.][]{2010ApJ...714..732P} 
then we can use equation \ref{los} to calculate the $d$ for each of our 9 clouds, 
since we already know all the other quantities. Once the value of $d$ is derived, 
we can determine the  line of sight displacement d$_{\rm{los}}$ of each cloud by 
using the projected separation from Sgr A* d$\rm{_{proj}}$ on the plane of the sky, 
(see Table \ref{capelli_reg}). Again there is an ambiguity as to whether the cloud
resides in front of or behind the plane containing Sgr A*. Employing this method
we find that the spread along the line of sight of the MCs is almost
an order of magnitude larger than the spread in the projected  distances.
Moreover, the inferred line of sight distribution of the MCs
is incompatible with the 3D distribution we derived from the
EW relation in Section \ref{metallicity}. The implication is that we must
tune (i.e. reduce) the luminosity of the ionizing source
to match the observed Fe-K$\alpha$ characteristics of the clouds.

Ideally, rather than assuming a particular L$_{\rm{X}}$ for the outburst,
we should build a model for  the X-ray activity of  Sgr A*
over the  last few  hundreds years,
which accounts  for all the Fe-K$\alpha$ variations measured in
the MCs in our sample. Unfortunately the uncertainty in 
the position of the clouds along our line of sight (relative to the plane containing
Sgr A*) hampers this endeavour.

In the general case, when employing a geometrical model such as the one shown in Fig.\ref{geometria},
the luminosity of the outburst is given as: 

\begin{equation}\label{lx-formula}
\mathrm{L_{X}=2.9\times10^{39}~\left[\frac{F_{64}~(d_{\rm{proj}}/\sin(\theta))^{2}}{R^{2}~\tau_{T}~Z_{Fe}}\right]~erg~s^{-1}},
\end{equation}

Also, the delay of the reflected signal relative to the direct light from Sgr A* is:

\begin{equation}\label{delay}
\mathrm{\Delta t=\frac{1}{c}~d_{\rm{proj}}~\left[\sqrt{1+\cot^{2}\theta} - \cot\theta\right]~s.}
\end{equation}

By allowing $\theta$ to vary over a particular range, one can use the above equations
to produce the track of L$_{\rm{X}}$ versus $\Delta$t matching the Fe-K$\alpha$ line emission
and other properties of the particular cloud.
Fig.\ref{lc_angles} shows the result of this analysis, where 
$\theta$ was constrained to vary within the interval $25^{\circ}-155^{\circ}$ as in Section \ref{metallicity}.

However, we can also use the results from Section \ref{metallicity} which are summarised in the
right-hand panel of Fig.\ref{geometry_ew} to add the further constraints in 
Fig.\ref{lc_angles}, represented by the vertical lines.

In Fig.\ref{lc_angles} we have split the clouds into subgroups depending on whether
they exhibit a decreasing flux (clouds B1 and F, top panel), an  increasing flux
(clouds B2, C and D, middle panel) or  a constant level (clouds  A, E, DS1 and DS2,
bottom  panel). In  each of  the three  panels  the
different colours refer to different clouds with the upper and lower limit
ranges on   L$_{\rm{X}}$ shown by the thinner curves of the same colour.
Note that in deriving these plots we use the Fe-K$\alpha$ fluxes from the
analysis of the stacked PN spectra (\textit{i.e.,} the plots are based on the average
Fe-K$\alpha$ flux of the cloud).

The challenge is to see whether there is a restricted pattern of variability in
Sgr A*, which plausibly matches all of the constraints implied by Fig.\ref{lc_angles}.
Specifically we are interested in those  L$_{\rm{X}}$-delay combinations
within each of sub-groups, which provide a consistent description
of the Sgr A* variability (\textit{i.e.,} where the L$_{\rm{X}}$ values for the different clouds
within a particular delay range are compatible with each other).
Of course, given the limitations we should bear in mind that there can be no guarantee
that any particular solution is unique.

\vspace{0.3cm}
\noindent\textit{The B1 and F clouds}: 
 
In  the top panel  of Fig.\ref{lc_angles}  we can  see that  the L$_{\rm{X}}$
curves for the  B1  and F  clouds  show a reasonable match around an  X-ray
luminosity  of  $\sim$4$\times$10$^{37}$ erg s$^{-1}$ 
and  a  time  delay around $\sim$100 years.
If these clouds are responding to the same decrease in the incident L$_{\rm{X}}$,
then we might infer that the scattering angle for the cloud F is
$\sim$95$^{\circ}$ and that of the cloud B1  near to 135$^{\circ}$. This places the
cloud F roughly on the same plane as Sgr A* and the cloud B1 15 pc {\textit{behind it} 
(\textit{ie} note that this approach can in principle
resolve the positional ambiguity alluded to earlier).  For both clouds the measured
excursion of the Fe-K$\alpha$ flux in their lightcurves  is about factor of two, 
implying a commensurate decline in the L$_{\rm{X}}$ of Sgr A* over 7  years. 
As shown in Fig.\ref{lc_angles}, the geometry inferred for these two MCs agrees remarkably
well  with   the   results  obtained earlier (see section 4.1).

In conclusion, a drop  in the X-ray luminosity of Sgr A* by a factor of 
two, from $\sim 4 - 2 \times$ 10$^{37}$~erg~s$^{-1}$ 
around 100 years ago may explain the decaying Fe-K$\alpha$
line flux currently observed from the B1 and F clouds.

\vspace{0.3cm}
%-----------------------------Figure Start------------------------------
\begin{figure}[!Ht]
\begin{center}
% un-comment the following line to include your fig1a.eps postscript file
\includegraphics[width=0.5\textwidth]{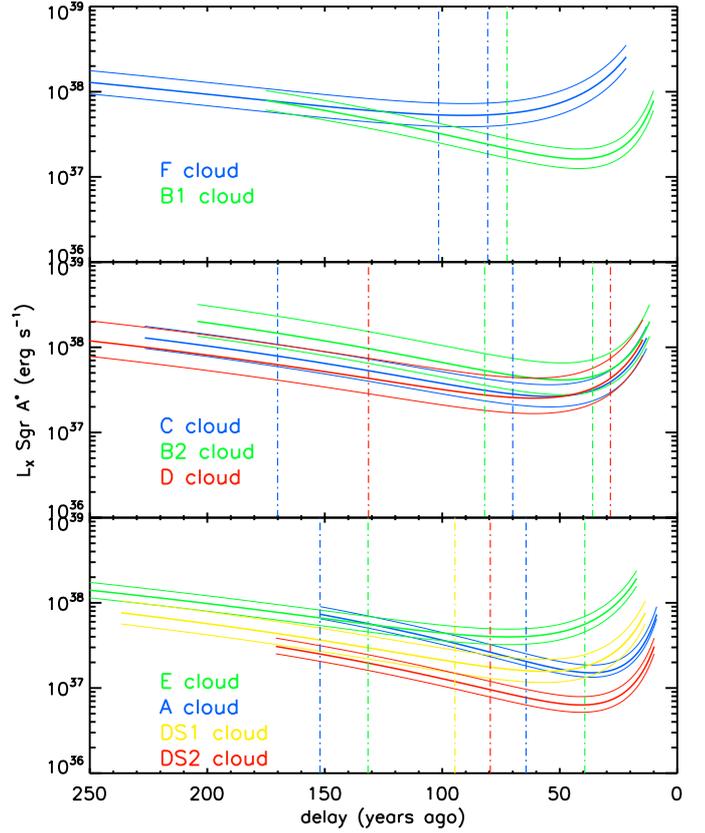}
\end{center}
\caption{X-ray luminosity as a function of the time delay
for the different MCs, as inferred from the measured cloud properties.  
We have divided the MCs into subgroups depending on the type of Fe-K$\alpha$
line variability exhibited. Also plotted are the 90\% confidence ranges on
the L$_{\rm{X}}$ calculation. The vertical dotted lines represent the constraints 
on $\theta$ derived in Section \ref{metallicity}. Where only one vertical
line is shown for a cloud, it represents the right-hand boundary of the allowed
range; in these cases the second boundary is situated to the left of
the Y axis (\textit{i.e.,} $\theta\geq$155 deg), and therefore not plotted. } 
\label{lc_angles}
\end{figure}
%-----------------------------Figure End--------------------------------

\noindent\textit{The B2, C  and D clouds}: 

In the same  way we can
try to put together a scenario which explains the steady increase
in the Fe-K$\alpha$ flux observed from clouds B2, C and D clouds.
(middle panel of Fig.\ref{lc_angles}). Here the situation is more
complex, since the tracks for the three clouds cluster together in
a fairly tight fashion along a broad extent of the time axis.  
In order to restrict the possibilities we need to bring in extra
information as follows. For example, we can perhaps safely 
exclude the time delays $\lesssim$ 50 years ago, since 
the requisite activity in Sgr A* might well have been observed
on this timescale in the early radio observations of the GC
(with X-ray observations in the 1980s certainly
excluding delays of less than 30 years). 
Another good argument which reinforces this assumption is the level of 6.4-keV
line emission  measured in the 50  km~s$^{-1}$ and 20 km~s$^{-1}$  MCs. 
These clouds
are most probably closer to  Sgr A* than the ones studied in  this paper, and should
at the present time show  a significant Fe-K$\alpha$ line  flux, if illuminated
by an L$_{\rm{X}}\sim$10$^{37-38}$ erg~s$^{-1}$ within the last 30 years.

%-----------------------------Figure Start------------------------------
\begin{figure*}[!Ht]
\begin{center}
%un-comment the following line to include your fig1a.eps postscript file
\includegraphics[width=0.9\textwidth]{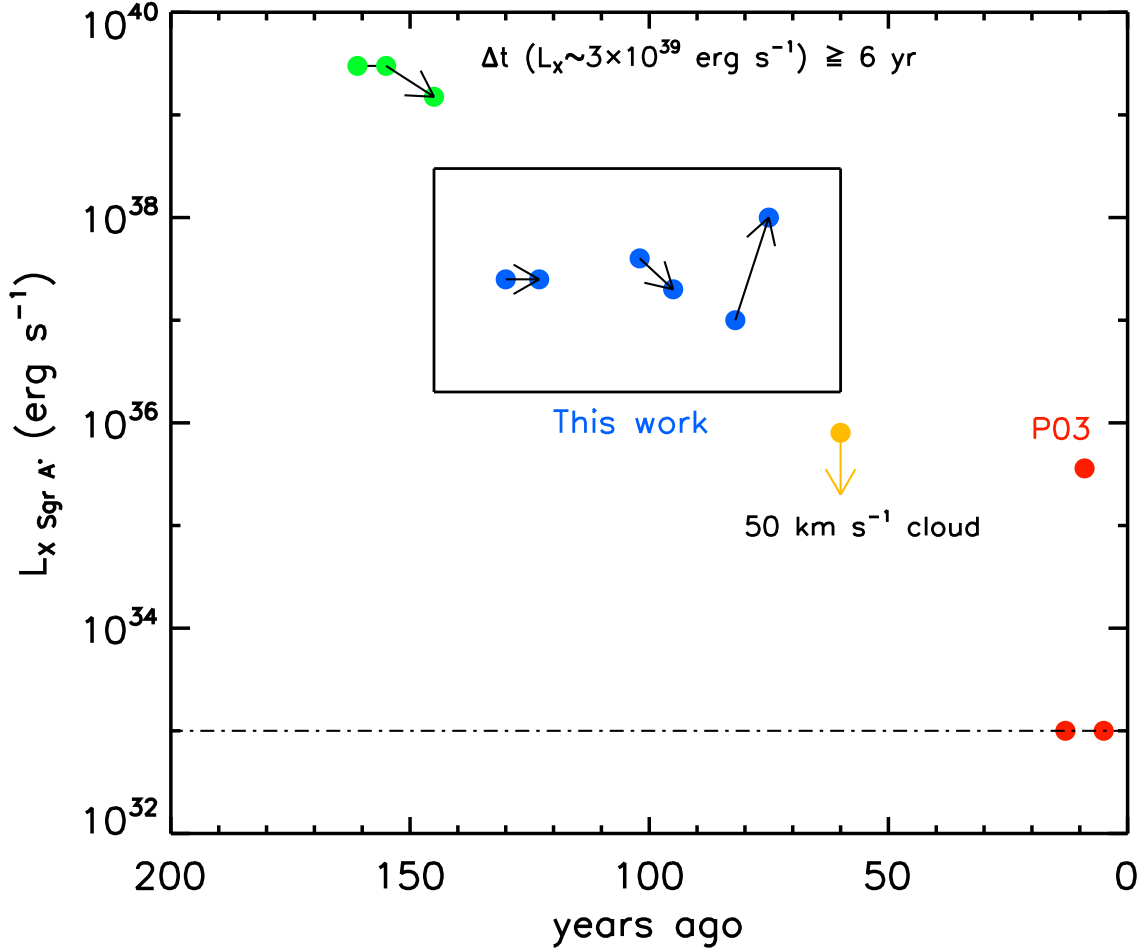}
\end{center}
\caption{X-ray light curve of Sgr A* over the past 250 years. The green points at 
L$_{X}\sim$10$^{39}$ erg s$^{-1}$ show the constraints on the Sgr A* activity inferred 
from the Fe-K$\alpha$ measurements in the Sgr B2 cloud \citep[see text,][]{2009PASJ...61S.241I,2010ApJ...719..143T}. 
The red points show the current quiescence level of Sgr A*, together with the brightest 
flare ever measured \citep[][P03]{2003A&A...407L..17P}. The blue points in the box show 
the results of this work.}
\label{lc_sgrastar}
\end{figure*}
%-----------------------------Figure End--------------------------------

Employing the same method as for the B1 and F clouds allows us to select a rather narrow range 
of time delay for the B2, C and D clouds; this is the intersection of the three areas in between the vertical boundaries
in the middle panel of Fig.\ref{lc_angles}. The lightcurves of the three regions match well with one another 
and with the results found in Section \ref{metallicity} for a time delay of between 70 and 82 years, 
for an X-ray luminosity in the range 10$^{37}$-10$^{38}$ erg s$^{-1}$. The corresponding angles, and therefore distances 
behind the plane of Sgr A*, for these complexes are about 110$^{\circ}$, 120$^{\circ}$ and 105$^{\circ}$ respectively, 
which translates into 
a distances of 6, 8.5 and 5 pc respectively.

We note that the B2 and D clouds may be subregions of a larger molecular
complex called the \textit{Bridge} by \citet[][]{2010ApJ...714..732P}.
These authors claimed to  have discovered a superluminal echo in
X-ray reflection, due to the  propagation of the ionising front within
a MC  located about 60 pc  behind the plane of Sgr A*.  If we
assume that the B2  and D clouds are part of one larger complex,  then to
explain the  superluminal Fe-K$\alpha$ line variability  we need to
locate the clouds behind Sgr A*, specifically at a distance greater
than about 8 pc (two times the sum of the linear sizes of the regions B2 and D).
There is therefore a slight disagreement between this requirement and
the position inferred for clouds B2 and D using the combined results from the 
spectral analysis and the EW-geometry relation. 

To conclude, the observed increase of the Fe-K$\alpha$ line flux in the B2, D and C clouds may be
well explained by a short enhancement of the X-ray luminosity of Sgr A* in the range 
10$^{37}$-10$^{38}$~erg~s$^{-1}$ about 80 years ago.

\vspace{0.3cm}
\noindent\textit{The A,  E, DS1 and  DS2 clouds}: 

The bottom  panel of Fig.\ref{lc_angles} shows  the L$_{\rm{X}}$ curves for the constant clouds. 
We can clearly see that  the different segments are  spread  over  a large  area  of  the  plot
and it  is  not straightforward to pick out one particular pattern of
activity in Sgr  A* as the most likely solution. In particular, the curve for the DS2 cloud
does not intersect the curves for the A and E clouds.  However, the measurements do at least
indicate an L$_{\rm{X}}$ range rather similar to that inferred from
the variable clouds.

From the analysis of the geometrical constraints derived in Section \ref{metallicity} and plotted in Fig.\ref{lc_angles},
we can try to derive a range for the time delay (and therefore position) for the clouds; by intersecting all the different regions,
we propose that the constant activity at L$_{\rm{X}}\sim$10$^{37}$ erg~s$^{-1}$ could have been experienced by Sgr A*
between 130 and 100 years ago. Assuming a time delay of 120 years ago, this translates into a distance behind the plane of 
Sgr A* for the clouds of 12, 16, 14 and 15 pc, respectively.

Finally, in the case of the clouds with constant Fe-K$\alpha$ flux, the displacement between the different
curves in Fig.\ref{lc_angles} might be explained  in terms of contribution
of other excitation processes to the total Fe fluorescence (see Section \ref{low_sb}).]

\subsection{The past lightcurve of Sgr A*}

The analysis of the previous section demonstrates very clearly that the Fe-K$\alpha$
bright clouds within 30pc of Sgr A* are not responding to the same outburst on Sgr A*
that appears to have illuminated Sgr B2. On the contrary, the observed
Fe-K$\alpha$ line fluxes and inferred geometry require an X-ray luminosity of Sgr A*
more typically in the range 10$^{37-38}$  erg~s$^{-1}$ over 
the last 100-150 years, in very good agreement with \citet[][]{2007ApJ...656L..69M}.  
Within this time frame, upwards and downwards
trends lasting from a few to ten years also seem to have occurred.

As a natural step forward, we have attempt to build a long term light
curve of the X-ray activity of  Sgr A*, incorporating both published
information and our new results - see Fig.\ref{lc_sgrastar}.
The first measurement (the green points in Fig.\ref{lc_sgrastar}) is
the  X-ray luminosity  required  to explain  the  strong 6.4-keV  line
emission in the Sgr B2 MC. Almost   since    its    discovery
\citep[][]{1996PASJ...48..249K},  there has  been a  general agreement
that the fluorescent of Sgr B2 requires a rather powerful X-ray flare on Sgr
A*,  reaching a  luminosity  of  a few $\times$10$^{39}$  erg~s$^{-1}$
a few hundred years  ago. The end  of  this  low luminosity  AGN
activity has recently been dated  back to 100$^{+55}_{-25}$
years ago  \citep[][]{2010ApJ...719..143T}.
Accordingly to \citet[][]{2009PASJ...61S.241I}, the 6.4-keV line flux from Sgr B2 was 
constant for at least six years (the time of the ASCA first measurement) before the onset of the 
decreasing trend in 2000. 
In Fig.\ref{lc_sgrastar} we highlight the drop in 
the X-ray activity through green points.

Between  the  end  of the ``high state''  activity  and  the present time, Sgr A*
appears to have entered a low to intermediate state. Fig.\ref{lc_sgrastar} shows, in a sketch form, 
the behaviour inferred from the measurements of the clouds within 30 pc of Sgr A*.
As discussed in the previous section, this variability pattern is not
unequivocally determined, although the inferred range of the X-ray luminosity
is on  a sounder footing.
The pattern of variability depicted in Fig.\ref{lc_sgrastar}, does allow the possibility
that the regions F and B1 (together with those showing a constant Fe-K$\alpha$ line emission) might be responding
to the declining phase of the flare which energised Sgr B2, whereas the other
XRN might be illuminated by a more recent smaller outburst.

Further evidence for the long-term downwards trend in the Sgr A* light curve
is provided by the limits on the Fe-K$\alpha$ flux from the 50 km~s$^{-1}$ MC. 
Given the proximity of this cloud to Sgr A* ($\sim$ 10 pc)
\citet[][]{2010ApJ...714..732P}  calculate that the mean  X-ray luminosity
of Sgr  A* in the  past 60 years must have been lower  than 8$\times$10$^{35}$
erg~s$^{-1}$.  
For comparison and completeness we have also plotted in Fig.\ref{lc_sgrastar}
the typical level of X-ray emission from the SMBH
(i.e., $\sim$10$^{33}$ erg~s$^{-1}$) and the brightest flare ever detected
\citep[L$_{\rm{X}}$=3.5$\times$10$^{35}$ erg~s$^{-1}$, ][]{2003A&A...407L..17P}.

In summary, in the framework of the XRN/Sgr A* scenario, we 
have found that the X-ray light curve of Sgr A* shows a
clear decreasing trend over the last 150 years.
This is not a completely smooth variation, with evidence
of occasional flaring or, more precisely, for periods of relative brightening 
which last seeming longer than $\sim$ 5 years.  Given that the light-curve is not
monotonic, a much more comprehensive study of the Fe-K$\alpha$ properties
of the MC within the central $\sim 100$ pc will be needed to piece together 
the full recent accretion history of the SMBH at the Galactic Center.

\section{The CR contribution to the Fe fluorescence}\label{low_sb}
\subsection{Method}

We have measured the 6.4-keV line flux from two extended regions to the 
East and West (EDE  and EDW)  of Sgr  A*, in order  to quantify  the difference
between the  low surface brightness emission at  positive and negative
galactic  longitudes.
For this  purpose we considered the two elliptical regions shown in 
Fig.\ref{diffuse}, the central coordinates and sizes of which can be
found in the  last two rows of Table \ref{capelli_reg}.
In the case of the EDE region, all the sky regions  coincident with the 
MCs previously studied have been excised, as have the 6.4-keV bright 
knots in the Arches cluster region  \citep[see][]{2011A&A...530A..38C},  
and  the Arches  cluster  itself  \cite[see][]{2011A&A...525L...2C}.

We have constructed MOS1\&2 spectra for both the remaining EDE area
and the EDW region by stacking all the data available.
Here we apply a background modelling technique in fitting the four
resultant spectra using the same spectral model and methodology as described  in
Section \ref{timing}.

\subsection{Results}

The results of this investigation are summarised in Table \ref{ede-edw-parameters}
and in Fig.\ref{diffuse-spectra} which shows the MOS2 spectra of the EDE 
and EDW regions and the corresponding best-fitting models. There are significant
residuals  near the high ionisation Si, S, Ar and  Ca
lines in the 2--4 keV region; these residuals may be  due to 
a range of temperatures and/or metallicities 
within the extended plasma which are not well matched by the simple
two-temperature model which has been applied. The Fe-K$\alpha$ line flux
is, however,  very  well determined in  all these spectra. 
The 6.4-keV line fluxes for the EDE and EDW regions are 3.44$\pm$0.06 and
1.91$\pm$0.06$\times$10$^{-4}$  photons~cm$^{-2}$~s$^{-1}$,  respectively.  The
average surface  brightness   calculated  from  these   values  are  therefore
1.71$\pm$0.03$\times$10$^{-6}$     and    7.8$\pm$0.3$\times$10$^{-7}$
photons~cm$^{-2}$~s$^{-1}$~arcmin$^{-2}$,  respectively. 
%[RENZO - JUST TO CHECK -
%DOES THE FORMER ALLOW FOR THE REDUCED AREA OF THE EDE REGION? - THE AREAS ARE
%THE SAME IN TABLE 8 - BUT MAYBE THIS DOESN't MATTER?]  
The substantial higher level of the \textit{underlying surface brightness}
to the east of Sgr A*, compared to the situation to the west, is
also apparent in Fig.\ref{diffuse}.

%-----------------------------Figure Start------------------------------
\begin{figure}[!Ht]
\begin{center}
% un-comment the following line to include your fig1a.eps postscript file
\includegraphics[width=0.45\textwidth]{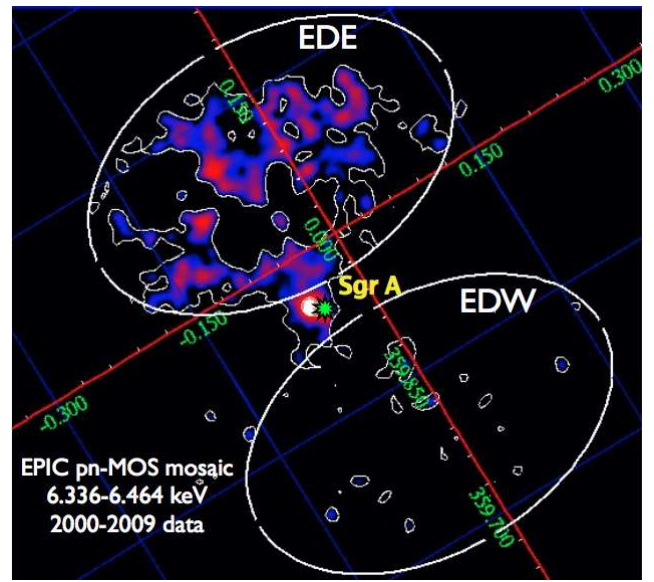}
\end{center}
\caption{A colour-coded and contoured map of the 6.4-keV line emission in GC region. 
The contours levels  are scaled in arbitrary units.
The position of Sgr A* is marked with the green star in the centre of the image, 
whereas the bright spot close to it is the Sgr A EAST SNR complex. 
The regions selected for the spectral analysis lie within the two ellipses marked as  
EDE and EDW. However, the bright MCs studied in this paper and in \citet[][]{2011A&A...530A..38C},
which are all located in the EDE region, have been blanked out.}
\label{diffuse}
\end{figure}
%-----------------------------Figure End--------------------------------

%*****************************************************************************
\begin{table}[!Ht]
\caption {Results from the spectral analysis of the regions EDE and EDW. 
The quoted parameter values are the weighted means of those obtained from 
individual MOS1 and MOS2 spectra. The units of N$_{H}$ are 10$^{22}$cm$^{-2}$, 
the temperatures of the two plasmas are quoted in keV, the normalization of the non-thermal
component has units of photons~cm$^{-2}$~s$^{-1}$~keV$^{-1}$ at 1 keV and 
the Fe-K$\alpha$ line flux has units of photons~cm$^{-2}$~s$^{-1}$. }
\begin{center}
\begin{tabular}{|c|cc|}
\hline
 & EDE & EDW \\
\hline
N$_{H}$ & 7.5$\pm$0.1 & 9.4$\pm$0.1 \\
kT$_{warm}$ & 0.91 $\pm$0.01 & 0.356$\pm$0.002 \\
%norm & 0.25$\pm$0.01 & 3.7$\pm$0.1 \\
kT$_{hot}$ & 6.4$\pm$0.1 & 6.1$\pm$0.2 \\
%norm & 3.11$\pm$0.01 & 2.65$\pm$0.01 \\
$\Gamma$ & 0.78$\pm$0.04 & 1.0$\pm$0.1 \\
norm & 5.5$\pm$0.1$\times$10$^{-3}$ & 1.61$\pm$0.01$\times$10$^{-2}$ \\
F$\rm{_{64}}$ & 3.44$\pm$0.06$\times$10$^{-4}$ & 1.91$\pm$0.06$\times$10$^{-4}$\\
EW & 0.37$\pm$0.02 & 0.077$\pm$0.002 \\
\hline
\end{tabular}
\end{center}
\label{ede-edw-parameters}
\end{table}
%**************************************************************************

%-----------------------------Figure Start------------------------------
\begin{figure*}[!Ht]
\begin{center}
% un-comment the following line to include your fig1a.eps postscript file
\includegraphics[width=0.35\textwidth,angle=-90]{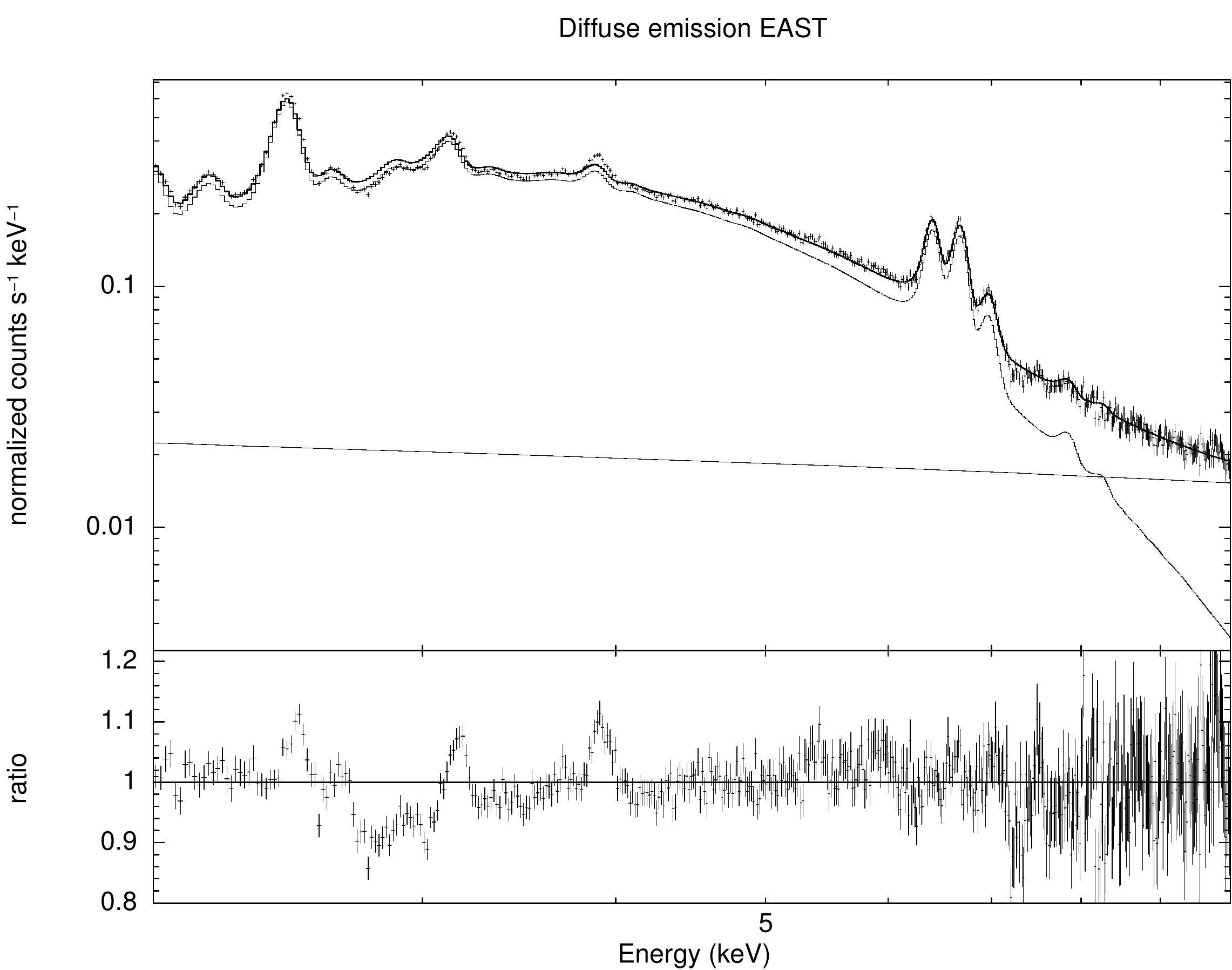}
\includegraphics[width=0.35\textwidth,angle=-90]{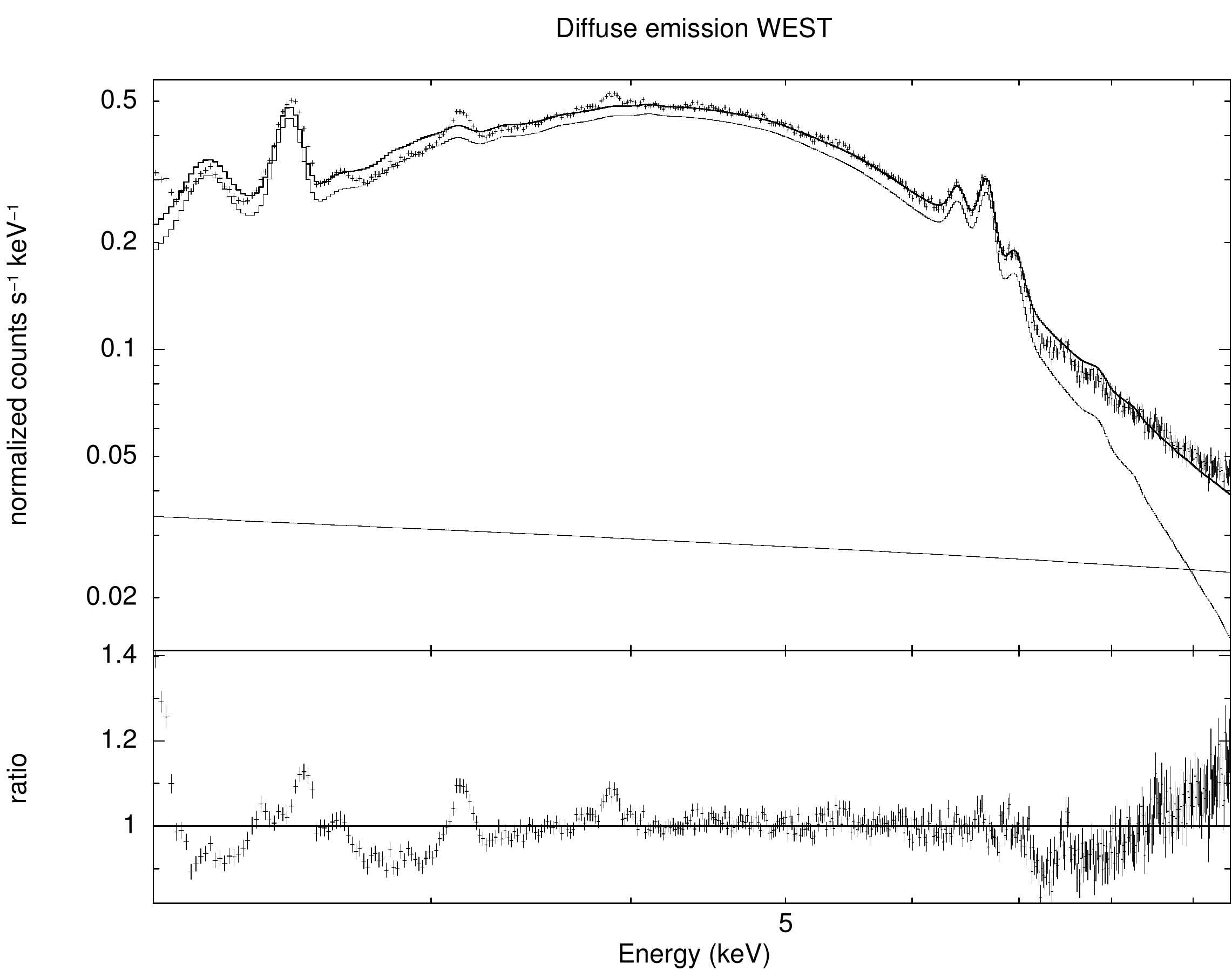}
\end{center}
\caption{$\textit{Left  panel}$: MOS2 spectrum of the EDE region, with the residuals 
in the bottom panel. The data have been collected stacking all the MOS2 spectra of
the region from all the observations. $\textit{Right  panel}$: same for the EDW region.}
\label{diffuse-spectra}
\end{figure*}
%-----------------------------Figure End--------------------------------

The fact that the underlying diffuse Fe-K$\alpha$ emission at positive Galactic
longitudes is roughly a factor two more intense than that measured at the corresponding negative
longitudes is an interesting result. In particular
it can be directly compared with the very-high energy $\gamma$-ray map of
the GC region obtained with the HESS telescope
\citep[][]{2006Natur.439..695A}. The TeV emission shows a strong enhancement to the East
of Sgr A*, with a peak close to the centre of our EDE region
\citep[see also][]{2007ApJ...665L.123Y}.

The TeV emission most likely arises from the hadronic interactions of
CR with  the molecular  material in the CMZ, which prompts the question
as to whether the underlying level of the 6.4-keV emission to
the east of the GC might similarly be the result of CR bombardment 
of the MCs (albeit by CR particles of very different energy to those
giving rise to the TeV signal). A further step is then to conjecture
that the non-varying emission pedestals seen in some of the
6.4-keV bright MCs might also be the result of the same
CR bombardment process.
Consider, in this context, the B2 and D regions (see
Fig.\ref{fluence}), which apparently compose the
region referred to by \citet[][]{2010ApJ...714..732P} as
the \textit{Bridge}.
The first three data points (covering the period 2002-2004) in their
6.4-keV light curves (see Table \ref{fluxes_final}
and Fig.\ref{pmulti_capelli}) are consistent with constant
levels of 1.1$\pm$0.1 and 0.8$\pm$0.1$\times$10$^{-5}$
photons~cm$^{-2}$~s$^{-1}$ respectively. These measurements translate to 
surface brightnesses of 5.2$\pm$0.5 and 4.5$\pm$0.6$\times$10$^{-6}$
photons~cm$^{-2}$~s$^{-1}$~arcmin$^{-2}$, which are
compatible with each other, but higher
than the mean 6.4-keV diffuse surface brightness calculated
for regions EDE and EDW.
If one then compares the 6.4-keV surface brightness of the B2 and D pedestals,
the EDE region (after excluding bright MCs) and the EDW region,
the ratio 6.4/2.2/1.0 is obtained. In a similar way, if we look
at the contours of the TeV emission map
\citep[fig.12a in][]{2007ApJ...665L.123Y}, we can see that there is a remarkable
agreement with this 6.4-keV surface brightness scaling.
The brightest TeV emission is detected in a region roughly matching the physical extent of the B2 
and D clouds \citep[innermost contour of Fig.12a in][]{2007ApJ...665L.123Y}, whereas the EDE region is 
well delimited by the third contour of the TeV surface brightness distribution, and the EDW region by the 
next two (lower) contour levels.

%*****************************************************************************
\begin{table}[!Ht]
\caption {Physical parameters for the MCs and other regions in the CMZ region which display
Fe-K$\alpha$ emission. The table lists the cloud column density 
(in units of 10$^{22}$ cm$^{-2}$), the Fe-K$\alpha$ line flux (in units of
10$^{-5}$ ph cm$^{-2}$ s$^{-1}$), and the sky area subtended by the MC or feature
(in arcmin$^{2}$).}
\begin{center}
\begin{tabular}{|cccc|}
\hline
Cloud & N$\rm{_{H_{C}}}$ & F$\rm{_{64}}$ & Area\\
\hline
\hline
A & 18.4$^{+1.4}_{-2.6}$ & 2.5$\pm$0.1 & 2.2 \\
B1 & 10.2$^{+2.3}_{-2.0}$ & 2.3$\pm$0.3 & 2.6 \\
B2 & 12.3$^{+3.0}_{-2.7}$ & 2.1$\pm$0.1 & 1.9 \\
C & 5.8$^{+2.3}_{-1.9}$ & 1.2$\pm$0.2 & 2.0 \\
D & 13.2$^{+5.0}_{-4.8}$ & 1.2$\pm$0.2 & 1.8 \\
E & 9.6$^{+1.7}_{-1.3}$ & 2.0$\pm$0.1 & 2.9 \\
F & 9.2$^{+2.7}_{-2.3}$ & 13.0$\pm$1.0 & 14.6 \\
DS1 & 15.5$^{+3.9}_{-3.3}$ & 1.5$\pm$0.1 & 3.7 \\
DS2 & 14.5$\pm$2.3 & 2.6$^{+0.2}_{-0.1}$ & 6.5 \\
EDE & 1.5 & 34.0$\pm$1.0 & 202.9 \\
EDW & 3.4 & 19.0$\pm$1.0 & 243.9 \\
\hline
\hline
N & 4.0 & 0.39$\pm$0.04 & 0.5 \\
S & 2.0 & 0.69$\pm$0.05 & 0.5 \\
SN & 2.0 & 0.46$\pm$0.04 & 1.0 \\
DX$_{CR}$ & 2.0 & 0.16$\pm$0.05 & 0.8 \\
DX$_{X}$ & 2.0 & 0.47$\pm$0.09 & 0.8 \\
\hline
\hline
Sgr B2$^{\dag}$ & 58.0$\pm$3.0 & 10.0$\pm$1.0 & 4.5 \\
50 km/s$^{\ddag}$ & 9.0 & $\leq$1.7 & 9.2 \\
G0.162$^{\S}$ & 2.0 & 0.2$\pm$0.1 & 3.1 \\
G0.174$^{\S}$ & 2.0 & 0.7$\pm$0.1 & 4.5 \\
G359.43$^{\dag\dag}$ & 10.0 & 0.6$\pm$0.1 & 7.1 \\
G359.47$^{\dag\dag}$ & 10.0 & 0.9$\pm$0.1 & 6.9 \\
\hline
\end{tabular}
\tablefoot{$\dag$ \citet[][]{2010ApJ...719..143T}. $\ddag$ \citet[][]{2010ApJ...714..732P}.
$\S$ \citet[][]{2009PASJ...61..593F}. $\dag\dag$ \citet[][]{2009PASJ...61S.233N}. 
The top section of the table is devoted to the MCs studied in this work, 
whereas the regions in the second panel are those in the vicinity of the Arches cluster
considered by \citet[][]{2011A&A...530A..38C}. 
Parameters for the clouds in the lower section are taken from the literature
(see the notes below). In the case of the MCs with a variable Fe-K$\alpha$ line flux, we list the
minimum value on record. Error ranges for the N$\rm{_{H_{C}}}$ values are not available in all 
the cases, specifically when the values are estimates or assumed values.}
\end{center}
\label{CR_table}
\end{table}
%**************************************************************************

We interpret the above as an indication that CR bombardment may well play a
role in exciting X-ray fluorescence in the GC region.  Specifically 
this process could potentially give rise
to the baseline levels (\textit{i.e.,} emission pedestals) noted in several of
clouds with variable 6.4-keV emission. In addition, for clouds lying
outside of the regions where the X-ray illumination is most intense,
CR bombardment might be the dominant source of
fluorescence excitation \citep[see also][]{2011A&A...530A..38C}.
Plausibly this might include some of the more diffuse structures
apparent in the Fe-K$\alpha$ maps.

Unfortunately, it is difficult to disentangle from X-ray studies alone, what contribution CR
bombardment makes to the total 6.4-keV line emission.
%an ideal approach would implement multiwavelength studies of the region targeted at the 
%measurement of CR densities, energetics, and interactions with the local ISM and photon field.
Nevertheless, we can investigate this issue by assuming (i)  that
CR electrons with energies in the range 10 keV - 1 GeV  are the cause of the underlying
X-ray fluorescence seen in the GC and that (ii) the concentration of these particles
is relatively uniform across the region of the CMZ studied in this paper.
Table \ref{CR_table} summarises the physical parameters of all the GC clouds and structures which
are known to be Fe-K$\alpha$ emitters plus some upper limits for several other prominent
clouds. For the variable clouds, we have considered the minimum 6.4-keV line flux recorded to date; for the Sgr B2 cloud we 
considered the Fe-K$\alpha$ flux value as extrapolated for the year 2010 from Fig.4 in \citet[][]{2010ApJ...719..143T}.

%-----------------------------Figure Start------------------------------
\begin{figure}[!Ht]
\begin{center}
% un-comment the following line to include your fig1a.eps postscript file
\includegraphics[width=0.5\textwidth]{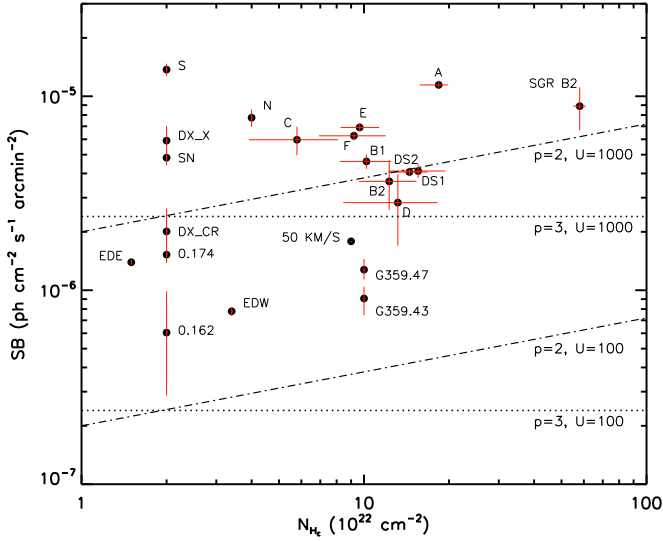}
\end{center}
\caption{Surface brightness of the Fe-K$\alpha$ line measured from the MCs in the CMZ, plotted against their
estimated column density, N$\rm{_{H_{C}}}$. The labels indicate the identity of each cloud, 
while the lines represent model predictions for CR electron densities of 100 and 1000 eV/cm$^{3}$
(lower and upper lines respectively),
for assumed slopes of the electron energy spectrum of p=2 (dotted lines) or p=3 (dashed lines).
\textit{CR contribution band} as described in the text.}
\label{cr_contribution}
\end{figure}
%-----------------------------Figure End--------------------------------

In Fig.\ref{cr_contribution}, we have plot the Fe-K$\alpha$ surface brightness of each cloud
(in photons cm$^{-2}$ s$^{-1}$ arcmin$^{-2}$) versus the estimated column density.
The contribution of CR electrons to the observed Fe-K$\alpha$ surface brightness of various
GC clouds has been discussed by \citet[][]{2007ApJ...656..847Y}. Using the same methodology
as \citet[][]{2007ApJ...656..847Y}
(in particular their Fig. 13 and equation 10), we obtain the estimates for the CR contribution shown in  
Fig.\ref{cr_contribution} as straight lines; here, U and p represent the energy density of the LECRe population
and the photon index of the power law energy distribution of the injected electrons (E$^{-p}$), respectively.
Fig.\ref{cr_contribution} has three main regions. The clouds with
column densities higher than $\sim$5$\times$10$^{22}$ cm$^{-2}$ and Fe-line intensities brighter than 
$\sim$5$\times$10$^{-6}$ photons cm$^{-2}$ s$^{-1}$ arcmin$^{-2}$, such as Sgr B2 and the
bright nebulae studied in the present work, are very likely excited by X-ray illumination,
\textit{i.e.,} are \textit{bona fide} XRN. At similar brightness levels, 
but somewhat lower N$\rm{_{H_{C}}}$, lie the MCs in the vicinity of the Arches cluster;
as we have shown in our previous study \citep[][]{2011A&A...530A..38C}, the Fe fluorescence within these 
clouds is most likely energised by particles emanating from the Arches cluster itself
(CR electrons or protons). The third region, lying between the predictions for 
CR electron energy densities of 100-1000 eV/cm$^{3}$, could also plausibly corresponded
to clouds primarily excited by CR bombardment.

A remarkable result shown by the plot in Fig.\ref{cr_contribution} is that, when considering the constant (minimum) 
base level of the variable Fe-K$\alpha$ line emission in regions B2 and D, this falls into the CR contribution region delimited 
by the straight lines in Fig.\ref{cr_contribution}; this strongly suggest that a CR component might be present in all 
lightcurves studied in this work, and it is of a strategic importance to carry out studies on the CR population in the 
innermost CMZ in order to 
definitely disentangle the contribution due to the two sources (CR and Sgr A*), in primis for better constraining the past
activity of the SMBH.

Due to the high uncertainties in the parameters plotted in this Figure, we do not want to rely too much on it in 
deriving all properties of the MCs we considered, nor putting a net separation between the two excitation 
processes (which could be present in tandem in some of the clouds). However, we think that some 
conclusions about the CR contribution to the Fe fluorescence observed in GC MCs can be drawn already at this stage
(see Fig.\ref{cr_contribution}): 

\begin{itemize}
\item{We can safely conclude that the 6.4-keV line emission measured in the A to F complexes is far too high to be only due to 
CR bombardment. Detections of the line variability and the absorption feature at the Fe-K edge give further 
confirmation of the XRN nature of these MCs, although a component to the Fe fluorescent seems to be present.}
\item{The EDE and EDW diffuse regions are most likely 
to be illuminated by LECRs, where the surface brightness gap may reflect the averaged dishomogeneity in the CR 
distribution.} 
\item{The Sgr C MC is very different 
from Sgr B2 and the XRN in the immediate vicinity of Sgr A*; the Fe-K$\alpha$ surface brighteness is about a factor 
20 lower than what measured in Sgr B2, and a factor of 10 lower than what measured in the A-F clouds. Assuming
the same illumination scenario as for the Sgr B2 cloud, we calculate that the total distance of the Sgr C clumps
G359-43 and G359.47 should be as high as 500 to 700 pc, which seems very unlikely. Moreover, all the X-ray
observations performed on Sgr C measured the same SB of the 6.4-keV line over about 10 years; we therefore 
propose that the description of the Fe fluorescence in the Sgr C cloud as due to LECRs bombardment as a likely possibility.}
\item{The complexes G0.162 and G0.174 well lie within the \textit{CR band}; as already proposed by 
\citet[][]{2009PASJ...61..593F}, the 6.4-keV line in the former is likely to be due to CR bombardment, 
given the relative vicinity to the radio arc. On the other hand, these autors argued that G0.174 is an XRN, 
whose origin is still not confirmed. For what concerns the plot in Fig.\ref{cr_contribution} we can see 
that while G0.162 is well within the \textit{CR band}, G0.174's surface brightness is significantly higher and other 
studies should be performed in order to prove the XRN nature of this MC. An Fe-K$\alpha$ EW of 0.9$\pm$0.2 keV
is not enough to disprove either hypothesis on the Fe fluorescence mechanism.}
\end{itemize}

Of course, the starting premise of a constant CR concentration is far from being correct, 
but a plot as the one shown in Fig.\ref{cr_contribution}, if nothing else, 
has the merit of illustrating the order of magnitude spread in surface brightness across all the GC clouds 
bright in the 6.4-keV line.

\section{Discussion}\label{discussion}

\subsection{X-ray Reflection Nebulae}

The XRN concept was    first   suggested   by   \citet[][]{1993ApJ...407..606S}   and
later developed by \citet[][]{1996PASJ...48..249K}, \citet[][]{1998MNRAS.297.1279S} and
\citet[][]{2000ApJ...534..283M}. In this scenario,  a powerful transient X-ray source
supplies  the primary  photons  needed  to produce  the fluorescence observed from 
the GC molecular clouds. The  whole phenomenology, both the line  flux and the topology
of the Fe-K$\alpha$  line emission, of the Sgr B2  giant molecular  cloud is  well
explained in this scenario.  In this context Sgr A* must have  been bright roughly
150 years ago (based on the projected separation of Sgr B2 and Sgr A*)
reaching an X-ray  luminosity of 2-5$\times$10$^{39}$ erg~s$^{-1}$ 
\citep[][]{2010ApJ...719..143T}.  This is about
10$^{5}$ times lower  than the Eddington luminosity for a SMBH  with the mass of
Sgr A*, and so is entirely plausible.

If Sgr A* is indeed the  primary source of the ionising photons for Sgr B2 then, 
by implication, the  Fe-K$\alpha$  emission observed in the more immediate 
vicinity of Sgr A* might similarly serve as a tracer of past outbursts in this 
source. \citet{2010ApJ...714..732P} have reported the complex nature of
the variability  seen in  the 6.4-keV bright filaments with 30 pc of Sgr A* and have
suggested a scenario in which these clouds are illuminated by the same outburst
responsible for the Sgr B2 Fe fluorescence.

This picture has to date been the favoured explanation of the 
6.4-keV line variability seen in the GC MCs. 
In broad terms our study of the MCs in the immediate vicinity of Sgr A*
serves also as a confirmation of many of the observational
details reported by \citet[][]{2010ApJ...714..732P}. However, our conclusion
is quite different, namely that the luminosity required to ionise the MCs 
studied in this paper must be at least one order of magnitude lower than
the outburst which is currently illuminating Sgr B2. Most importantly, the derived X-ray 
lightcurve of Sgr A* over the last $\sim$150 years appears to exhibit
occasional episodes of brightening, which last at least 5 years,  
superimposed on a general decreasing trend.

The main breakthrough that sets our paper apart from earlier
contributions on this topic is the direct measurement of the 
column density and the Fe abundance of the MCs in our sample, 
which are the two fundamental parameters for the study of the XRN behaviour.
Using this approach we were able to study the 3D distribution
of the MCs in the innermost CMZ in two independent ways,
which returned consistent results and therefore reinforced
our findings.

First, in Section \ref{metallicity} we used our measurements of the EW of the
6.4-keV line in order to  calculate, within a theoretical framework, the
average Fe abundance within the selected MCs. Our results show that the
metallicity of  the  MCs  in our study is  supersolar with
Z$\rm{_{Fe}}$=1.6$\pm$0.1 Z$_{\odot}$, a value which is in very good agreement
with previous estimates \citep[e.g.][]{2010PASJ...62..423N,2011ApJ...739L..52N}.
Secondly, in Section \ref{NH_calculation} we determine the column densities of the
MCs in our sample directly from the fitting of their X-ray spectra.
Our X-ray derived  N$\rm{_{H_{C}}}$ values are consistently higher than those
estimated from CO-CS measurements. Of course, measurements based on molecular lines
require the assumption of a thermal temperature for the molecular gas which 
is not known a priori, and do not take into account abundance variations
in molecules such as CS. Another possible way to estimate the N$\rm{_{H_{C}}}$ of a
molecular cloud is the comparison with mm continuum maps of dust emission;
however, here one meets severe difficulties in picking out relatively
small clouds against the general continuum brightness of the GC region.
Also the gas-to-dust ratio in the GC environment may differ
from that found in other regions of the Galaxy.

Both the high value of the EW of the 6.4 keV Fe-K$\alpha$ line and 
the strong suppression of the non-thermal X-ray continuum at the
Fe-K edge (at 7.1 keV) supports the view that the brightest
clouds in our sample are true XRN.  Our determination of the spectral slope
of the ionising continuum to be $\Gamma \approx$ 1.9 also matches the
likely spectral form of the continuum X-ray emission emanating from Sgr A*,
which illuminates the Sgr B2 cloud \citep[e.g.][]{2004A&A...425L..49R}.

In the context of the activity of Sgr A* in the recent past, we have two important new results: 

\begin{itemize}
\item{In a pure reflection scenario where Sgr A* supplies the primary photons at a constant
luminosity, one might expect the 6.4-keV surface brightness of the densest clouds
in the inner 30 pc region to be between 20-50 times higher that of Sgr B2,
commensurate with the inverse square fall-off in the illuminating intensity.
In fact there is no evidence for such an effect; on the contrary none of
the central region clouds have 6.4-keV surface brightness values in excess
of that on Sgr B2 (see Fig. 9).  More quantitatively, we show that the Fe
fluorescence of the MCs in our sample can be explained in terms of the X-ray
luminosity L$_{X}$ of Sgr A* reaching levels of between 10$^{37}$ and
10$^{38}$ erg~s$^{-1}$ over the last 150 years.  This is more than an order of magnitude
lower than that required to explain the observed fluorescence of the Sgr B2 cloud.}

 \item{Within the last 150 years the X-ray activity of the primary source can be characterised
as a declining trend upon which occasional episodes of brightening, lasting at least
5-years, are superimposed.  A scenario in which the observed pattern of activity might arise
is suggested by the very recent discovery of a dust cloud
on its way towards its likely final accretion onto Sgr A* \citep[][]{2012Natur.481...51G}.
Unfortunately, there is no detailed knowledge of the physical parameters of this cloud, 
and the X-ray luminosity which might ensue following this accretion event is hard
to predict with any certainty. Nevertheless, on timescales of 10's to 100's of years,
accretion events triggered by the capture of passing clouds would seem to be a very likely
occurrence, particularly since the amount of interstellar gas in the accretion flow onto
the SMBH at the GC is sufficient, in principle, to produce L$_{X}\sim$10$^{40}$ erg~s$^{-1}$
\citep[][]{2005MNRAS.360L..55C}.} 

\end{itemize}

\subsubsection{Association of the X-ray transient XMMU J174554.4-285456 with region C}\label{Cassociation}

The X-ray  transient source  XMMU J174554.4-285456 was  first detected
with   \textit{XMM-Newton}   during   an   outburst   on   2002   October   3rd
\citep[][]{2005A&A...430L...9P}.  Repeated   observations  of  the  GC
region performed with Chandra later  revealed this source to be bright
in June-July 2004
\cite[][]{2005ApJ...622L.113M,2006MNRAS.371...38W}. A  closer
inspection to  the \textit{XMM-Newton} dataset  employed in this  work revealed
the presence of  this source in the March 2004  data. So far, the
nature of this peculiar X-ray  binary has not yet been ascertained. The
ratio  between  the  X-ray  luminosity  at the  outburst  and  in  the
quiescence state  is about  10$^{4}$. For this  reason, we  decided to
investigate whether the  measured fluxes of the 6.4-keV  line from the
region C could be due (in  part) to the flaring activity of this X-ray
binary.

The observed 6.4-keV  line flux lightcurve from the  region C shows a
step  behaviour  (see  Fig.\ref{pmulti_capelli}).  Dividing the  six
epochs into two, we obtain weighted means of 0.8$\pm$0.1    and   1.2$\pm$0.1$\times$10$^{-5}$
photons~cm$^{-2}$~s$^{-1}$,  respectively.  A closer  inspection  to the  full
band X-ray image from observations 0202670501 and 0202670601 shows
the presence of  the X-ray  transient source XMMU  J174554.4-285456 within
the bounds of region C 
\citep[see Fig. 15,][]{2005A&A...430L...9P}.  The  best  spectral model  has  been
found to  be an absorbed  powerlaw, with a $\Gamma$=1.7$\pm$0.2  and a
very high column density (N$\rm{_{H_{C}}}$) of
14.1$^{+1.0}_{-1.4}\times$10$^{22}$cm$^{-2}$.  The flaring  X-ray luminosity  of this
source   was  1.5$\times$10$^{35}$erg~s$^{-1}$   in  the   0.5-10   keV  band
\citep[][]{2005A&A...430L...9P}.

%-----------------------------Figure Start------------------------------
\begin{figure}[!Ht]
\begin{center}
% un-comment the following line to include your fig1a.eps postscript file
\includegraphics[width=0.4\textwidth]{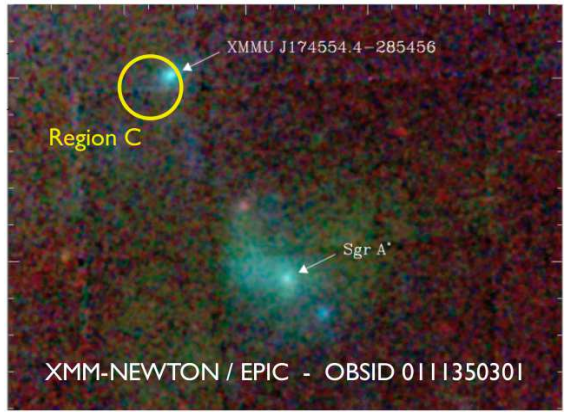}
\end{center}
\caption{XMM Newton / EPIC colour image of the X-ray transient 
XMMU J174554.4-285456 \citep[][]{2005A&A...430L...9P}. The colour coding is as follows:
red (0.2-2 keV), green (2-5 keV) and blue (5-10 keV). 
The yellow circle on the top of the image shows region C. }
\label{transient}
\end{figure}
%-----------------------------Figure End--------------------------------

Assuming the enhancement in the Fe-K$\alpha$ light curve
is due to the brightening of XMMU J174554.4-285456, we calculate
the requisite distance  of the binary to be of the order of 2 pc.
Given the high  N$\rm{_{H_{C}}}$ value measured
towards this source, we favour the geometry in which the MC lies in the
line of sight between us and XMMU J174554.4-285456. In this setting,
the measured time delay  between the  first reported flare on the binary 
(October 2002) and the onset of enhanced 6.4-keV flux (4.5 years), 
the size of the cloud (3.7 pc) and the distance of the X-ray source (2 pc)
represent a consistent picture.  

In summary the Fe fluorescence of region C may, in principle, be explained
without recourse to the XRN/Sgr A* paradigm.  Its base (or pedestal)
level of Fe-K$\alpha$ emission might arise from CR particle bombardment
(see Section \ref{low_sb}) upon which is superimposed a component  due to
X-ray  irradiation of the cloud by a (very) nearby transient source
XMMU J174554.4-285456. Potentially this is  the second XRN
positively associated with an X-ray  transient source in the
GC region \citep[][]{2011A&A...530A..38C}.

\subsection{CR Particle bombardment}

This     model    has     been    proposed     and     discussed    by
\citet[][]{2000ApJ...543..733V}, \citet[][]{2003AN....324...73P},
\citet[][]{2007ApJ...656..847Y}  and  \citet[][]{2009PASJ...61..901D}.  In  this
context,  the primary source  of the  Fe fluorescence  are CR particles,
potentially  with  a  variety  of  compositions,  origins  and
energies. While the photoionisation  cross-section is a steep function
of the energy of the primary photon, the cross-section for collisional
ionisation  is a  relatively  smooth  function of  the  energy of  the
incident    particle.  For electrons with energies in the range 10-100 keV and 
protons with energy between 10 and 100 MeV, the cross section for collisional 
ionisation of Fe remains relative close to its maximum value, i.e.
these are the energy ranges over which the particle bombardment most likely
occurs.  However, the number of particles will fall sharply with energy so one
might not expect the major contribution to be due to the highest energy
component.  
%[ELECTRONS WILL NOT PENETRATE VERY FAR INTO THE CLOUD, SO THE EFFICIENCY OF
%OF FE K ALPHA PRODUCTION IS QUITE COMPLICATED - LEAVE OUT THE NEXT FEW WORDS??]
%but rather to particles at approximately 30 keV (electrons) and 30 MeV (protons).
%[NB YUSEF-ZADEH APJ 656 p 865 - < 100 keV electrons contribute <20% of the Fe-K FLUX]

Although electrons have been historically favoured as the source of ionisation
of MCs in the CMZ, electrons with a kinetic energy of some tens of keV can barely penetrate
the target clouds since they are stopped by a column density 
of only $\sim 10^{21}$cm$^{-2}$. As a consequence, fluorescent emission via
electron bombardment most likely takes place within the surface layers of a dense cloud.
On the other hand, protons have a much larger penetrating power than electrons; 
a 30 MeV proton is stopped by a column density higher than 10$^{22}$cm$^{-2}$, 
a value approaching the column densities found in the GC region. 
Accordingly, protons may be as, or more, important than electrons as
a source of ionisation of MCs, although peculiar locations (the Radio Arc, 
the Arches cluster) might reveal different mechanisms at work. 

Very  recently,  \citet[][]{2011A&A...530A..38C}  showed that  in  the
Arches cluster region  there are MCs similar to  those studied in this
paper in terms of their Fe-K$\alpha$ surface brightness, size and inferred 
density. In fact, it is very likely 
that the bulk of the underlying fluorescence in these clouds is induced 
by particle bombardment, with an estimated CR
energy  density  of about  50-100  eV/cm$^{3}$,  that is  two orders  of
magnitude higher than that estimated for the Galactic plane in general
\citep[see][]{2011A&A...530A..38C};  the  Arches cluster is, of course,
the most likely site of the requisite particle acceleration. 

Our results confirm that a  major contribution 
to the Fe-K$\alpha$ line emission in the MCs in our sample comes from X-ray
photoionisation. 
However, we argue that a significant contribution from CR bombardment might be present in
the spectra of these XRN - as  discussed in  Section \ref{low_sb}.

To support this hypothesis, we presented in Fig.\ref{cr_contribution} possible evidence for the underlying 
particle ionisation component. In a plot of the Fe-K$\alpha$ surface brightnesses against the 
cloud column density of the MCs, we were able to distinguish different regions of parameter space
in which X-ray illumination and CR particle bombardment might operate as the primary source 
of the Fe fluorescence.  However, much more information is required in relation
to the CR energy density and spatial distribution in the innermost CMZ is required
if we are to understand the relative importance of the various excitation
processses. 

The Fe-K$\alpha$ line emission from particle bombardment is generally expected to show
a constant lightcurve.  However, it is at least conceivable that time variable magnetic fields
in the filaments and threads and/or time variable CR fluxes might induce measureable
changes over timescales upwards of a few years in some compact molecular structures.
Evidence for such behaviour could come from variability studies of the radio continuum
in narrow radio emitting filaments confined by magnetic pressure, although of course in
such cases the particles will be electrons with relativistic energies
emitting synchrotron radiation.

\cite{2010arXiv1002.1526O} discovered high  velocity compact clouds in
the GC region which show  a CO J=3-2/J=1-0 intensity ratio higher than
that typically found  in the Galactic disc. These  features are likely
to be knots of shocked  molecular gas, heated by energetic events such
as supernovae explosions. The  distribution of these clumpy structures
correlates well with  the location of the molecular  filaments we have
studied in this  work. The same is true for the  Sgr C molecular cloud
\cite[][]{2010arXiv1002.1526O}.   This  correlation  suggests a  close
relation   between   these  shocked   molecular   complexes  and   the
Fe-K$\alpha$ line emission, since shocks internal to the clouds act
as   particles  accelerators.    Moreover,  \cite{2008ApJ...673..251M}
showed that there are many  physical processes active in the GC region
which  appear to produce  diffuse X-ray  features. For  example, these
authors discovered  $\approx$ 20 candidate pulsar  wind nebulae within
20 pc  of Sgr A*,  which are among  the best established sites  of CR
acceleration.    The  particle   bombardment  hypothesis
places  few direct constraints on the cloud/particle flow geometry,
given that particle will travel along the local field lines.
Nevertheless this scenario does provide at least a plausible
explanation of why the brightest Fe fluorescence regions are sometimes located
on the farside of the cloud as viewed from Sgr  A*
(for example, in Sgr C (\citealt{2007ApJ...656..847Y}).

\section{Conclusions}\label{Sect:conclusions}

We have studied the X-ray properties of selected molecular clouds and filaments to the
east of Sgr A* at a projected distance of $\approx$8-30 pc. 

\begin{itemize}

\item{\textit{Fe-K$\alpha$ topology}: we have demonstrated that the Fe-K$\alpha$ emission
delineates both compact and more diffuse structures. 
We have studied the temporal and spectral
properties of 9 clouds; three of which are newly studied (regions C, DS1 and DS2).} 

\item{\textit{Fe-K$\alpha$ variability}: significant  variability  is  seen  from
some of the molecular  clouds, although the pattern is by no means a  simple  one.  
Our Fe-K$\alpha$  flux and variability measurements
agree reasonably well with previous published results, although in the case of one cloud (B1)
we measured a decrease of the Fe-K$\alpha$ line flux, in contrast with what found previously.} 

\item{\textit{Fe-K$\alpha$ surface brightness}: we measure the surface  brightness of the
6.4-keV line to  be of the same order of magnitude in all the molecular filaments we have studied. 
Typically the observed variability only involves a factor of two change. In the Sgr A*/XRN scenario, 
the surface brightness of the filaments might  be expected to decline with distance from Sgr A*.
This issue is resolved by our finding that the X-ray luminosity of Sgr A* has declined substantial
over the last 150 years.}

\item{\textit{Fe-K$\alpha$ EW and Fe-K edge measurements}: we measured a high value of the
EW of the 6.4-keV line from all the MCs in our sample, consistent with an origin of
the bulk of the fluorescence in photoionisation. Moreover, all the clouds show an absorption feature
at the Fe-K edge energy of 7.1 keV.} 

\item{\textit{Spectral hardness of the reflected continuum}: the spectral shape of the reflected/ionising
X-ray continuum emission associated with the Fe fluorescence is found to be $\Gamma \approx$1.9, a result
compatible with the XRN/Sgr A* scenario, since this is a value rather typical of that
found in AGN.}

\item{\textit{Fe abundance}: we use the measured EW values of the Fe-K$\alpha$ line
in our sample of MCs to calculate the average Fe abundance (relative to solar). 
We show that Z$\rm{_{Fe}}$=1.6, a result which is consistent with the general finding that a
higher than solar metallicity characterises all the ISM phases in the GC region.}

\item{\textit{Column density through the MCs}: we have employed a relatively direct method to calculate
the hydrogen column density, N$\rm{_{H_{C}}}$, within the MCs. This is based on the joint modelling of the low-energy
absorption and the absorption at the Fe-K edge imprinted on  the incident continuum in its
passage through the cloud (both up to and after the point of scattering).  This approach benefits
from prior knowledge of the relatively iron abundance, Z$\rm{_{Fe}}$, in the cloud.
The column densities so calculated are close to 10$^{23}$ cm$^{-2}$ and, in most cases, significantly higher 
than those inferred in previous studies, which made use of radio molecular line measurements.} 

\item{\textit{The past X-ray activity of Sgr A*}: based on our study of the 6.4-keV line emission properties of the MCs 
in the inner 30 pc of the GC region, we outline a model for the X-ray emission of Sgr A* 
encompassing the last $\sim$150 years. Over this period the X-ray luminosity has declined from an apparent peak
of $\sim$10$^{39}$ erg~s$^{-1}$ roughly 150 years ago, to 10$^{37-38}$ erg~s$^{-1}$ perhaps 100 years ago,
down to typically 10$^{33-35}$ erg~s$^{-1}$ at the present time.  
This decreasing long-term trend has, however, been punctuated by counter-trend episodes of brightening by factors
of a few over timescales in excess of $\sim$ 5 years.}

\item{\textit{Cloud C - an XRN energised by a transient source?}: we have found that the 6.4-keV line
flux variability measured from cloud C could also be associated with the transient X-ray source XMMU J174554.4-285456. 
In this scenario, the Fe fluorescence in this complex is composed of a steady non-zero level, possibly 
produced by the interaction of CR with the ambient gas plus a superimposed contribution due to the
localised variable X-ray irradiation. If this picture is confirmed, this is the second XRN found in the GC 
region which has been illuminated  by a transient X-ray source \citep[][]{2011A&A...530A..38C}.}

\item{\textit{Low surface brightness 6.4-keV line emission}: we have found a good 
correlation between the  TeV emission and the Fe-K$\alpha$ line emission observed on arcminute 
scales in the inner CMZ. Specifically we measured diffuse, low surface brightness emission to the east of Sgr A*, 
which seems to permeate the whole region between the Sgr A* and the giant Radio Arc. 
A pedestal (\textit{i.e.} constant) fluorescence component might in fact be present in  all the lightcurves which show a 
varying Fe-K$\alpha$ line flux. We have also shown that this diffuse low surface 
brightness 6.4-keV line emission correlates quite well with the contours  of the brightest
features  seen in TeV $\gamma$-rays. This suggests that CRs may contribute to
the ionisation and fluorescence of cold gas in this region.}

\item{\textit{CR contribution}: we presented arguments in favour of a CR contribution to the total Fe
fluorescence in the GC MCs. In particular, LECR may be responsible for the constant of 6.4-keV line emission observed 
in several structures (including the Bridge), as well as in other regions of the CMZ, like the Arches cluster, 
the Radio Arc and Sgr C. We suggest that further study
of the Fe-K$\alpha$ surface brightness versus N$\rm{_{H_{C}}}$ relation for GC 
clouds will help reveal the contribution of CR to the Fe-K fluorescence observed in the GC region.}

\end{itemize}

\noindent We are far from unveiling the full mystery of the Fe  fluorescence 
observed in the GC region. It will be of  
great importance in the future to continue to monitor the Fe-K$\alpha$ line emission from the MCs 
in the CMZ 
so as to better characterise the past activity of Sgr A*. Future X-ray observations can also
complement  studies in different energy  domains, from  the radio band (continuum  and  lines)
up to the TeV  regime,  which aim to fully chart the influence of CR particles on the GC environment.

\begin{acknowledgements}

\noindent This work is based on observations performed with the \textit{XMM-Newton} satellite, 
an ESA mission with contributions funded by ESA Member Countries and NASA. We thank the referee for useful comments and suggestions.
R.C. would like to thank all people involved in this work. 
\end{acknowledgements}

% =============================================================================
%\bibliographystyle{aa}
%\bibliography{paper}

% =============================================================================

\begin{appendix}

\section{Dealing with the EPIC detector background}\label{background_appendix}

Within an  \textit{XMM-Newton} EPIC dataset, two
main background components  can be defined, namely  particle-induced events
and photon-induced events. Solar-flare  particles and CRs are  
responsible for  the continuum element of the non X-ray
Background   (NXB),   as   well   as   for   the   strong   instrumental   lines
\cite[][]{2008A&A...486..359L}.  A further minor  component within  the 
particle-induced background is  due to Quiescent Soft Protons  (QSP). These are 
particles which enter the optics and are focused onto the detector; they are mostly filtered
during  the GTI selection,  but is  likely that  there is  a low-level residual
component in the screened event file.  In the case of the photon background from
the sky, the main components are the  hard X-ray emission along the plane of the
Galaxy (the  Galactic Ridge emission both  in the foreground and extending beyond
the GC), and also the extragalactic Cosmic  X-ray Background (CXB).  We did not  
take into account  the Galactic
Halo     since    its     emission     is    negligible     above    2.0     keV
\cite[][]{2008A&A...486..359L}.  The  difficulty of  carrying  out a  systematic
analysis with a careful characterization  of the background is that the particle-induced  
background, which is  the dominant  component in  the hard  X-ray band,
changes both  spatially on the detector  and in time throughout  the duration of
the mission. Given the focus of our investigation, there was a clear requirement
that all the datasets should be  analysed in a very systematic fashion. This led
us to consider the advantages and disadvantages of the standard methods of
dealing with the background in \textit{XMM-Newton} observations.

\vspace{0.1cm}

\noindent  $\textit{Local Background subtraction}$:  this procedure  consists of
selecting a region in the same image/dataset close to the source of interest; it
is certainly the best choice for the analysis of a bright point source. The fact
that we are dealing with extended sources, which are close to  each other, forces us
to select a  region well separated from the position of the filaments under study. 
This may, potentially, produce a large  discrepancy between the intensity of  
the instrumental lines
in the  source and in  the background spectra,  especially in the PN  camera 
(\textit{e.g.} at the Cu line  at 8.05 keV),  generating  strong  systematic  errors which are difficult  
to  deal
with.  As a  further difficulty,  the presence  of the  Galactic  Ridge Emission
throughout the whole FOV prevents us from selecting a region for the background
with a low surface brightness.

\vspace{0.1cm}

\noindent $\textit{Blank Sky  Fields}$: Blank sky files are  constructed using a
superposition   of  pointed  observations   of  regions   of  the   sky  without
contamination  by bright  sources  \cite[][]{2007A&A...464.1155C}. They  contain
most of  the background components listed  above, like the  particle induced NXB
and the CXB. To study an extended  source, we tried to cast the blank event file
onto the  sky at the position  of our observations  using the $\textit{skycast}$
script. In doing so,  we are  able to  select the  same detector  region  for the
production of  both the source and  the background spectra. This  method has the
great  advantage of  completely  avoiding the possibility of having  different
instrumental lines intensities, but the various components of the background can
all vary in different  ways over long time periods. So far,  the blank sky files
available are for observations  done before revolution $\approx$1200, that is
early 2007 (private  communication, A.  Read). We  tested the  public blank  
fields  in our spectral  analysis. While  for revolutions  earlier than  1200  the instrumental
background subtraction was feasible, the  spectra obtained with later data still
show  strong contamination, which is easily  noticeable in  the intense  
Cu K$\alpha$
line  at  8.05  keV in  the  PN  camera.  Since we are using \textit{XMM-Newton}
observations taken over a period  of $\approx$10 years, the blank sky event
files technique cannot be used for our purposes.

\vspace{0.1cm}

\noindent  $\textit{Background  modeling}$: this  technique  requires a  careful
characterization of all the background components.  The strong advantage of
modelling the background  rather than subtracting  it, is  that 
many of the systematics may thereby be avoided.     
\cite{2004A&A...419..837D}    and \cite{2008A&A...486..359L}  studied in  
detail  the properties  of the  particle-induced background  
for the  MOS cameras,  giving a spectral  shape to  both the
continuum  and the  line  emission. Modeling  the  background, instrumental  and
cosmic components we can, in principle, account for all the photons collected  
by the instrumentation and make every photon count.

\vspace{0.1cm} 

On the basis of the above, we decided  to follow the background modelling technique
for measuring the temporal variability of  the Fe-K$\alpha$ emission.
With  this choice we did not consider further the   EPIC-PN  data, but
worked  only   with  MOS  cameras.  Up to the present, a
self-consistent and  exhaustive characterization  of the background  for EPIC-PN
camera has not been developed, mainly because the fraction of out-of-time events 
in the PN CCDs is not negligible and the Out FOV region is much smaller 
compared to the
one  in the  MOS  detectors. Moreover,  MOS  cameras are  better suited 
to the study  of extended sources  because of the  lower level of the
instrumental  background (which is most likely a consequence of
the  different CCD  structure). However, for the analysis of the stacked spectra
the excellent hard response of the PN camera outweighed these considerations
and for this study we employed the PN spectra and local background
subtraction.

\end{appendix}

\end{document}